\documentclass[article]{aa} 

\usepackage{graphicx}
\usepackage{caption}
\usepackage{subcaption}
\usepackage{epstopdf}
\usepackage{amsmath}
\usepackage[english]{babel}
\usepackage{amssymb}
\usepackage{appendix}
\usepackage{textcomp}
\usepackage{multirow}
\bibpunct{(}{)}{;}{a}{}{,}
\usepackage{array}
\usepackage{ragged2e}
\usepackage{adjustbox}
\usepackage{txfonts}
\usepackage{xcolor}
\usepackage{float}
\usepackage{soul}
\usepackage[bookmarksnumbered=true]{hyperref} 
\hypersetup{
    colorlinks = true,
    linkcolor = blue,
    anchorcolor = blue,
    citecolor = blue,
    filecolor = blue,
    urlcolor = blue
    }

\begin{document}

   \title{Jetted narrow-line Seyfert 1 galaxies breaking the jet paradigm -- a comprehensive study of host galaxy morphologies} 
 \titlerunning{Jetted NLS1s breaking the jet paradigm}

   \author{I. Varglund
          \inst{1,2}
          \and
          E. Järvelä\inst{3}
          \and 
          A. Lähteenmäki\inst{1,2}
          \and 
          M. Berton\inst{4,5,1}
          \and
          S. Ciroi\inst{6}
          \and
          E. Congiu\inst{7}
          }

    \institute{Aalto University Metsähovi Radio Observatory, Metsähovintie 114, FI-02540 Kylmälä, Finland
         \and
             Aalto University Department of Electronics and Nanoengineering, P.O. Box 15500, FI-00076 AALTO, Finland
             \and
             European Space Agency (ESA), European Space Astronomy Centre (ESAC), Camino Bajo del Castillo s/n, 28692 Villanueva de la Cañada, Madrid, Spain
         \and
             European Southern Observatory (ESO), Alonso de C\'ordova 3107, Casilla 19, Santiago 19001, Chile
        \and
             Finnish Centre for Astronomy with ESO (FINCA), University of Turku, Vesilinnantie 5, FI-20014 University of Turku, Finland
             \and
            Dipartimento di Fisica e Astronomia "G. Galilei", Università di Padova, Vicolo dell’Osservatorio 3, 35122 Padova, Italy
        \and
            Departamento de Astronomía, Universidad de Chile, Camino del Observatorio 1515, Las Condes, Santiago, Chile;}
        
   \date{Received }

  \abstract
{Narrow-line Seyfert 1 (NLS1) galaxies are unevolved active galactic nuclei (AGN) that exist predominantly in spiral galaxies. However, mostly due to the small number of sources studied, it has been under debate whether also the hosts of jetted NLS1 galaxies, a particular subclass of these sources hosting a relativistic jet, are disk-like, or elliptical, as the hosts of more powerful jetted AGN. We studied the host morphologies of 14 NLS1 galaxies, 11 of which have been detected at 37~GHz indicating that these sources harbour relativistic jets. The \textit{J}- and \textit{Ks}-band data used in this study were obtained with the Nordic Optical Telescope (NOT). We performed the photometric decomposition of the host galaxy using the band that gave a better fit, and additionally, created colour maps of all sources that had both a \textit{J}- and a \textit{Ks}-band observation. We were able to successfully model 12 sources, nine of which most likely have disk-like morphology. Of the remaining sources, one source could possibly be hosted either in a disk-like or a dwarf galaxy, and in two cases the results are inconclusive. Only one of our sources shows clear signs of interaction, but the colour maps of most of our sources hint at ample dust in the nuclei, possibly indicating earlier minor mergers, that can go unnoticed due to the limited resolution of these observations. Our results further support disk-like galaxies as the predominant host type of jetted NLS1 galaxies. Most importantly, with the number of modelled hosts of jetted NLS1s now exceeding 50, with only a few elliptical hosts, it seems to be safe to conclude that also disk-like galaxies are able to launch and maintain relativistic jets, and that the traditional jet paradigm stating that only massive elliptical galaxies are capable of hosting relativistic jets is severely outdated. 
}


   \maketitle
%

\section{Introduction}
\label{sec:intro}

\begin{table*}
\caption[]{Basic properties of the sample and the observations.}
\centering
\scalebox{0.9}{
\begin{tabular}{l l l l l l l l l}
\hline\hline
Source                   & z      & Scale   & RA          & Dec         & Exp (\textit{J}) & Exp (\textit{Ks}) & Date*    & Seeing\\
                         &        & (kpc$/$\arcsec    ) & (J2000.0)   & (J2000.0)   & (s)   & (s)    &  (month/day)              & (\arcsec    )  \\ \hline
6dFGS gJ042021.7-053054 & 0.1991 & 3.166   & 04:20:21.77 & -05:30:54.36 & 2250     &   2250     & 02/07 & 0.7 \\      
6dFGS gJ044739.0-040330  & 0.0815 & 1.475   & 04:47:38.95 & -04:03:30.43 & 2250    & 2160       & 02/08-10 & 0.8 \\ 
FBQS J0744+5149  & 0.4600 & 5.647   & 07:44:02.28 & 51:49:17.53 & 5400     & 2970     & 02/09 & 0.8 \\ 
SDSS J080535.16+302201.7  & 0.5506 & 6.222   & 08:05:35.17 & 30:22:01.63 & 4050    &  3240    & 02/07-08 & 0.9 \\ 
6dFGS gJ084510.2-073205   & 0.1036 & 1.829   & 08:45:10.24 & -07:32:05.24 & 450   &  2250    & 02/08-11 & 0.9 \\ 
SDSS J084516.17+421129.9  & 0.5248 & 6.070  & 08:45:16.17 & 42:11:29.85 & 3600     & 5400   & 02/11 & 0.9 \\ 
SDSS J090113.23+465734.7  & 0.4300  & 5.429  & 09:01:13.23 & 46:57:34.69 & 3600    &  3240   & 02/10 & 0.9 \\ 
SDSS J093712.32+500852.1  & 0.2755  & 4.048  & 09:37:12.31 & 50:08:52.04 &  2700    &   2160  & 02/07 & 0.8 \\
SDSS J102906.68+555625.2  & 0.4510  & 5.583  & 10:29:06.68 & 55:56:25.22 &   N/A  & 3510    & 02/08 & 1.3 \\ 
SDSS J103123.73+423439.2  & 0.3764  & 5.003  & 10:31:23.73 & 42:34:38.68 & 2700    &   2970  & 02/11 & 0.9 \\ 
SDSS J122844.81+501751.2  & 0.2628  & 3.912  & 12:28:44.79 & 50:17:51.39 & 3150    & 2700    & 02/07 & 0.8 \\ 
SDSS J123220.11+495721.8  & 0.2620  & 3.902  & 12:32:20.07 & 49:57:21.62 & 3600    & 2970    & 02/07-10 & 0.8 \\
SDSS J125635.89+500852.4  & 0.2447  & 3.711  & 12:56:35.74 & 50:08:53.06 & 3600    & 3240    & 02/10 & 1.0 \\
SDSS J150916.18+613716.7  & 0.2014  & 3.195  & 15:09:16.18 & 61:37:16.74 & 3150    & 3240    & 02/10 & 0.9 \\   
\hline
\end{tabular}
}
\tablefoot{Columns: (1) Source name, (2) Redshift, (3) Scale at the redshift of the source, (4) Right ascension, (5) Declination, (6) Total exposure time in J-band, (7) Total exposure time in Ks-band, (8) Observation dates, (9) Average seeing during the observations measured from the images in the band used for modelling. \\ *All observations were carried out in 2020.}
\label{tab:sample}
\end{table*}

Narrow-line Seyfert 1 \citep[NLS1, ][]{1985osterbrock1} galaxies are a class of active galactic nuclei (AGN) principally identified from the characteristics of their optical spectra. By definition, the full width at half maximum (FWHM) of the H$\beta$ is less than 2000 km s$^{-1}$ \citep[]{1989goodrich1}. Narrow-line Seyfert 1 galaxies have faint [O~\textsc{III}] emission with respect to H$\beta$ ($\rm F([O~\textsc{III}]\lambda5007) / F(H\beta) < 3$, \citealt{1985osterbrock1}). In the optical spectra it is possible to often find Fe~\textsc{II} multiplets \citep[]{1992boroson}. Narrow-line Seyfert 1 galaxies can also have blueshifted line profiles, indicative of outflows, though mostly in high-ionisation lines \citep[e.g.,][]{2002zamanov,2005boroson, 2021berton2}. 

Narrow-line Seyfert 1 galaxies are generally believed to be powered by low-mass central supermassive black holes ($M_{\text{BH}} \sim 10^{6} M_{\sun}$ - $10^{8} M_{\sun}$; \citealp[e.g.,][]{2011peterson, 2015jarvela, 2016cracco1, 2018chen}) and therefore they are considered to be in an early evolutionary stage \citep{2000mathur}. Some studies suggest that the low black hole mass is due to an orientation effect \cite{2008decarli}. These studies claim that the pole-on orientation of the broad-line region (BLR) hinders the possibility of seeing Doppler broadening in the emission lines that come from the BLR. Furthermore, if the BLR is flattened, then we can only see the velocity component that points toward us, leading to the velocity vector being shortened from our point of view \citep[e.g.,][]{2008decarli}. However, reverberation mapping campaigns suggest that the black hole masses of NLS1 galaxies is intrinsically low \citep[e.g.,][]{2016wang, 2018du, 2019du1}. The Eddington ratio of NLS1 galaxies is high, between 0.1 and 1, possibly even higher \citep{1992boroson, 2018marziani, 2022tortosa}. 

Narrow-line Seyfert 1 galaxies are one of the three classes of AGN known to host beamed relativistic jets \citep[e.g.,][]{2006komossa1, 2006zhou,  2008yuan, 2011foschini1, 2015foschini1, 2017lahteenmaki1, 2018lahteenmaki}. The other two classes being flat-spectrum radio quasars (FSRQs) and BL Lacertae objects (BL Lacs) \citep[e.g.,][]{2009abdo2, 2010foschini}. Furthermore, NLS1 galaxies are also capable of emitting $\gamma$-rays, further proving the presence of relativistic jets \citep[e.g.,][]{2009abdo1, 2018paliya}. This contradicts the paradigm that only galaxies with the most massive supermassive black holes ($M_{\text{BH}} > 10^{8} M_{\sun}$) are capable of forming and maintaining fully evolved powerful relativistic jets \citep{2000laor1}. 

Of all NLS1 galaxies, only a very small percentage,  $\sim16\%$, are known to be detectable at radio frequencies \citep{2006komossa1}. Among this 16\%, $\sim$6$\%$ are radio-quiet ($R <$ 10) and $\sim$10$\%$ are radio-loud{\footnote{Radio-loudness, abbreviated $R$, in astronomy is defined as the ratio of the radio flux density at 5 GHz and the optical flux density in the B-band \cite{1989kellermann1}}} ($R > 10$) \citep{2006komossa1}. All other NLS1 galaxies have yet to be detected at radio frequencies, thus they are currently considered radio-silent. For some AGN, the definition of radio-silent can, however, change as there have been detections of some formerly radio-silent NLS1 galaxies at 37 GHz \citep{2018lahteenmaki}. A detection at 37~GHz strongly indicates the presence of relativistic jets. All but two of our sources, detected at 37~GHz, have multiple detections. With recurrent detections and strong variability, star formation is not a likely culprit of the radio emission. Furthermore, star formation is unable to produce radio emission at the detected flux levels at 37~GHz. The use of the radio-loudness parameter is questionable at best, since \textit{R} is often calculated using non-simultaneous data, thus discounting temporal variation. In other words, it is possible for a source to be deemed radio-quiet one day and radio-loud on another day. Furthermore, a relativistic jet with high inclination, with respect to the line of sight can be weak due to the lack of relativistic boosting, leading to a wrong classification. Finally, a population of relativistic radio jets, almost totally absorbed at low radio frequencies and thus deemed radio-silent/-quiet, has recently emerged \citep{2018jarvelaphd, 2021berton2}. These issues with the radio-loudness remain true even at high redshifts \citep{2021sbarrato}. At low frequencies ($<$ 10 GHz) a source can in some cases be classified as radio-loud even if it does not have relativistic jet, or as radio-silent even if it does host relativistic jets. In the former case, a non-jetted source could be deemed radio-loud due to the star formation at low radio frequencies being so significant \citep{2015caccianiga,2019ganci, 2020berton}. In the latter case, a source with a small-scale jet may be deemed radio-silent due to the jet being completely absorbed through either synchrotron self-absorption or free-free absorption, consequently hiding the radio emission \citep{2020berton}. Due to the ambiguity of the radio-loudness parameter, we prefer classification based on the physical properties of a source, and thus will use the terms non-jetted and jetted, the latter referring to sources that host relativistic jets.

The AGN and its host affect each other \citep[e.g.,][]{2010vandeven1, 2012fabian1, 2012povic1} through, for example, the regulation of the influx of gas from the host galaxy to the central black hole. The AGN has several feedback mechanisms through which it affects the host, such as jets, winds/outflows, and the ionising photons originating from the central engine. The AGN feedback can cause the star formation to either increase or decrease, depending on if the feedback is negative or positive \citep[e.g.,][]{2016carniani, 2018jarvela}. Furthermore, extending away from the host, the circumgalactic environments can alter both the nuclear activity and the properties of the galaxy. In dense environments mergers, major or minor, as well as close encounters are more likely to occur than in less populated environments. These interactions can cause a galaxy to undergo changes, for example, in the galaxy morphology and/or its star formation. There is currently no clear consensus on how much a merger or an interaction affects the AGN activity levels. There are studies supporting no connection between nuclear action and minor or major mergers \citep[e.g.,][]{2000corbin1, 2011cisternas1}, and studies indicating that mergers do play a role in the onset of nuclear activity, or the triggering of the jets \citep[e.g.,][]{2008urrutia1, 2008barth1}.

The majority of non-jetted NLS1 galaxies are found in spiral galaxies \citep{2003crenshaw}. Two of the most common features in NLS1 host galaxies are nuclear dust spirals, found in 83\% of NLS1 galaxies \citep{2006deo1}, and large-scale stellar bars, found in 65\% to 80\% of NLS1 galaxies \citep{2003crenshaw}. The majority of the nuclear dust spirals in NLS1 galaxies are of grand-design, meaning that they have two symmetric, long, spiral arms \citep{2006deo1}.  Furthermore, near the core, NLS1 galaxies are known to exhibit increased star formation \citep{2010sani1, 2022winkel}. 

Earlier host galaxy studies have focused on either large-sample studies or individual source analyses, with a majority of the earlier studies being on the individual side \citep{2001krongold, 2008anton, 2011orban, 2012mathur,2014leontavares, 2016kotilainen, 2017dammando,  2017olguiglesias,2018dammando, 2018jarvela, 2019berton, 2020olguiglesias, 2021hamilton}. The studies with large samples have had sample sizes varying from roughly 10 to roughly 30 NLS1 galaxies and they target mostly non-jetted NLS1 galaxies. All of the individual source analysis studies have been on jetted NLS1 galaxies, with $\sim70\%$ of the sources being interacting. Nearly all of the studied NLS1 galaxies have been identified as disk-like, with only a few cases of elliptical hosts \citep{2017dammando, 2018dammando}. Furthermore, a couple of large sample studies have taken a look at the possible connection between interaction and powerful relativistic jets. The previous studies highlight the possibility of interaction being a key driving factor of relativistic jets in NLS1 galaxies \citep[e.g.,][]{2018jarvela, 2020olguiglesias}. \citet{2020olguiglesias} speculate that in $\gamma$-ray-emitting NLS1 galaxies nuclear activity is caused by minor mergers, while the nuclear activity behind radio-loud non-$\gamma$ NLS1 galaxies is due to secular evolution. Furthermore, evolution of NLS1 galaxies has been hypothesised to be at least a partial cause of the heterogeneity of the NLS1 population. This theory is supported by studies that have found that the jetted NLS1 galaxies favour more dense large-scale environments compared to their non-jetted counterparts \citep{2017jarvela, 2018jarvela}. With NLS1 galaxies being predominantly located in spiral galaxies or mergers of spiral galaxies, several host galaxy studies have argued against the current jet paradigm that claims that only old elliptical galaxies are capable of harbouring powerful relativistic jets \citep[][]{2018jarvela, 2019berton, 2020olguiglesias}. 

In this paper we study the morphology of 14 NLS1 galaxies, 11 of which have been detected at 37~GHz and are considered jetted. Six of the sources have been detected at 1.4~GHz in the Faint Images of the Radio Sky at Twenty-Centimeters (FIRST) survey. In an upcoming paper, we will expand the sample by analysing new data of Southern NLS1 galaxies obtained with the Magellan telescopes. The cosmological parameters used in this paper are H$_{0}$ = 73 km s$^{-1}$ Mpc$^{-1}$, $\Omega_{\text{matter}}$ = 0.27 and $\Omega_{\text{vacuum}}$ = 0.73 \citep{2007spergel1}. In Section~\ref{sec:obs} we go through the sample selection, the observations, and the data reduction. The data analysing process is described in Section~\ref{data}. The results for each individual source with a successful fit can be found in Section~\ref{sec:individual} and are discussed in Section~\ref{discussion}. The two galaxies that we were unable to fit can be found in Appendix~\ref{compromised-fits}.

\begin{table*}
\caption[]{Radio properties of the sample, and the black hole mass estimates. }
\centering
\begin{tabular}{l l l l l l l l l l l l l l}
\hline\hline
Source              & $S_{\mathrm{1.4~GHz}}$ & $S_{\mathrm{37~GHz, max}}$ & \textit{RL} &log $M_{\text{BH}}$ & $\gamma$-ray detected  \\
                    & (mJy)                  & (Jy)     & & ($M_{\odot}$) & (Yes/No) \\ \hline
6dFGS gJ042021.7-053054  &           & 1.15$^b$ &           & 5.52$^6$   & No  \\
6dFGS gJ044739.0-040330  &           &          &           & 7.04$^6$   & No \\
FBQS J0744+5149          & 11.9$^c$  &          & 59$^c$    & 8.42$^3$   & No \\
SDSS J080535.16+302201.7 & 60.81$^a$ & 1.44$^b$ & 638$^a$   & 7.11$^2$   & No  \\
6dFGS gJ084510.2-073205  &           &          &           & 5.11$^6$   & No   \\
SDSS J084516.17+421129.9 &           & 0.38$^b$ &           & 7.75$^2$   & No  \\
SDSS J090113.23+465734.7 & 1.55$^a$  & 0.27$^b$ & 21$^a$    & 7.15$^2$   & No \\
SDSS J093712.32+500852.1 & 166.6$^5$ & 0.92$^b$ &           & 7.56$^5$   & Yes \citep{2018romano} \\
SDSS J102906.68+555625.2 &           & 0.52$^b$ &           & 7.33$^4$   & No \\
SDSS J103123.73+423439.2 & 16.95$^a$ & 0.30$^a$ & 239$^a$   & 8.40$^2$    & No \\
SDSS J122844.81+501751.2 &           & 0.53$^b$ &           & 6.84$^4$   & No \\
SDSS J123220.11+495721.8 &           & 0.59$^b$ &           & 7.30$^4$   & No \\
SDSS J125635.89+500852.4 & 209.08$^a$& 0.62$^b$ & 3203$^a$  & 6.94$^2$   & No \\
SDSS J150916.18+613716.7 &           & 1.02$^b$ &           & 6.66$^4$   & No \\
\hline
\end{tabular}
\tablefoot{Columns: (1) Source name, (2) 1.4 GHz FIRST maximum flux density level, (3) Metsähovi 37 GHz maximum flux density level, (4) Radio-loudness (5) Black hole mass, (6) $\gamma$-ray detection. \\ (a) Values from \cite{2015jarvela} \\ (b) New data from Metsähovi Radio Observatory \\ (c) Values from \cite{2011foschini1} \\ Black hole masses from (1) \cite{2012macleod}, (2) \cite{2015jarvela}, (3) \cite{2017foschini}, (4) \cite{2018lahteenmaki}, (5) \cite{2019paliya} , and (6) \cite{2018chen}}

\label{tab:additional}
\end{table*}

\section{Observations and data reduction}
\label{sec:obs}

Our original sample included all NLS1 galaxies detected in $\gamma$-rays or at 37~GHz at $\sim$Jy levels, thus being likely to host relativistic jets, that were observable from La Palma during the semester of our observations (Oct 2019 - Mar 2020, proposal ID 60-005, PI E. Järvelä). Based on our earlier experience with NOTCam, we applied a redshift cut of $z <$ 0.6 to ensure good quality data, and excluded sources which have published results. After applying these criteria we were left with a total of 20 jetted NLS1 galaxies. Some observing time was lost due to cloudy weather, and in the end we were able to observe only 11 sources from the original sample. In addition, dusty, strong wind forced us to observe toward a certain direction for extended periods of time, and due to the lack of small RA sources in our original sample, we instead observed one source from \cite{2017lahteenmaki1} and two sources from the Southern NLS1 sample by \cite{2018chen}, which were then included in the sample. The 37~GHz detection values are all from the Metsähovi Radio Observatory AGN monitoring programme. All but one value is an unpublished and thus new value. Furthermore, one of our sources, SDSS J093712.32+500852.1, has been detected at $\gamma$-rays. The detection was made with the Large Area Telescope of the \textit{Fermi} Gamma-ray Space Telescope.

Table~\ref{tab:sample} lists the properties of our sample as well as some basic characteristics of the observations. The observations were carried out with the Nordic Optical Telescope (NOT) near-infrared Camera (NOTCam) during five nights, from February 7th to February 12th, 2020. We observed the full sample with NOTCam wide-field (WF) mode. The CCD in NOTCam is 1024 $\times$ 1024 pixels. The field of view (FOV) of wide-field imaging is 4' $\times$ 4', giving a scale of 0.234\arcsec px$^{-1}$. The weather during our observations was challenging, as discussed earlier.

We used the frame mode and 9-point dithering. The dithering step was 10\arcsec and the skew was 2\arcsec. We observed all but one source (SDSS J102906.68+555625.2) in both \textit{J}- and \textit{Ks}- bands. The total exposure time for each source can be found in Table~\ref{tab:sample}. For some sources, a bright point spread function (PSF) star was observed in both bands (for details of the PSF stars, see Section~\ref{data}).

We used the IRAF NOTCam data reduction package{\footnote{\url{http://www.not.iac.es/instruments/notcam/}}}. First we combined the dithered images to obtain one image for each dithering sequence. After that, we combined the dithered images of one night to produce one final image. This was done for each filter. We omitted observations where the seeing was very bad or clearly worse than in the other observations to obtain the best possible final image. For those sources that a PSF star had been observed for, the data reduction and the image combination process was the same. The final data reduction step was calculating a zero point magnitude for all the images. This was done by using the Graphical Astronomy and Image Analysis Tool\footnote{\url{http://star-www.dur.ac.uk/~pdraper/gaia/gaia.html}} (GAIA). To obtain a zero point magnitude, we compared the magnitude of stars in the 2MASS catalog to values we obtained from our images using GAIA. When taking into account also the exposure time, we were able to obtain a zero point magnitude.

\section{Data analysis}
\label{data}

The following images and results presented in this paper are the results for the band with which we were able to obtain a better fit. In most cases, this meant the observation with the better seeing. In some cases the seeing was so similar that the changes to the image quality were low. We used GALFIT version 3 \citep{2010peng1} to complete a 2D photometric decomposition of our images. With GALFIT we were able to complete a photometric decomposition for all of our images. The first thing to take into account with GALFIT is having a properly modelled PSF. With a good PSF it is possible to remove the contamination caused by the AGN, whereas a bad PSF will not model the AGN contamination properly, making it impossible to achieve an accurate model for the host galaxy. 

We used three different PSF techniques to obtain the best possible PSF. In the first technique, we searched for bright, isolated and not saturated stars in the FOV of each image. If such a star was present, we extracted a $\sim$ 100 $\times$ 100 pixel image (the same size used to extract a cutout around the target galaxy), we subtracted the background, by averaging the measurements obtained in eight different location of the cutout and use it as a model of the PSF. In the second technique if a specific PSF star had been observed, we extracted a cutout of this star using otherwise the same procedure as in the first technique. The third PSF technique was to use the DAOPHOT package in IRAF \citep{1987stetson} to build a detailed model of the PSF using a list of suitable PSF stars chosen from a 3D plot. When using DAOPHOT, there is no need to subtract the background sky as it has been subtracted during the PSF building process. We always chose the PSF technique that gave us the best fit of the star profile. The fitting process was the same for all sources.

The first fit we did on any given source was a clean PSF fit, with the PSF being the only component being fitted. After visual inspection of the residuals of the PSF fit, we made a decision on whether or not more components were needed. Each new component was added one at a time, inspecting the residuals after each successful run to determine once again if more components were needed. We also followed how the reduced $\chi^2$ value changed with the component additions. The initial fitting parameters were varied to ensure that the values the fit was converging to were correct and stable. We also varied the number of components to obtain the best, reasonable result. We used two different types of components to obtain the best fit: PSF and a S\'{e}rsic profile. The S\'{e}rsic profile can be given as: \begin{equation} I(r) = I_\text{e} \texttt{exp} \Bigg[ -\kappa_n \Bigg( \bigg( \frac{r}{r_\text{e}} \bigg)^{1/n} -1 \Bigg) \Bigg] , \end{equation} where $I(r)$ is the surface brightness at radius $r$, $I_\text{e}$ is the surface brightness at half-light radius, $r_\text{e}$, and $\kappa_n$ is a parameter related to the S\'{e}rsic index, $n$ \citep{2005graham1}.

We chose the S\'{e}rsic profile for fitting because of its versatility. By changing the S\'{e}rsic index, it is possible to model both elliptical and disk-like light distributions as well as classical bulges and pseudo-bulges. Common key S\'{e}rsic indices are \textit{n} = 0.5, 1, and 4. A Gaussian profile can be obtained with \textit{n} = 0.5. An exponential profile, often used for modelling disks, can be obtained with \textit{n} = 1. A de Vaucouleurs profile, often used to model the light distribution of elliptical galaxies, can be obtained with \textit{n} = 4.  Smaller S\'{e}rsic indices typically indicate both a disk-like morphology and pseudo-bulges of galaxies. In these cases the core flattens more rapidly at $r < r_\text{e}$ and the intensity decreases quicker at $r > r_\text{e}$ \citep{2008fisher}. However, even if S\'{e}rsic indices of $n$ = 1 and $n$ = 2 typically model disks and pseudo-bulges, respectively, spiral galaxies have been proven to exhibit a range of S\'{e}rsic indices, often varying from approximately 0.2 to approximately 3, with some spiral galaxies even having as high S\'{e}rsic indices as 4 \citep[e.g.,][]{2014elmegreen, 2015salo}.

With the exception of the sky background parameter, all parameters were kept unfrozen to maintain as unbiased a fit as possible. The sky background was estimated by picking several evenly distributed blank locations in the sky and calculating the average value of the sky background in them. To obtain the best results, we maintained as large of a cutout as we possibly could. In a few cases there were some other nearby sources inside the cutout. In these cases, we fitted the secondary sources as well. The goodness of fit was determined through a numerical and a visual inspection as well as through ensuring that the obtained parameters were logical and reasonable. The visual inspection meant looking at the residuals and trying to determine if everything had been properly fitted. Numerical inspection meant studying the reduced $\chi^2_{\nu}$ parameter. If the  $\chi^2_{\nu}$ parameter was significantly larger than one or less than one, we did not accept the fit and continued the fitting procedure until we received a value close to one.

After getting a satisfactory fit, we extracted the radial surface brightness profiles. We extracted the profile from the observed image, the model image, and the separate component images. We used the IRAF task \textit{ELLIPSE} for the extraction, setting the values of the different parameters recovered from GALFIT as initial values for the process. For the radial surface brightness profiles of the model and observed images we used the same initial values to ensure similar radii of the ellipses. The other components had in most cases unique initial values compared to the model and observed image initial values. After a successful run, we plotted the results for each source, including all the possible components of the given source, into one figure. 

To account for the minor change induced by the sky variation to the Galfit results we ran a $\pm$1$\sigma$ value fit for all the sources. This gave us the errors for the surface brightness at the half-light radius, $r_e$, the S\'{e}rsic index, $n$, the axial ratio, and the position angle (PA). For the magnitude, we used error propagation for determining the error since we had three error sources: sky error, 2MASS catalog error, and the zero point magnitude error. These errors have been taken into account when plotting the results from \textit{ELLIPSE}. Individual component errors, taking into account all error sources, can be found in the best fit parameter tables of each individual source.

We also created colour maps of all sources that had both \textit{J}- and \textit{Ks}- band observations. The only source that we do not have a colour map of is SDSS J102906.68+555625.2. Due to the PSF of the \textit{J}- and \textit{Ks}- band images having different widths, we convolved the image with a smaller PSF with a Gaussian of a tailored size to try and get the PSF of both images to be of the same size. We obtained a satisfying result for all but SDSS J150916.18+613716.7. For this source, a simple convolution with a Gaussian was not enough to correct the issue with the PSF sizes. For the other sources, we were able to obtain physically correct colour maps. We have presented the colour map of SDSS J150916.18+613716.7 in this paper, and when analysing it, we have taken into account the issue with the PSF sizes.

\section{Individual source analysis}
\label{sec:individual}

\subsection{6dFGS gJ042021.7-053054}

This source has been detected at 37~GHz at $\sim$ 1~Jy level (see Table~\ref{tab:additional}), indicating the presence of relativistic jets. There are no earlier radio detections.

We opted to manually choose a PSF star from the image to have the best possible quality. The best fit for this source is presented in Table~\ref{tab:g0420}, with one PSF and one S\'{e}rsic component. Based on the best fit parameters, the component fitted with a S\'{e}rsic function is likely a disk (\textit{n} = 1.19$\substack{+0.11 \\ -0.07}$). There were two nearby sources close this target. These two sources were modelled successfully with one PSF component each. 

\begin{figure*}
\centering
\adjustbox{valign=t}{\begin{minipage}{0.35\textwidth}
\centering
\includegraphics[width=1\textwidth]{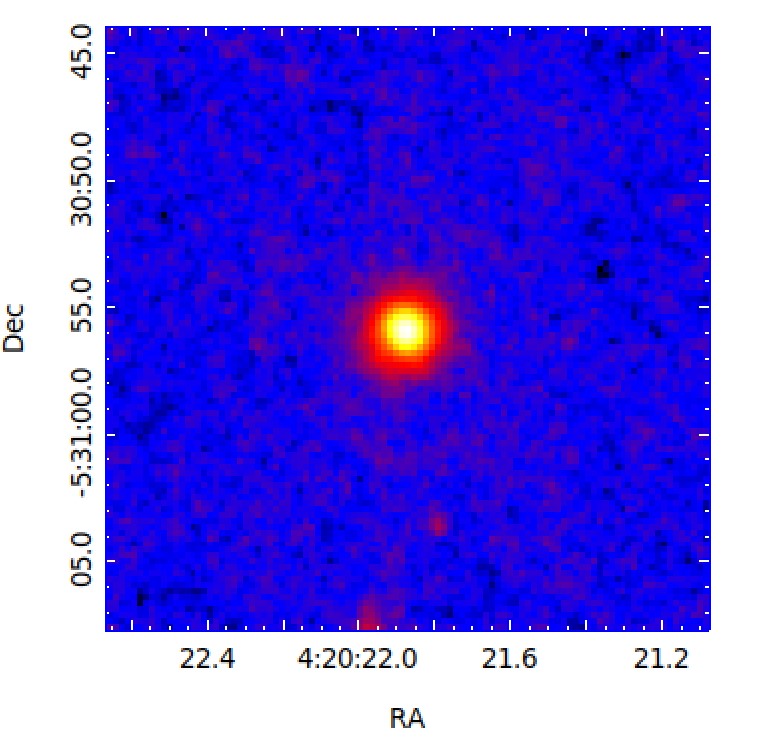}
\end{minipage}}
\adjustbox{valign=t}{\begin{minipage}{0.31\textwidth}
\centering
\includegraphics[width=0.95\textwidth]{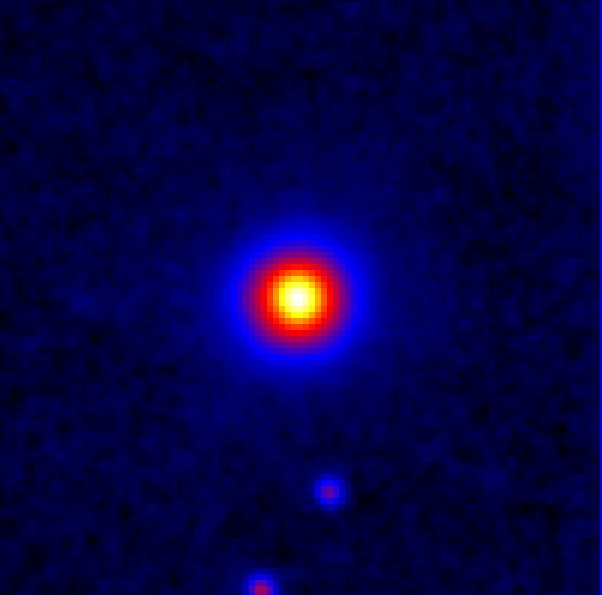}
\end{minipage}}
\adjustbox{valign=t}{\begin{minipage}{0.31\textwidth}
\centering
\includegraphics[width=0.95\textwidth]{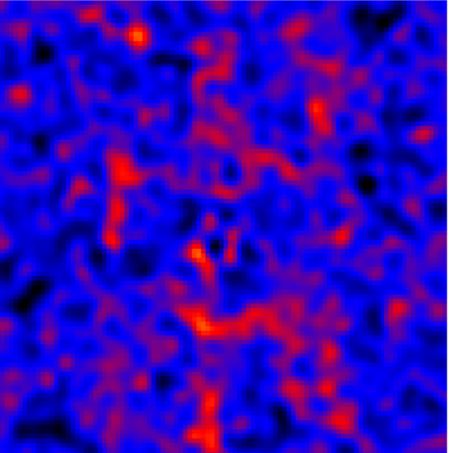}
\end{minipage}}
\hfill
    \caption{\textit{J}-band images of 6dFGS gJ042021.7-053054. The FoV is 23.4\arcsec    $/$ 74.1~kpc in all images. \emph{Left panel:} observed image, \emph{middle panel:} model image, and \emph{right panel:} residual image, smoothed over 3px.}  \label{fig:g0420}

\adjustbox{valign=t}{\begin{minipage}{0.49\textwidth}
\centering
\includegraphics[width=0.95\textwidth]{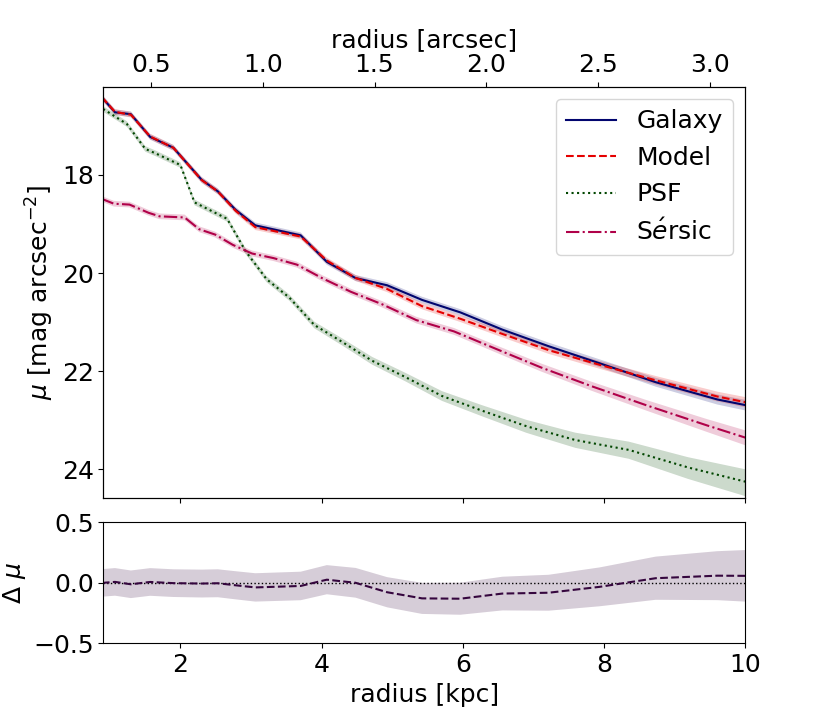}
\caption{Radial surface brightness profile plot of 6dFGS gJ042021.7-053054. The galaxy component is plotted with a blue line, the model component is plotted with a dashed red line, the PSF component is plotted with a dotted green line, and the S\'{e}rsic component is plotted with a dashed pink line. The shaded area surrounding each profile curve depicts the error linked to each component.}
\label{fig:g0420comps}
\end{minipage}}
\hfill
\adjustbox{valign=t}{\begin{minipage}{0.49\textwidth}
\centering
\includegraphics[width=0.95\textwidth]{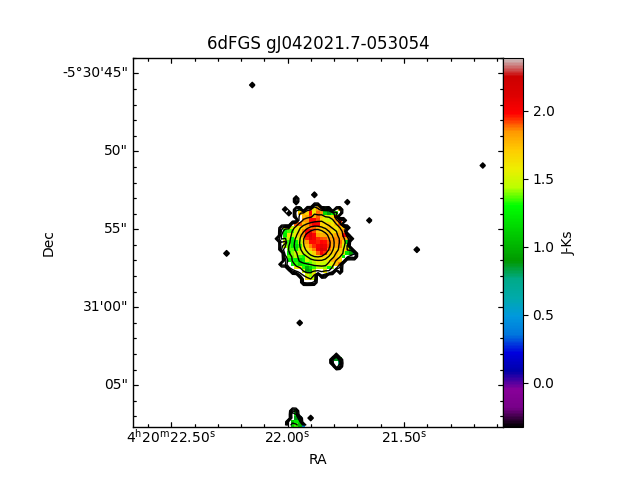}
\caption{\textit{J - $Ks$} colour map of 6dFGS gJ042021.7-053054.}
\label{fig:g0420color}
\end{minipage}}
\hfill
\end{figure*}

The observed, model, and the residual images of the source can be seen in Fig.~\ref{fig:g0420}. We had some residuals left, but the $\chi^2_{\nu}$ value is good. The residuals left in the image do not directly hint on any specific component missing, and it is entirely possible that they are noise. However, it is also possible that the remaining residuals are remains of a ring-like structure. The axial ratio of approximately 1 supports this theory. The ring-structure is also visible in the observed image. Adding more S\'{e}rsic components did not improve the results and would not have been physically justifiable. 

The radial surface brightness profile is presented in Fig.~\ref{fig:g0420comps}. The model and galaxy curves in the plot are nearly identical, suggesting that our fit obtained with GALFIT is good. The \textit{J - $Ks$} colour map of the source is shown in Fig.~\ref{fig:g0420color}. In general, the magnitude difference between $J$ and $Ks$ magnitudes is approximately between 0 and 2.5 for Seyfert galaxies \citep{2000jarrett}. With this in mind, the colour map presented here is in agreement with what is expected to be seen in a Seyfert 1 galaxy. Furthermore, galaxies with a difference in the $J$ and $Ks$ magnitude of more than 1.3 are considered very red. On average, the red colour in AGN is typically dust extinction. The size of the galaxy is very small, however, the colour of the central region indicates a possibility of dust extinction. The map also suggest a possible bar structure. Otherwise, the colours of the galaxy are mostly uniform.

\begin{table*}
\caption[]{Best fit parameters of 6dFGS gJ042021.7-053054. $\chi^2_{\nu}$ = 1.145 $\substack{+0.01 \\ -0.01}$. }
\centering
\begin{tabular}{l l l l l l l}
\hline\hline
Function & Mag & $r_\text{e}$ & $n$ & Axial & PA & Notes \\
 & & (kpc) & & ratio & (\textdegree) & \\ 
\hline
PSF 1 & 16.64 $\substack{+0.05 \\ -0.05}$ & & & & & \\
S\'{e}rsic & 17.06 $\substack{+0.05 \\ -0.06}$ & 2.49 $\substack{+0.01 \\ -0.01}$ & 1.19 $\substack{+0.11 \\ -0.07}$ & 0.98 $\substack{+0.00 \\ -0.00}$ & 9.85  $\substack{+0.09 \\ -0.37}$ & Disk \\
PSF 2 & 20.78 $\substack{+0.05 \\ -0.05}$ & & & & &  Nearby source \\
PSF 3 & 21.54 $\substack{+0.06 \\ -0.06}$ & & & & &  Nearby source \\
\hline   
\end{tabular}
\tablefoot{Columns: (1) Function used for modelling, (2) Magnitude, (3) Effective radius, (4) S\'{e}rsic index, (5) Axial ratio, (6) Position angle, (7) Additional notes.}
\label{tab:g0420}
\end{table*}

\subsection{FBQS J0744+5149}

Despite earlier radio detections at 1.4~GHz, this source has not been detected at 37~GHz \citep{2017lahteenmaki1}. Observations with the Karl. G Jansky Very Large Array (JVLA) at 5~GHz \citep{2018berton} show an unresolved steep core, with total luminosity comparable to compact steep-spectrum (CSS) sources. We observed it as an extra due to the weather constraints mentioned in Section~\ref{sec:obs}. 

We used DAOPHOT to obtain the best possible PSF. The best fit was obtained with one PSF and two S\'{e}rsic components; the corresponding  best fit parameters can be found in Table~\ref{tab:j0744}. Based on the parameters of the S\'{e}rsic 1 component, such as its unusually small axial ratio (0.13$\substack{+0.00 \\ -0.01}$) and the effective radius (0.92$\substack{+0.00 \\ -0.03}$ kpc) that is considerably smaller than the seeing disk, we believe that this component is not fully resolved, which might be due to the moderately high redshift of this source, or an incomplete PSF model. The S\'{e}rsic 2 seems to favour a disk-like morphology, rather than an elliptical morphology, however, also this component might not be fully resolved. The galaxy seems to be very small, with an effective radius of roughly 2~kpc. However, this would be quite low even for a disk-like galaxy, so we cannot rule out that the host is some kind of a dwarf galaxy. FBQS J0744+5149 has the highest black hole mass in the whole sample, exceeding 10$^8 M_{\odot}$. This seems to be in contradiction with the small size of its host. However, the black hole mass was originally estimated using magnitudes \citep{2017foschini}, which is not as accurate as the more commonly used methods, such as those based on the continuum luminosity and the emission line widths, so the estimate might not be very reliable.

\begin{figure*}
\centering
\adjustbox{valign=t}{\begin{minipage}{0.35\textwidth}
\centering
\includegraphics[width=1\textwidth]{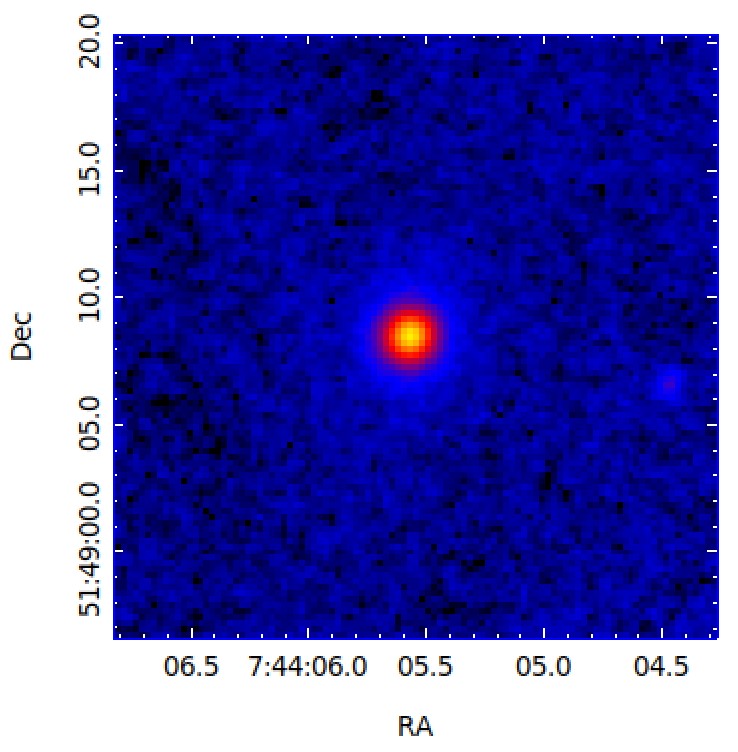}
\end{minipage}}
\adjustbox{valign=t}{\begin{minipage}{0.31\textwidth}
\centering
\includegraphics[width=0.95\textwidth]{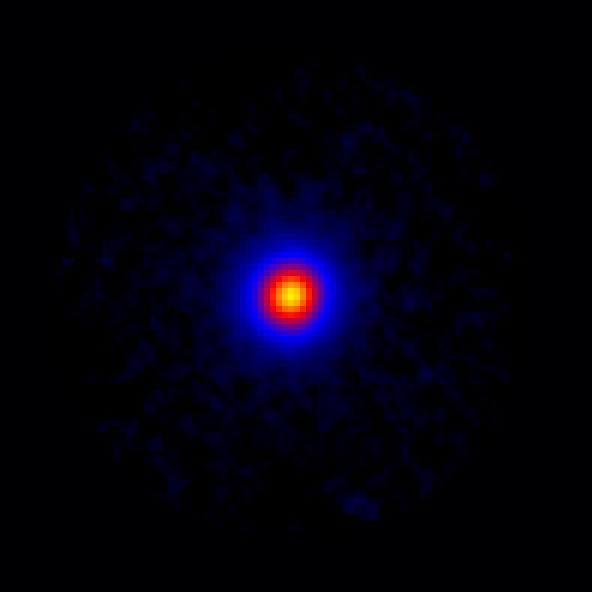}
\end{minipage}}
\adjustbox{valign=t}{\begin{minipage}{0.31\textwidth}
\centering
\includegraphics[width=0.95\textwidth]{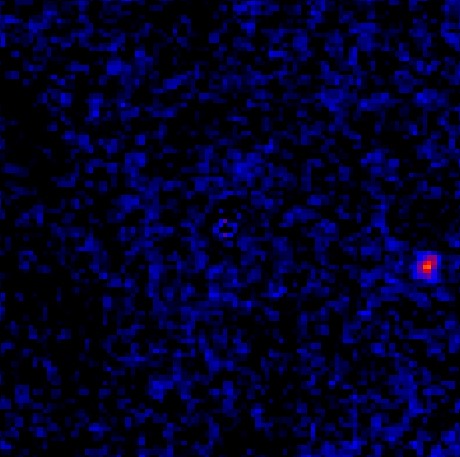}
\end{minipage}}
\hfill
    \caption{\textit{J}- band images of FBQS J0744+5149. The FoV is 15.0\arcsec $/$ 84.6~kpc in all images. \emph{Left panel:} observed image, \emph{middle panel:} model image, and \emph{right panel:} residual image, smoothed over 3px.}  
    \label{fig:j0744}

\adjustbox{valign=t}{\begin{minipage}{0.49\textwidth}
\centering
\includegraphics[width=0.95\textwidth]{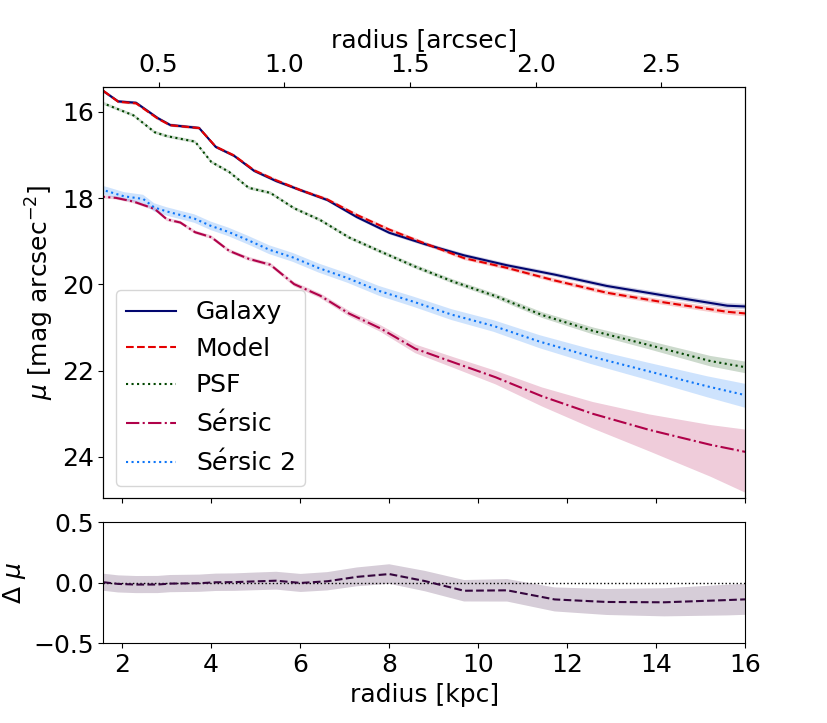}
\caption{Radial surface brightness profile plot of FBQS J0744+5149. The galaxy component is plotted with a blue line, the model component is plotted with a dashed red line, the PSF component is plotted with a dotted green line, the S\'{e}rsic 1 component is plotted with a dashed pink line, and finally the S\'{e}rsic 2 component is plotted with a dotted light blue line. The shaded area surrounding each profile curve depicts the error linked to each component.}
\label{fig:j0744comps}
\end{minipage}}
\hfill
\adjustbox{valign=t}{\begin{minipage}{0.49\textwidth}
\centering
\includegraphics[width=0.95\textwidth]{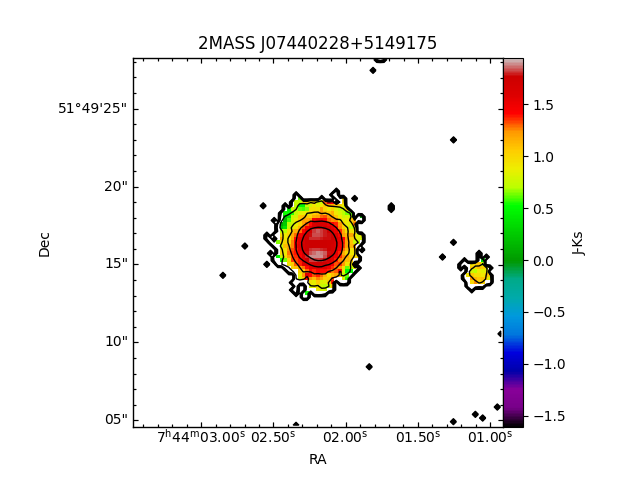}
\caption{\textit{J - $Ks$} colour map of FBQS J0744+5149.}
\label{fig:j0744color}
\end{minipage}}
\hfill
\end{figure*}

The observed, model, and the residual image of the source can be seen in Fig.~\ref{fig:j0744}. Based on the very clean residuals and the good reduced $\chi^2_{\nu}$ value, it seems that the galaxy has been modelled nearly completely. The radial surface brightness profile is presented in Fig.~\ref{fig:j0744comps}. In the plot, the model and galaxy curves are nearly identical, which supports the idea of our GALFIT fit being good. The \textit{J - $Ks$} colour map of the source is displayed in Fig.~\ref{fig:j0744color}. The \textit{J - $Ks$} magnitude is $\sim$1.5 and thus the galaxy can be deemed as extremely red. The colours in the central region suggest possible dust extinction. The difference in the $J$ and $Ks$ magnitude is within the accepted range for a Seyfert 1 galaxy.

\begin{table*}
\caption[]{Best fit parameters of FBQS J0744+5149. $\chi^2_{\nu}$ = 1.070 $\substack{+0.03 \\ -0.07}$.}
\centering
\begin{tabular}{l l l l l l l}
\hline\hline
Function       & Mag                             & $r_\text{e}$                          & $n$                              & Axial & PA & Notes               \\
             &                                   & (kpc)                            &                                  & ratio & (\textdegree) &  \\ \hline
PSF          & 15.37 $\substack{+0.03 \\ -0.03}$ &                                  &                                  &       &               &   \\

S\'{e}rsic 1  & 17.35 $\substack{+0.03 \\ -0.04}$ & 0.92 $\substack{+0.00 \\ -0.03}$ & 1.04 $\substack{+0.26 \\ -0.00}$ & 0.13 $\substack{+0.00 \\ -0.01}$ & 29.45  $\substack{+0.00 \\ -1.84}$ & \\

S\'{e}rsic 2  & 16.84 $\substack{+0.03 \\ -0.04}$ & 2.10 $\substack{+0.03 \\ -0.01}$ & 2.62 $\substack{+0.42 \\ -0.23}$ & 0.65 $\substack{+0.03 \\ -0.01}$ & 73.63  $\substack{+1.81 \\ -0.00}$ &   \\
  \hline   
\end{tabular}
\tablefoot{Columns: (1) Function used for modelling, (2) Magnitude, (3) Effective radius, (4) S\'{e}rsic index, (5) Axial ratio, (6) Position angle, (7) Additional notes.}
\label{tab:j0744}
\end{table*}

\subsection{SDSS J080535.16+302201.7}

This jetted source, detected at 1.4 GHz, can be fairly bright (approximately 1.4 Jy when flaring) also at 37 GHz (see Table~\ref{tab:additional}).

For this source, we were able to locate an isolated star directly from the image for the PSF model. We obtained the best fit with one PSF and one S\'{e}rsic component presented in Table~\ref{tab:j0805}.  

\begin{figure*}[ht!]
\centering
\adjustbox{valign=t}{\begin{minipage}{0.35\textwidth}
\centering
\includegraphics[width=\textwidth]{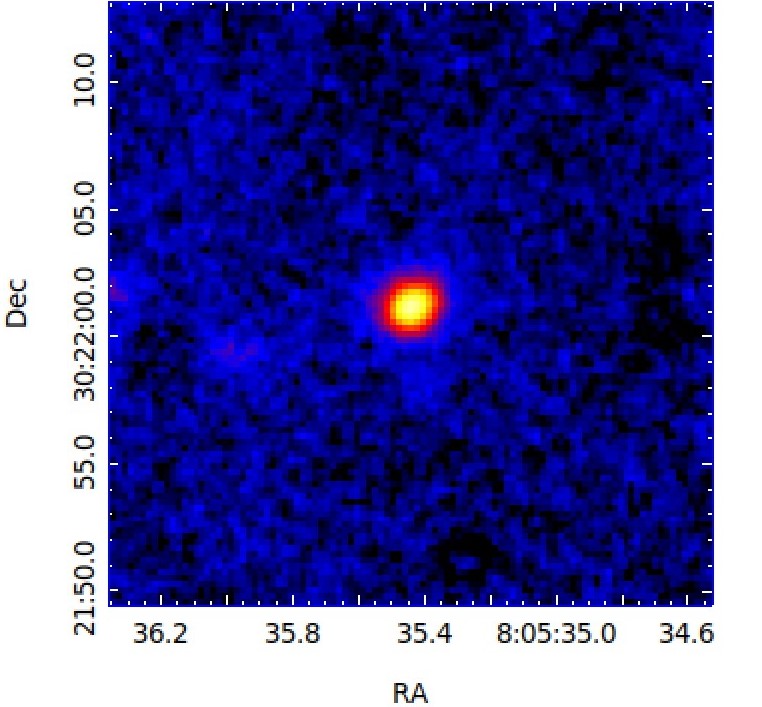}
\end{minipage}}
\adjustbox{valign=t}{\begin{minipage}{0.31\textwidth}
\centering
\includegraphics[width=0.95\textwidth]{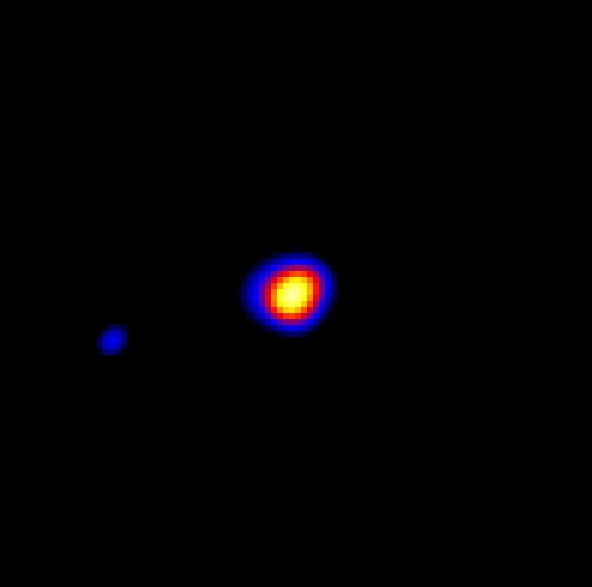}
\end{minipage}}
\adjustbox{valign=t}{\begin{minipage}{0.31\textwidth}
\centering
\includegraphics[width=0.95\textwidth]{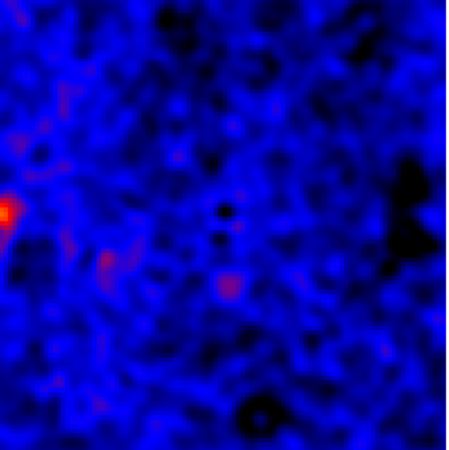}
\end{minipage}}
\hfill
    \caption{\textit{J}-band images of SDSS J080535.16+302201.7. The FoV is 23.4\arcsec $/$    145.6~kpc in all images. \emph{Left panel:} observed image, \emph{middle panel:} model image, and \emph{right panel:} residual image, smoothed over 3px.}  \label{fig:j0805}

\adjustbox{valign=t}{\begin{minipage}{0.49\textwidth}
\centering
\includegraphics[width=0.95\textwidth]{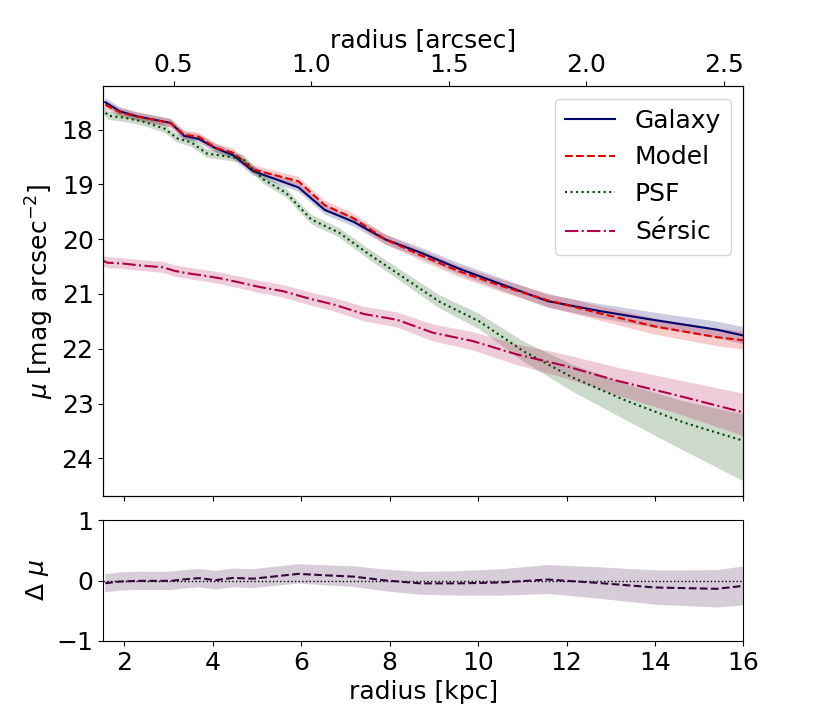}
\caption{Radial surface brightness profile plot of SDSS J080535.16+302201.7. The galaxy component is plotted with a blue line, the model component is plotted with a dashed red line, the PSF component is plotted with a dotted green line, and the S\'{e}rsic component is plotted with a dashed pink line. The shaded area surrounding each profile curve depicts the error linked to each component.}
\label{fig:j0805comps}
\end{minipage}}
\hfill
\adjustbox{valign=t}{\begin{minipage}{0.49\textwidth}
\centering
\includegraphics[width=0.95\textwidth]{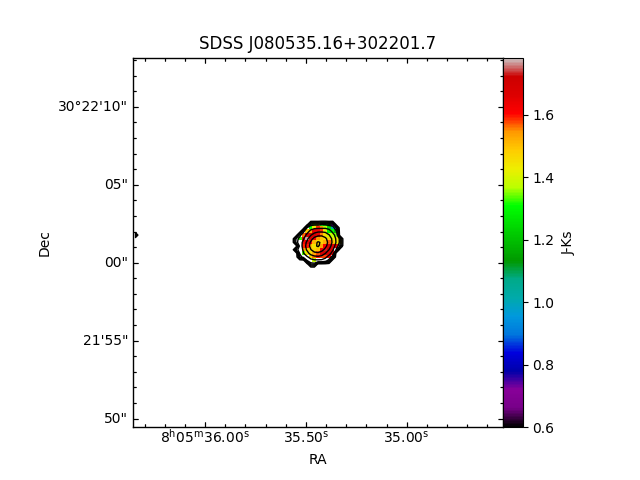}
\caption{\textit{J - $Ks$} colour map of SDSS J080535.16+302201.7}
\label{fig:j0805color}
\end{minipage}}
\hfill
\end{figure*}

The observed, model, and the residual image of the source can be seen in Fig.~\ref{fig:j0805}. Some leftover residuals are visible. We tried various approaches but could not get a physically meaningful fit that would have modelled the residuals completely. It is possible that some of the residuals that remain on the southern side of the source are due to a background source. Based on the best fit parameters, the host galaxy is most likely disk-like due to the low S\'{e}rsic index ($n$ = 1.32$\substack{+0.69 \\ -0.41}$). The best fit parameters and the residuals indicate that we have modelled the whole galaxy. Since there was a second source close to our target, we used one PSF to model that one to obtain as clean a residual image as possible.

The radial surface brightness profile can be found in Fig.~\ref{fig:j0805comps}. The curves of the model and observed image are very similar, with minor deviation towards the higher radii. The deviation could possibly be explained by the residuals remaining southward of the galaxy, however, we cannot say if the deviation is related to our source or a background/foreground object. The curves give us a strong reason to believe that our best fit is good. The \textit{J - $Ks$} colour map of the source is shown in Fig.~\ref{fig:j0805color}. The angular size of the galaxy is very small due to the high redshift and therefore its structure is difficult to distinguish. The galaxy has a \textit{J - $Ks$} colour of about 1.5 and with that, counts as a very red galaxy.

\begin{table*}[ht!]
\caption[]{Best fit parameters of SDSS J080535.16+302201.7. $\chi^2_{\nu}$ = 1.138 $\substack{+0.10 \\ -0.05}$.}
\centering
\begin{tabular}{l l l l l l l}
\hline\hline
Function       & Mag                             & $r_\text{e}$                          & $n$                              & Axial & PA    & Notes           \\
             &                                   & (kpc)                            &                                  & ratio & (\textdegree) &  \\ \hline
PSF 1         & 17.11 $\substack{+0.04 \\ -0.06}$ &                                  &                                  &       &               &   \\

S\'{e}rsic  & 18.57 $\substack{+0.58 \\ -0.13}$ & 6.66 $\substack{+2.74 \\ -1.09}$ & 1.32 $\substack{+0.69 \\ -0.41}$ & 0.69 $\substack{+0.02 \\ -0.05}$ & 12.93  $\substack{+3.75 \\ -4.21}$ & Bulge and disk \\

PSF 2          & 20.24 $\substack{+0.07 \\ -0.07}$ &                                  &                                  &       &               &  Nearby source \\

  \hline   
\end{tabular}
\tablefoot{Columns: (1) Function used for modelling, (2) Magnitude, (3) Effective radius, (4) S\'{e}rsic index, (5) Axial ratio, (6) Position angle, (7) Additional notes.}
\label{tab:j0805}
\end{table*}

\subsection{SDSS J084516.17+421129.9}

There are no previous radio detections of SDSS J084516.17+421129.9, however, it has been detected at 37 GHz (see Table~\ref{tab:additional}) and is therefore considered jetted.

We obtained the best PSF with DAOPHOT. The best fit was achieved with one PSF and one S\'{e}rsic; the corresponding parameters are listed in Table~\ref{tab:j0845+}. The S\'{e}rsic index, \textit{n} = 1.56$\substack{+1.16 \\ -0.01}$, suggests a disk-like morphology. Due to the effective radius being so large, we believe that we have fitted the whole galaxy at once. We fitted a PSF model on two nearby sources, to obtain as clean residuals as possible. Our PSF model was able to completely model the two sources. Due to this, it is possible that the two nearby sources are stars. The nearby sources are very faint, with magnitudes fainter than 18~mag, and thus they are quite invisible in the plots.

\begin{figure*}[ht!]
\centering
\adjustbox{valign=t}{\begin{minipage}{0.35\textwidth}
\centering
\includegraphics[width=1\textwidth]{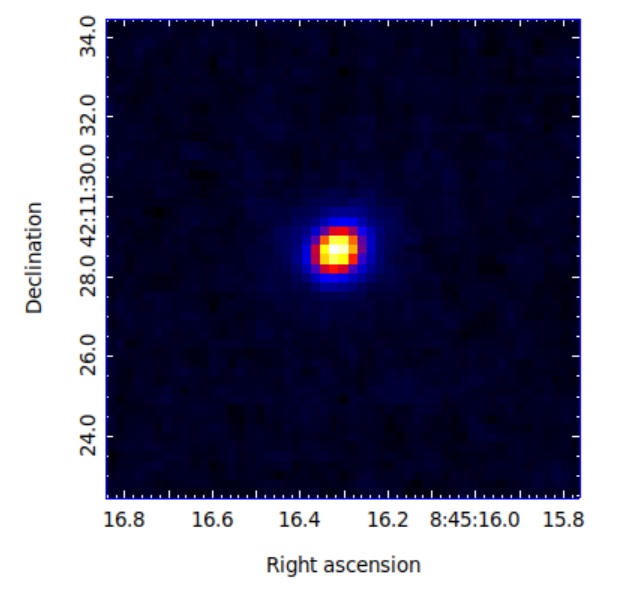}
\end{minipage}}
\adjustbox{valign=t}{\begin{minipage}{0.31\textwidth}
\centering
\includegraphics[width=0.95\textwidth]{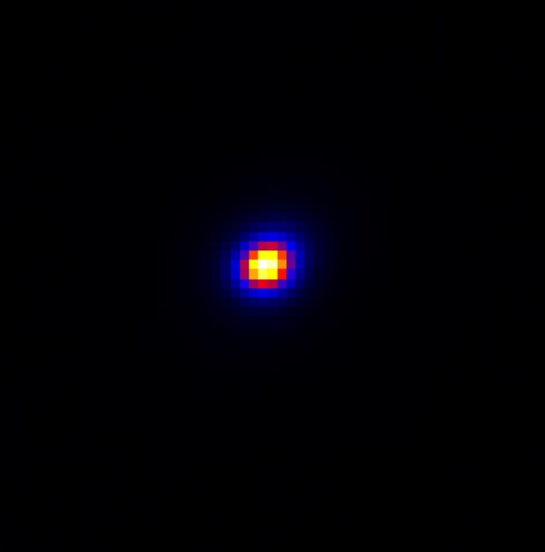}
\end{minipage}}
\adjustbox{valign=t}{\begin{minipage}{0.31\textwidth}
\centering
\includegraphics[width=0.95\textwidth]{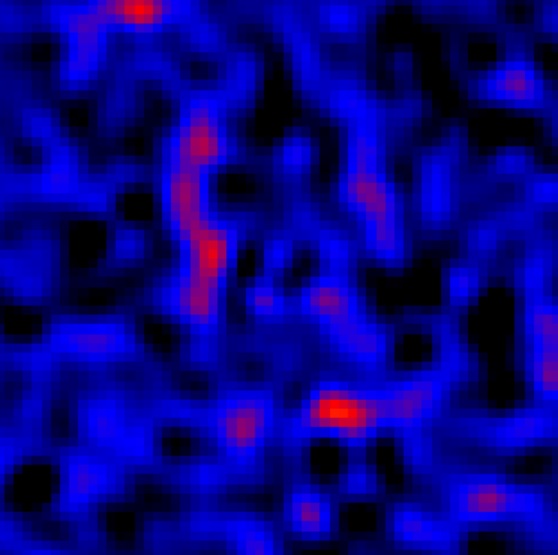}
\end{minipage}}
\hfill
    \caption{\textit{Ks}-band images of SDSS J084516.17+421129.9. The FoV is 5.9\arcsec $/$    35.5~kpc in all images. \emph{Left panel:} observed image, \emph{middle panel:} model image, and \emph{right panel:} residual image, smoothed over 3px.}  \label{fig:j0845+}

\adjustbox{valign=t}{\begin{minipage}{0.49\textwidth}
\centering
\includegraphics[width=0.95\textwidth]{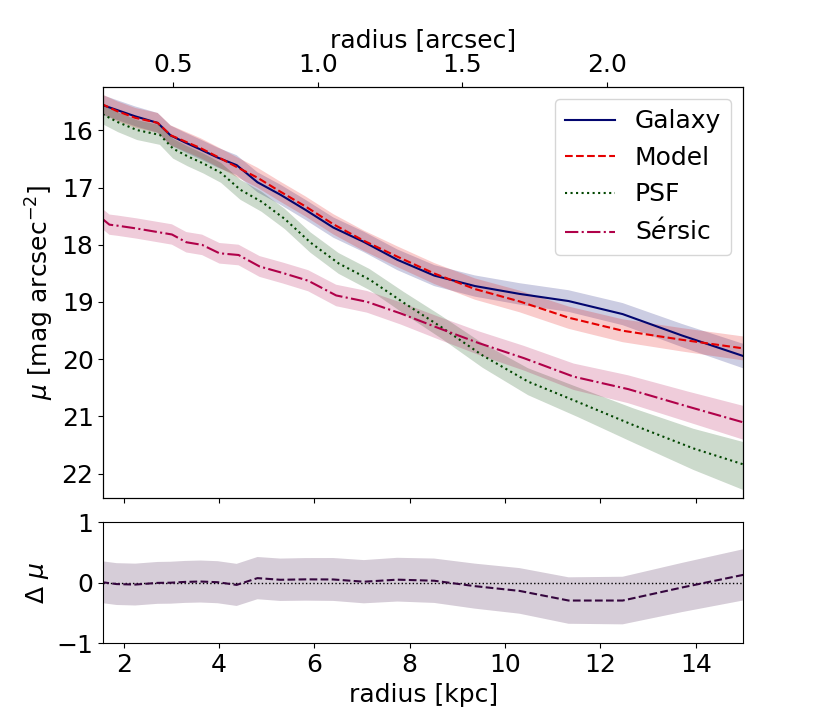}
\caption{Radial surface brightness profile plot of SDSS J084516.17+421129.9. The galaxy component is plotted with a blue line, the model component is plotted with a dashed red line, the PSF component is plotted with a dotted green line, and the S\'{e}rsic component is plotted with a dashed pink line. The shaded area surrounding each profile curve depicts the error linked to each component.}
\label{fig:j0845+comps}
\end{minipage}}
\hfill
\adjustbox{valign=t}{\begin{minipage}{0.49\textwidth}
\centering
\includegraphics[width=0.95\textwidth]{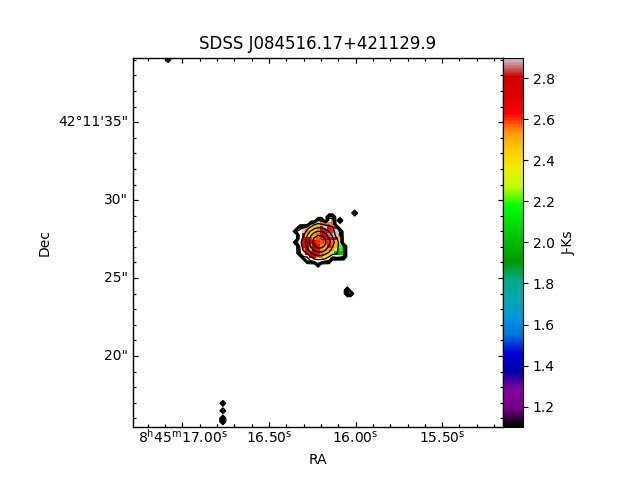}
\caption{\textit{J - $Ks$} colour map of SDSS J084516.17+421129.9.}
\label{fig:j0845+color}
\end{minipage}}
\hfill
\end{figure*}

The observed, model, and the residual image of the source are presented in Fig.~\ref{fig:j0845+}. The residuals are nearly nonexistent and in combination with a good reduced $\chi^2_{\nu}$ value support a good fit. The radial surface brightness profile can be seen in Fig.~\ref{fig:j0845+comps}. The curves of the model and observed image are very similar, with some slight deviation near the higher radii. The similarity in the curves supports the idea of a good fit. The \textit{J - $Ks$} colour map of the source is shown in Fig.~\ref{fig:j0845+color}. The difference in the \textit{J - $Ks$} magnitude is on the upper-end of generally accepted Seyfert 1 galaxies, but still remains inside the range. With the galaxy being on the upper end of the range, it counts as very red.

\begin{table*}[ht!]
\caption[]{Best fit parameters of SDSS J084516.17+421129.9. $\chi^2_{\nu}$ = 1.12 $\substack{+0.00 \\ -0.03}$.}
\centering
\begin{tabular}{l l l l l l l}
\hline\hline
Function       & Mag                             & $r_\text{e}$                          & $n$                              & Axial & PA  & Notes             \\
             &                                   & (kpc)                            &                                  & ratio & (\textdegree) &  \\ \hline
PSF 1          & 15.43 $\substack{+0.33 \\ -0.34}$ &                                  &                                  &       &               &   \\

S\'{e}rsic  & 16.53 $\substack{+0.33 \\ -0.41}$ & 5.65 $\substack{+0.00 \\ -0.48}$ & 1.56 $\substack{+1.16 \\ -0.01}$ & 0.69 $\substack{+0.01 \\ -0.00}$ & 43.78  $\substack{+2.43 \\ -0.00}$ &  \\

PSF 2         & 18.49 $\substack{+0.34 \\ -0.34}$ &                                  &                                  &       &               & Nearby source   \\
PSF 3         & 18.85 $\substack{+0.33 \\ -0.33}$ &                                  &                                  &       &               & Nearby source   \\

\hline   
\end{tabular}
\tablefoot{Columns: (1) Function used for modelling, (2) Magnitude, (3) Effective radius, (4) S\'{e}rsic index, (5) Axial ratio, (6) Position angle, (7) Additional notes.}
\label{tab:j0845+}
\end{table*}

\subsection{SDSS J090113.23+465734.7}

This source has been previously detected at 1.4~GHz, and it has also been detected at 37~GHz (see Table~\ref{tab:additional}).

We created the best PSF model with DAOPHOT and then obtained the best fit with one PSF and one S\'{e}rsic component. The best fit values can be found in Table~\ref{tab:j0901}. The result suggest that we are most likely modelling the whole galaxy. The S\'{e}rsic index, $n$ = 0.42$\substack{+0.10 \\ -0.09}$, is low for a disk and closer to values usually associated with bars, but rules out an elliptical morphology. We also fitted a PSF on a nearby source. The PSF modelled the very faint nearby source completely. Due to the faintness of the source, the nearby source is virtually invisible in the intensity plots.

\begin{figure*}[ht!]
\centering
\adjustbox{valign=t}{\begin{minipage}{0.35\textwidth}
\centering
\includegraphics[width=1\textwidth]{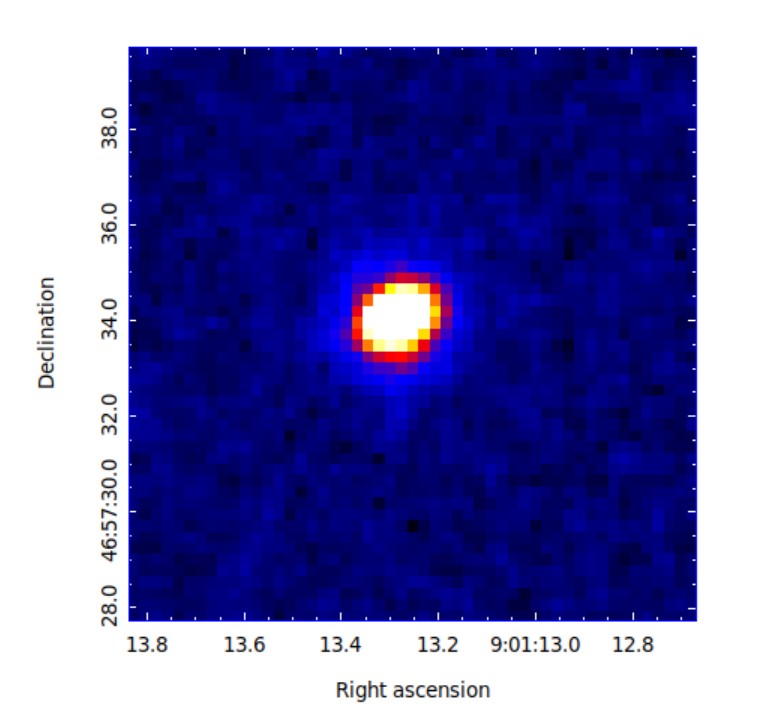}
\end{minipage}}
\adjustbox{valign=t}{\begin{minipage}{0.31\textwidth}
\centering
\includegraphics[width=0.95\textwidth]{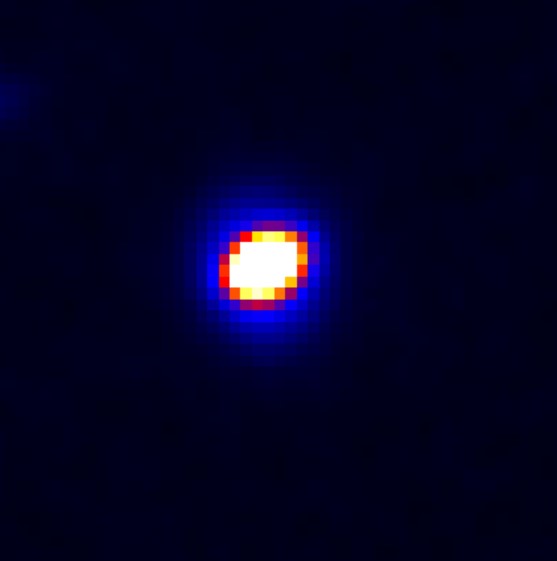}
\end{minipage}}
\adjustbox{valign=t}{\begin{minipage}{0.31\textwidth}
\centering
\includegraphics[width=0.95\textwidth]{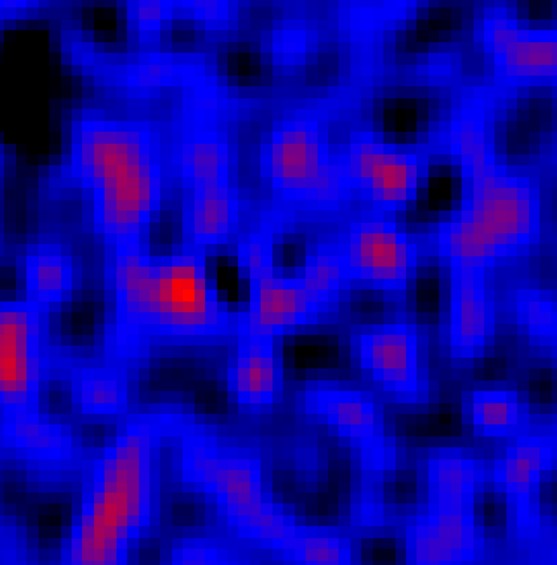}
\end{minipage}}
\hfill
    \caption{\textit{Ks}- band images of SDSS J090113.23+465734.7. The FoV is 5.9\arcsec $/$ 31.8~kpc in all images. \emph{Left panel:} observed image, \emph{middle panel:} model image, and \emph{right panel:} residual image, smoothed over 3px.}  \label{fig:j0901}

\adjustbox{valign=t}{\begin{minipage}{0.49\textwidth}
\centering
\includegraphics[width=0.95\textwidth]{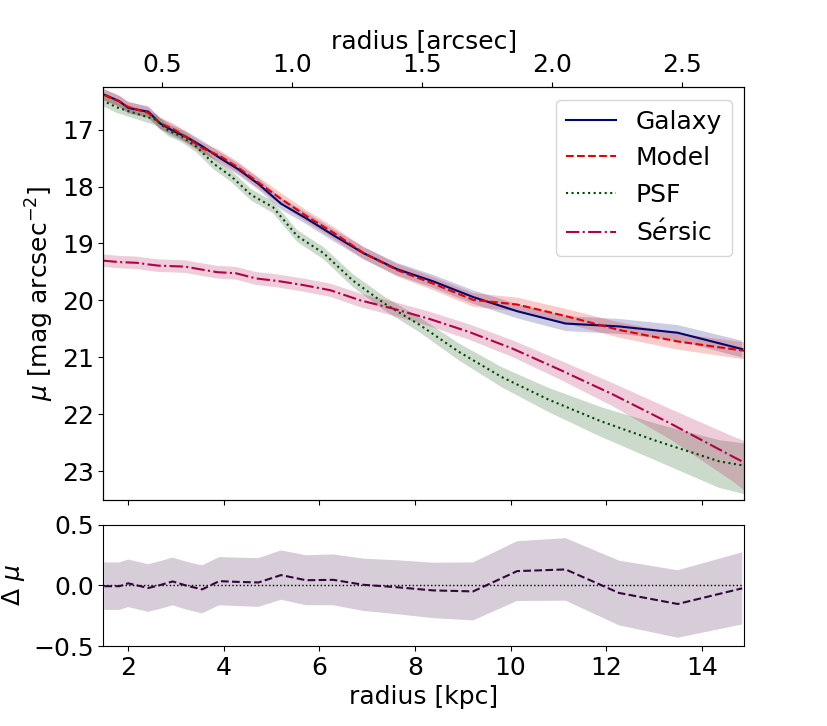}
\caption{Radial surface brightness profile plot of J090113.23+465734.7.  The galaxy component is plotted with a blue line, the model component is plotted with a dashed red line, the PSF component is plotted with a dotted green line, and the S\'{e}rsic 1 component is plotted with a dashed pink line. The shaded area surrounding each profile curve depicts the error linked to each component.}
\label{fig:j0901comps}
\end{minipage}}
\hfill
\adjustbox{valign=t}{\begin{minipage}{0.49\textwidth}
\centering
\includegraphics[width=0.95\textwidth]{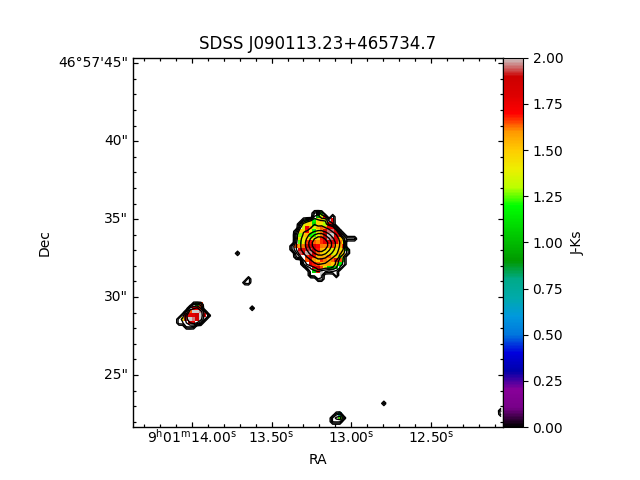}
\caption{\textit{J - $Ks$} colour map of J090113.23+465734.7}
\label{fig:j0901color}
\end{minipage}}
\hfill
\end{figure*}

The observed, model, and the residual image of the source can be seen in Fig.~\ref{fig:j0901}. There are only minor residuals and the reduced $\chi^2_{\nu}$ value of our best fit is acceptable. Based on the residuals and the best fit parameters, especially the high effective radius, we believe that we are modelling the entire galaxy at once. The radial surface brightness profile can be found in Fig.~\ref{fig:j0901comps}. The observed image and model have very similar radial surface brightness profile curves. For the lower radii, the curves are nearly identical. There is some larger deviation toward the higher radii. Our results suggest that the best fit that we have obtained is good. The \textit{J - $Ks$} colour map of the source is shown in Fig.~\ref{fig:j0901color}. Due to the small size of the galaxy, we cannot say anything about the structure. The colours are otherwise in line with what is expected.

\begin{table*}
\caption[]{Best fit parameters of SDSS J090113.23+465734.7. $\chi^2_{\nu}$ = 1.172 $\substack{+0.07 \\ -0.05}$.}
\centering
\begin{tabular}{l l l l l l l}
\hline\hline
Function       & Mag                             & $r_\text{e}$                          & $n$                              & Axial & PA & Notes              \\
             &                                   & (kpc)                            &                                  & ratio & (\textdegree) &  \\ \hline
PSF 1         & 16.15 $\substack{+0.10 \\ -0.10}$ &                                  &                                  &       &               &   \\

S\'{e}rsic  & 17.38 $\substack{+0.11 \\ -0.13}$ & 5.89 $\substack{+0.20 \\ -0.03}$ & 0.42 $\substack{+0.10 \\ -0.09}$ & 0.66 $\substack{+0.00 \\ -0.01}$ & -69.85  $\substack{+0.84 \\ -0.00}$ &  \\

PSF 2          & 18.74 $\substack{+0.10 \\ -0.10}$ &                                  &                                  &       &               & Nearby source  \\

  \hline   
\end{tabular}
\tablefoot{Columns: (1) Function used for modelling, (2) Magnitude, (3) Effective radius, (4) S\'{e}rsic index, (5) Axial ratio, (6) Position angle, (7) Additional notes.}
\label{tab:j0901}
\end{table*}

\subsection{SDSS J093712.32+500852.1}

This source has been detected at 1.4 and 37 GHz (see Table~\ref{tab:additional}), indicating the presence of a jet. This source is the only one of our sources that has also been detected at $\gamma$-rays \citep{2018romano}.

For the PSF model, we used a suitable star from the same image as the source. The best fit was obtained with one PSF and one S\'{e}rsic function, and the corresponding fit values can be found in Table~\ref{tab:j0937}. The S\'{e}rsic function shows properties of both a bulge and a bar component, possibly fitting a combination of the two. Our belief stems from the best fit parameters, especially the small half-light radius, and the colour map. We also fitted a PSF on a nearby source, to have as clean residuals as possible. The PSF modelled the nearby source well, thus suggesting that our PSF model was good.

\begin{figure*}[ht!]
\centering
\adjustbox{valign=t}{\begin{minipage}{0.35\textwidth}
\centering
\includegraphics[width=1\textwidth]{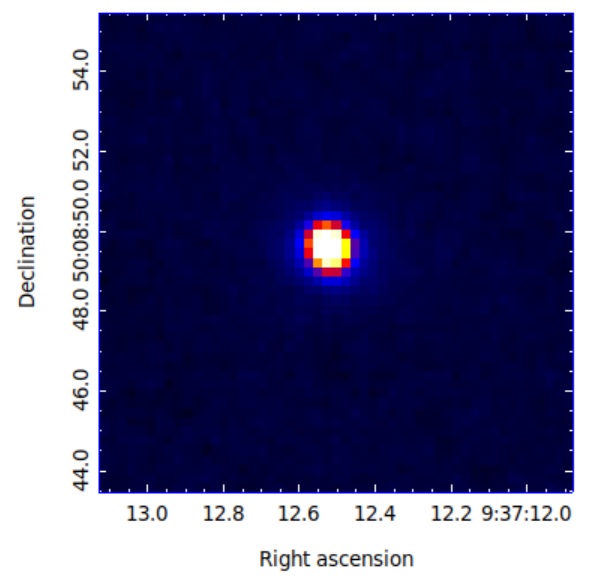}
\end{minipage}}
\adjustbox{valign=t}{\begin{minipage}{0.31\textwidth}
\centering
\includegraphics[width=0.95\textwidth]{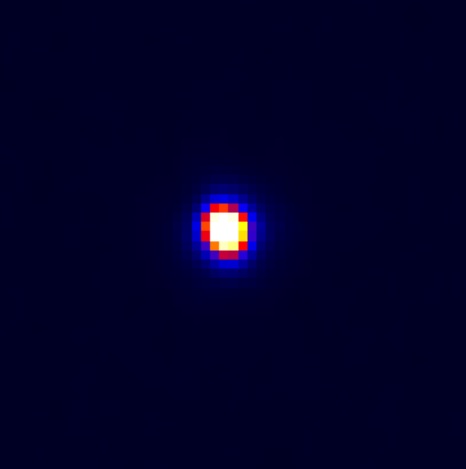}
\end{minipage}}
\adjustbox{valign=t}{\begin{minipage}{0.31\textwidth}
\centering
\includegraphics[width=0.95\textwidth]{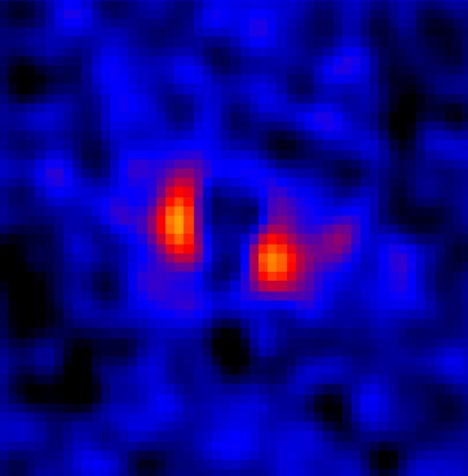}
\end{minipage}}
\hfill
    \caption{\textit{Ks}- band images of SDSS J093712.32+500852.1. The FoV is 5.9\arcsec $/$ 23.7~kpc in all images. \emph{Left panel:} observed image, \emph{middle panel:} model image, and \emph{right panel:} residual image, smoothed over 3px.}  \label{fig:j0937}

\adjustbox{valign=t}{\begin{minipage}{0.49\textwidth}
\centering
\includegraphics[width=0.95\textwidth]{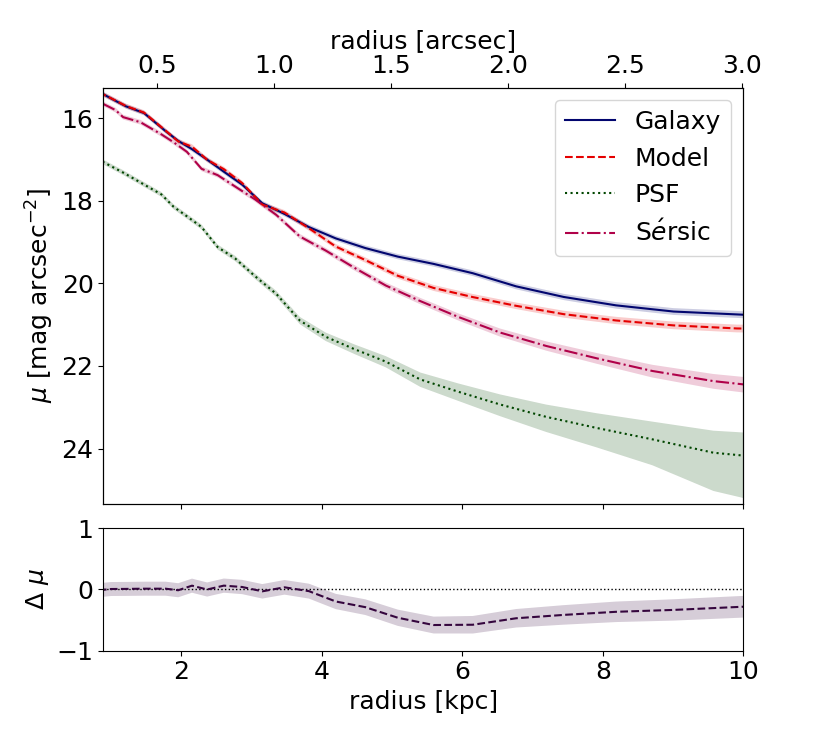}
\caption{Radial surface brightness profile plot of SDSS J093712.32+500852.1.  The galaxy component is plotted with a blue line, the model component is plotted with a dashed red line, the PSF component is plotted with a dotted green line, and the S\'{e}rsic component is plotted with a dashed pink line. The shaded area surrounding each profile curve depicts the error linked to each component.}
\label{fig:j0937comps}
\end{minipage}}
\hfill
\adjustbox{valign=t}{\begin{minipage}{0.49\textwidth}
\centering
\centering
\includegraphics[width=0.95\textwidth]{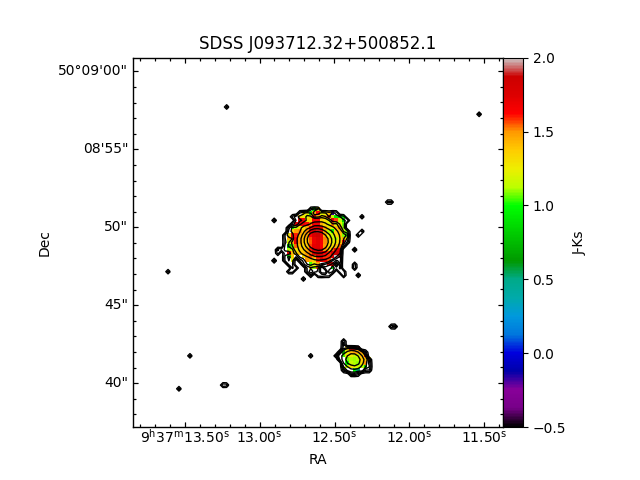}
\caption{\textit{J - $Ks$} colour map of SDSS J093712.32+500852.1}
\label{fig:j0937color}
\end{minipage}}
\hfill
\end{figure*}

The observed, model, and the residual image of the source can be seen in Fig.~\ref{fig:j0937}. There are some residuals left, on both sides of the galaxy. If the S\'{e}rsic component modelled the bulge/bar combination, it would indicate that the residuals are from the disk component that we were unable to model. Either way, the results point to this galaxy having at least a bar, and most probably disk-like morphology.

The reduced $\chi^2_{\nu}$ value is of the higher end of our sample, however, the value is still good. The radial surface brightness profile can be seen in Fig.~\ref{fig:j0937comps}. The model and observed images curves are, as can be guessed by the reduced $\chi^2_{\nu}$ value, are slightly different. The \textit{J - $Ks$} colour map of the source is shown in Fig.~\ref{fig:j0937color}. The colour map supports the results of the modelling, as a bar-like structure, with a PA similar to the S\'{e}rsic component, can be seen. The colour map is within the acceptable range for a Seyfert 1 galaxy.

\begin{table*}
\caption[]{Best fit parameters of SDSS J093712.32+500852.1. $\chi^2_{\nu}$ = 1.249 $\substack{+0.02 \\ -0.01}$. }
\centering
\begin{tabular}{l l l l l l l}
\hline\hline
Function       & Mag                             & $r_\text{e}$                          & $n$                              & Axial & PA  & Notes             \\
             &                                   & (kpc)                            &                                  & ratio & (\textdegree) &  \\ \hline
PSF 1         & 17.16 $\substack{+0.06 \\ -0.06}$ &                                  &                                  &       &               &   \\

S\'{e}rsic  & 15.50 $\substack{+0.05 \\ -0.05}$ & 1.00 $\substack{+0.00 \\ -0.00}$ & 1.15 $\substack{+0.00 \\ -0.03}$ & 0.40 $\substack{+0.04 \\ -0.00}$ & 83.05  $\substack{+0.01 \\ -0.05}$ &  Bulge and bar \\

PSF 2          & 18.37 $\substack{+0.06 \\ -0.05}$ &                                  &                                  &       &               & Nearby source  \\

  \hline   
\end{tabular}
\tablefoot{Columns: (1) Function used for modelling, (2) Magnitude, (3) Effective radius, (4) S\'{e}rsic index, (5) Axial ratio, (6) Position angle, (7) Additional notes.}
\label{tab:j0937}
\end{table*}

\subsection{SDSS J102906.68+555625.2}

This source is a jetted NLS1, detected only at 37~GHz (see Table~\ref{tab:additional}). The detection at a high radio frequency strongly suggests the presence of powerful relativistic jets. Due to the aforementioned issues with the observations, we observed this source only in \textit{Ks}- band. The observations of this source struggled with the worst seeing of any of our observations. Nearly half of the exposures were deemed unusable due to significantly higher seeing. 

We used DAOPHOT to model our PSF. The best fit values can be found in Table~\ref{tab:j1029}. The components fitting the galaxy are reasonable, but the error in the S\'{e}rsic index is relatively high ($\sim$+2). Due to the error, our findings are inconclusive regarding the morphology. We suspect that the error of the S\'{e}rsic index, $n$ = 2.86$\substack{+2.09 \\ -0.00}$, is due to the turbulent sky that resulted in poor seeing. The effective radius is 2.14~kpc, and thus S\'{e}rsic might not be fully resolved. A nearby source was modelled completely with a single PSF. 

\begin{figure*}
\centering
\adjustbox{valign=t}{\begin{minipage}{0.35\textwidth}
\centering
\includegraphics[width=1\textwidth]{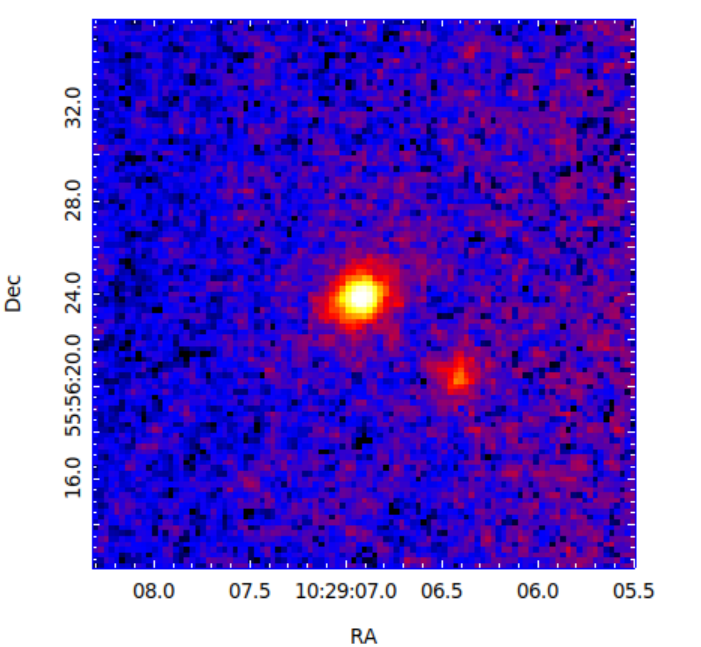}
\end{minipage}}
\adjustbox{valign=t}{\begin{minipage}{0.31\textwidth}
\centering
\includegraphics[width=0.95\textwidth]{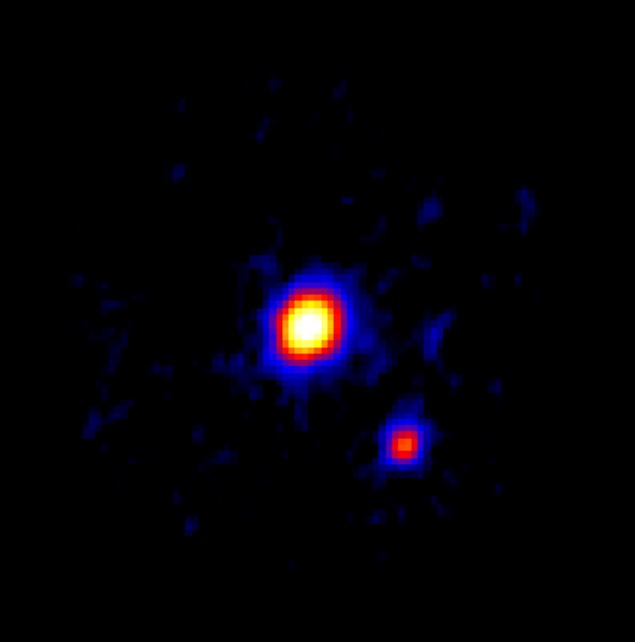}
\end{minipage}}
\adjustbox{valign=t}{\begin{minipage}{0.31\textwidth}
\centering
\includegraphics[width=0.95\textwidth]{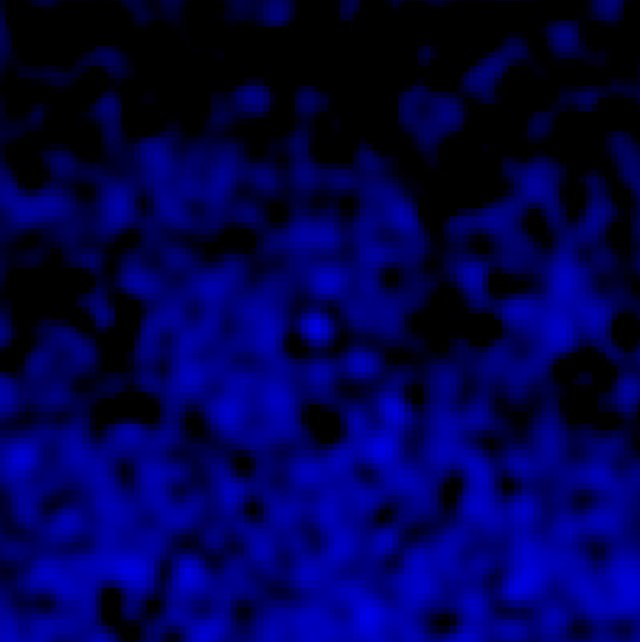}
\end{minipage}}
\hfill
    \caption{\textit{Ks}- band images of SDSS J102906.68+555625.2. The FoV is 8.4\arcsec $/$ 47.0~kpc in all images. \emph{Left panel:} observed image, \emph{middle panel:} model image, and \emph{right panel:} residual image, smoothed over 3px.}  \label{fig:j1029}

\adjustbox{valign=t}{\begin{minipage}{0.49\textwidth}
\centering
\includegraphics[width=1.0\textwidth]{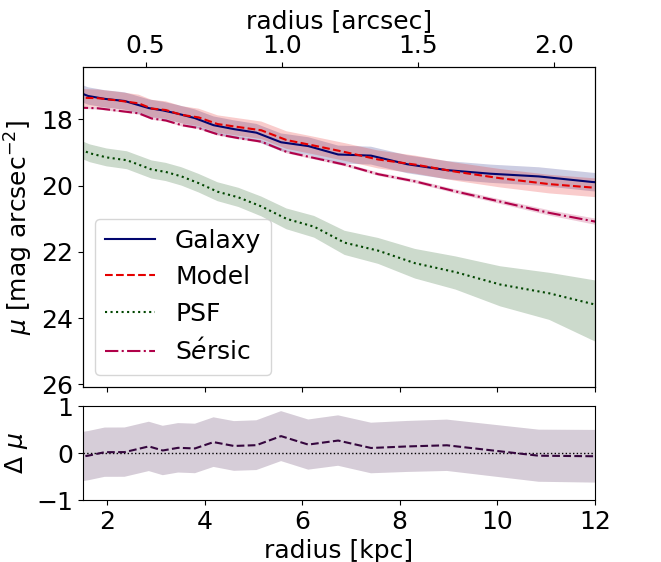}
\caption{Radial surface brightness profile plot of SDSS J102906.68+555625.2.  The galaxy component is plotted with a blue line, the model component is plotted with a dashed red line, the PSF component is plotted with a dotted green line, and the S\'{e}rsic component is plotted with a dashed pink line. The shaded area surrounding the profile curve depicts the error linked to the component.}
\label{fig:j1029comps}
\end{minipage}}
\hfill
\end{figure*}

The observed, model, and the residual image of the source can be seen in Fig.~\ref{fig:j1029}. The reduced $\chi^2_{\nu}$ value is good and there are no visible residuals left. The radial surface brightness profile is in Fig.~\ref{fig:j1029comps}. The light curves for all but the PSF have small errors and even the error in the PSF is acceptable. We have no colour map of this galaxy, however, due to the small size of the galaxy and the high redshift, the colour map would most likely not have helped us to determine the morphology or any components of the galaxy.

\begin{table*}
\caption[]{Best fit parameters of SDSS J102906.68+555625.2. $\chi^2_{\nu}$ = 1.194 $\substack{+0.01 \\ -0.05}$.}
\centering
\begin{tabular}{l l l l l l l}
\hline\hline
Function       & Mag                             & $r_\text{e}$                          & $n$                              & Axial & PA & Notes              \\
             &                                   & (kpc)                            &                                  & ratio & (\textdegree) &  \\ \hline
PSF 1          & 18.39 $\substack{+0.17 \\ -0.21}$ &                                  &                                  &       &               &   \\

S\'{e}rsic  & 16.58 $\substack{+0.00 \\ -0.13}$ & 2.14 $\substack{+0.84 \\ -0.22}$ & 2.86 $\substack{+2.09 \\ -0.00}$ & 0.57 $\substack{+0.00 \\ -0.02}$ & 31.97  $\substack{+1.22 \\ -4.28}$ &  \\

PSF 2          & 18.51 $\substack{+0.01 \\ -0.05}$ &                                  &                                  &       &               &  Nearby source  \\

  \hline   
\end{tabular}
\tablefoot{Columns: (1) Function used for modelling, (2) Magnitude, (3) Effective radius, (4) S\'{e}rsic index, (5) Axial ratio, (6) Position angle, (7) Additional notes.}
\label{tab:j1029}
\end{table*}

\subsection{SDSS J103123.73+423439.2}

Earlier detected at 1.4~GHz, this source has also been detected at 37 GHz (see Table~\ref{tab:additional}). According to \cite{2018berton}, who observed the object with the JVLA at 5 GHz, the source could probably be considered as a CSS source, due to its compact structure. 

A visual inspection of this source suggests that the morphology is elongated. The PSF model was created with DAOPHOT. The best fit was obtained with one PSF and one S\'{e}rsic function. The best fit values can be found in Table~\ref{tab:j1031}. Based on the best fit parameters, we believe that we have modelled the whole galaxy with these functions. This is supported by the effective radius being of a reasonable size for a galaxy with disk-like morphology. Though the S\'{e}rsic index, \textit{n} = 2.57$\substack{+1.13 \\ -0.00}$, is on the higher side for disk-like galaxies, the elongated morphology still suggests disk-like morphology.

\begin{figure*}
\centering
\adjustbox{valign=t}{\begin{minipage}{0.35\textwidth}
\centering
\includegraphics[width=1\textwidth]{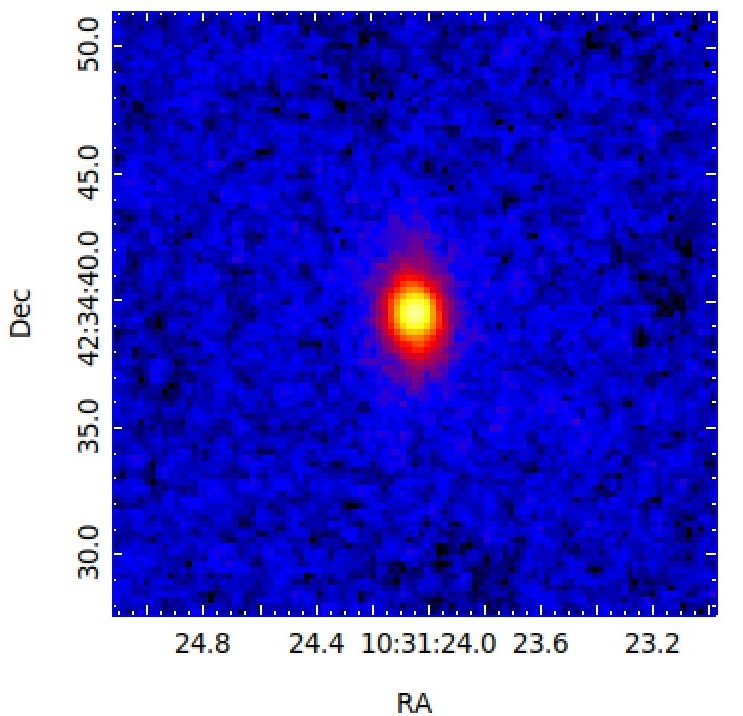}
\end{minipage}}
\adjustbox{valign=t}{\begin{minipage}{0.31\textwidth}
\centering
\includegraphics[width=0.95\textwidth]{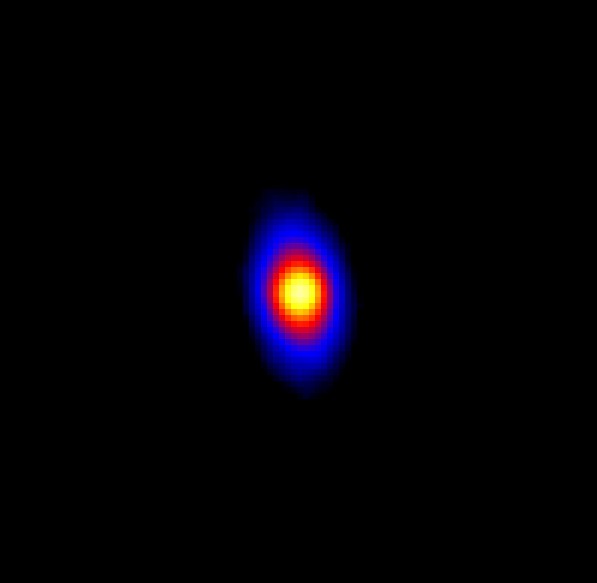}
\end{minipage}}
\adjustbox{valign=t}{\begin{minipage}{0.31\textwidth}
\centering
\includegraphics[width=0.95\textwidth]{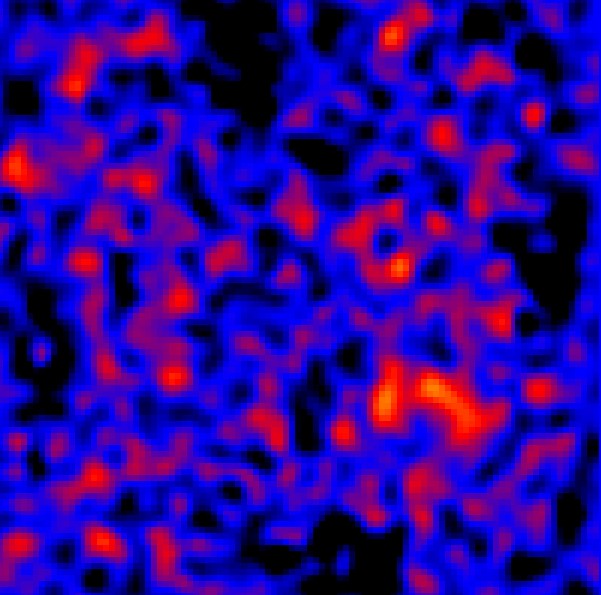}
\end{minipage}}
\hfill
    \caption{\textit{Ks}- band images of SDSS J103123.73+423439.2. The FoV is 23.4\arcsec$/$117.1~kpc in all images. \emph{Left panel:} observed image, \emph{middle panel:} model image, and \emph{right panel:} residual image, smoothed over 3px.}  \label{fig:j1031}

\adjustbox{valign=t}{\begin{minipage}{0.49\textwidth}
\centering
\includegraphics[width=0.95\textwidth]{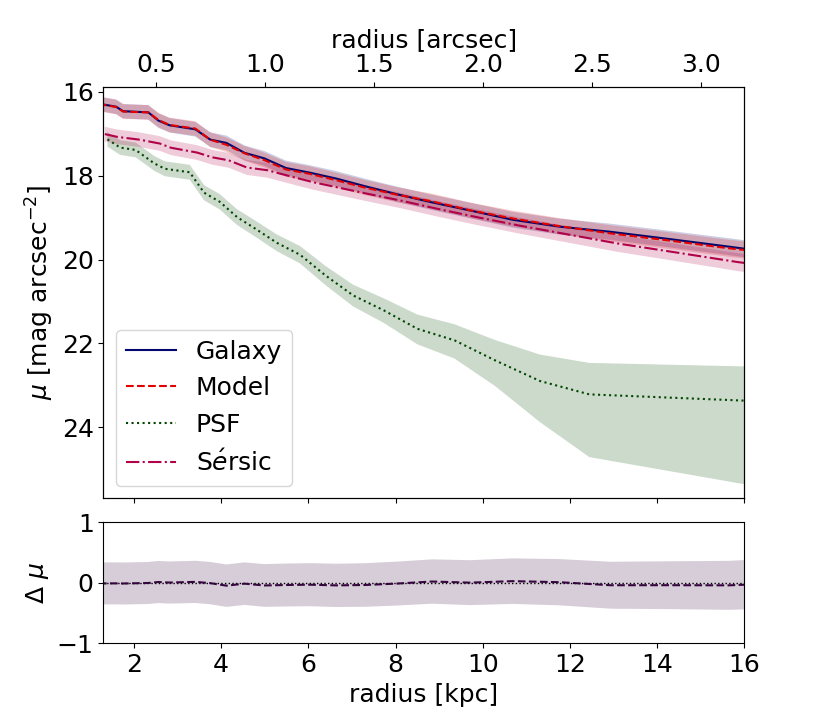}
\caption{Radial surface brightness profile plot of SDSS J103123.73+423439.2.  The galaxy component is plotted with a blue line, the model component is plotted with a dashed red line, the PSF component is plotted with a dotted green line, and the S\'{e}rsic component is plotted with a dashed pink line. The shaded area surrounding each profile curve depicts the error linked to each component.}
\label{fig:j1031comps}
\end{minipage}}
\hfill
\adjustbox{valign=t}{\begin{minipage}{0.49\textwidth}
\centering
\includegraphics[width=0.95\textwidth]{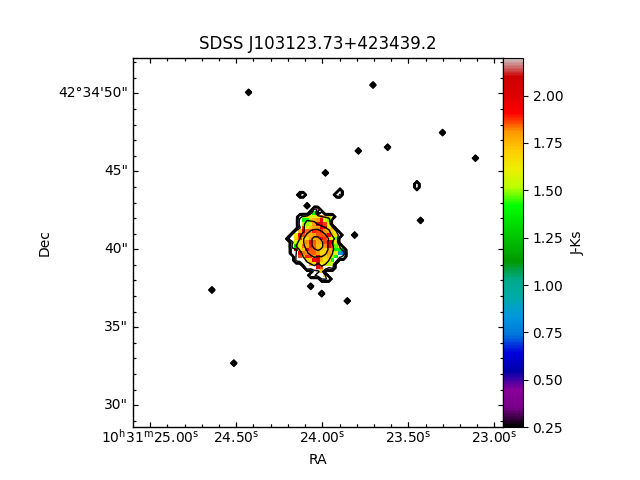}
\caption{\textit{J - $Ks$} colour map of SDSS J103123.73+423439.2.}
\label{fig:j1031color}
\end{minipage}}
\hfill
\end{figure*}

The observed, model, and the residual image of the source can be seen in Fig.~\ref{fig:j1031}. The residual image is very clean, with little to no residuals visible in the image. The residuals and the reduced $\chi^2_{\nu}$ value suggest that the source has been modelled well. This is supported by the model and observed image radial surface brightness profiles, seen in Fig.~\ref{fig:j0937comps}, being nearly identical. In the radial surface brightness profile (Fig.~\ref{fig:j1031comps}) the S\'{e}rsic curve is very close to the model and observed image curves, dominating the overall light profile, and indicating that the AGN is quite faint. The PSF curve has a relatively large error toward the higher radii, due to being near the sky level. The \textit{J - $Ks$} colour map of the source is shown in Fig.~\ref{fig:j1031color}. This is another very small galaxy and its structure cannot be determined from this colour map. In general it is in agreement with what is expected to be seen in a Seyfert 1 galaxy.

\begin{table*}
\caption[]{Best fit parameters of SDSS J103123.73+423439.2. $\chi^2_{\nu}$ = 1.039 $\substack{+0.01 \\ -0.00}$. }
\centering
\begin{tabular}{l l l l l l l}
\hline\hline
Function       & Mag                             & $r_\text{e}$                          & $n$                              & Axial & PA  & Notes             \\
             &                                   & (kpc)                            &                                  & ratio & (\textdegree) &  \\ \hline
PSF          & 16.66 $\substack{+0.73 \\ -0.71}$ &                                  &                                  &       &               &   \\

S\'{e}rsic   & 15.18 $\substack{+0.71 \\ -0.72}$ & 7.60 $\substack{+0.86 \\ -0.00}$ & 2.57 $\substack{+1.13 \\ -0.00}$ & 0.46 $\substack{+0.01 \\ -0.00}$ & -81.46  $\substack{+0.18 \\ -0.12}$ &   \\

  \hline   
\end{tabular}
\tablefoot{Columns: (1) Function used for modelling, (2) Magnitude, (3) Effective radius, (4) S\'{e}rsic index, (5) Axial ratio, (6) Position angle, (7) Additional notes.}
\label{tab:j1031}
\end{table*}

\subsection{SDSS J122844.81+501751.2}

This source has been detected at 37 GHz (see Table~\ref{tab:additional}), however, there are no other detections in the radio regime.

Based on visual inspection it appears that there is another source seemingly close to our galaxy. However, without a spectrum of the possible companion, it is impossible to say if it is at the same redshift as our source, or instead a background or a foreground source. The morphologies of the two sources seem undisturbed, indicating that if they are at the same redshift, any interaction has not taken place yet. We reference to this source as tail in the best fit parameter table, Table~\ref{tab:j1228}. A separate PSF star gave us the best PSF. We obtained the best fit values with one PSF and two S\'{e}rsic functions. Based on a visual analysis of the model of S\'{e}rsic 1, it appears that the function mainly models the companion, however, it is possible that the function also models a the western part of the main galaxy. The S\'{e}rsic index of the companion, $n_\text{S\'{e}rsic 1}$ = 1.21$\substack{+0.00 \\ -0.08}$, suggests a disk-like morphology. The S\'{e}rsic 2, modelling the NLS1, has an index of $n_\text{S\'{e}rsic 2}$ = 2.01$\substack{+0.59 \\ -0.12}$ and an effective radius of 1.15~kpc, indicating that it may represent a bulge component, which, however, might not be fully resolved. There was a nearby source that required two components, a PSF and a S\'{e}rsic, to be modelled completely.

\begin{figure*}
\centering
\adjustbox{valign=t}{\begin{minipage}{0.35\textwidth}
\centering
\includegraphics[width=1\textwidth]{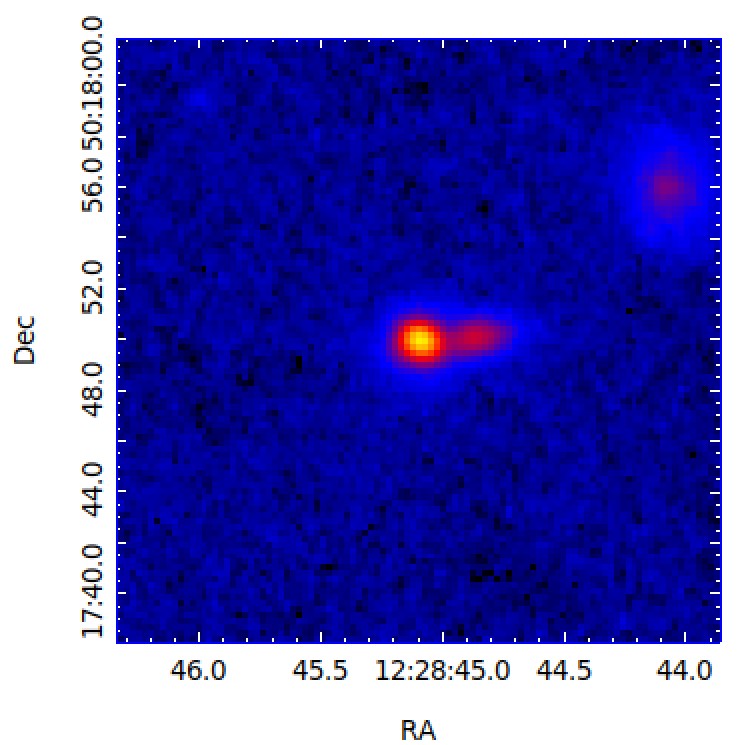}
\end{minipage}}
\adjustbox{valign=t}{\begin{minipage}{0.31\textwidth}
\centering
\includegraphics[width=0.95\textwidth]{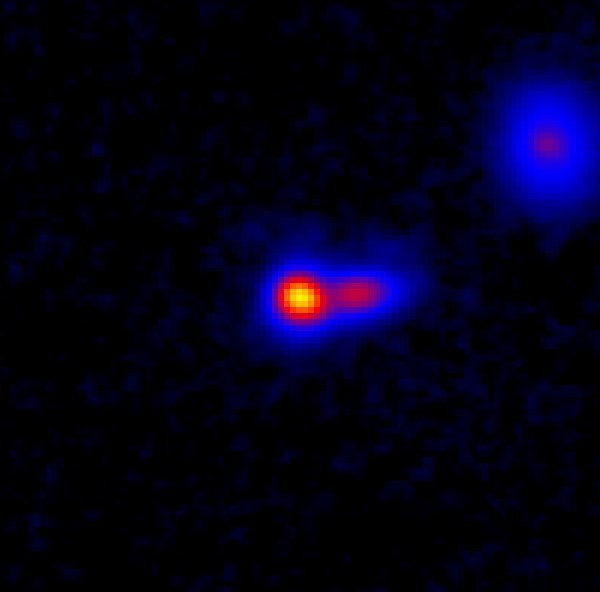}
\end{minipage}}
\adjustbox{valign=t}{\begin{minipage}{0.31\textwidth}
\centering
\includegraphics[width=0.95\textwidth]{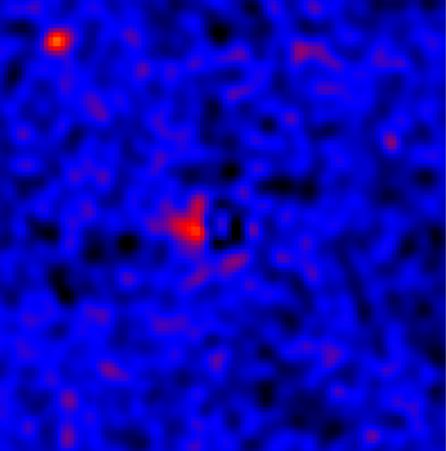}
\end{minipage}}
\hfill
    \caption{\textit{Ks}- band images of SDSS J122844.81+501751.2. The FoV is 23.4\arcsec    $/$ 91.5~kpc in all images. \emph{Left panel:} observed image, \emph{middle panel:} model image, and \emph{right panel:} residual image, smoothed over 3px.}  \label{fig:j1228}

\adjustbox{valign=t}{\begin{minipage}{0.49\textwidth}
\centering
\includegraphics[width=0.95\textwidth]{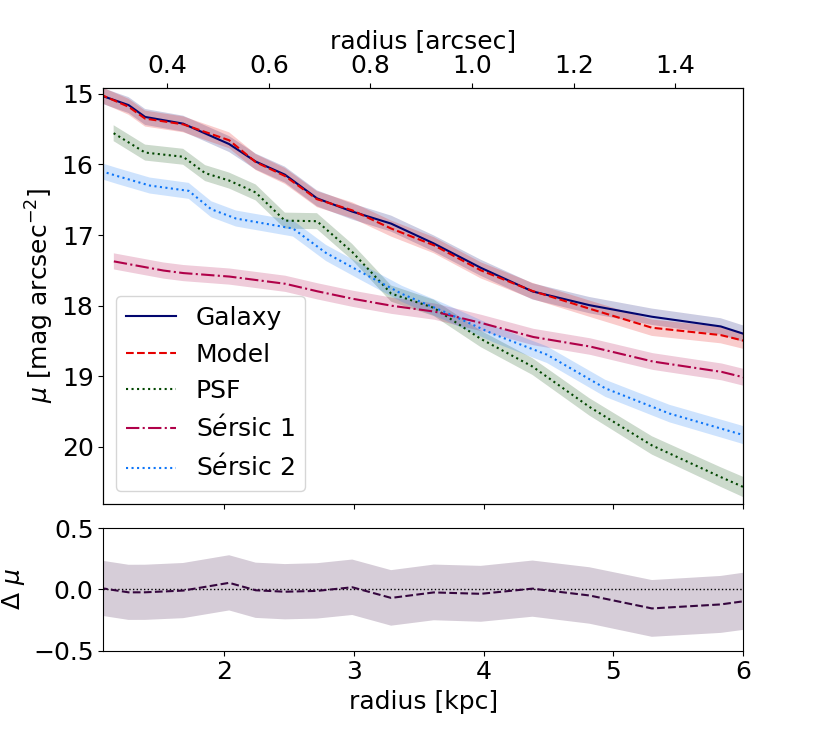}
\caption{Radial surface brightness profile plot of SDSS J122844.81+501751.2. The galaxy component is plotted with a blue line, the model component is plotted with a dashed red line, the PSF component is plotted with a dotted green line, the S\'{e}rsic 1 component is plotted with a dashed pink line, and finally the S\'{e}rsic 2 component is plotted with a dotted light blue line. The shaded area surrounding each profile curve depicts the error linked to each component.}
\label{fig:j1228comps}
\end{minipage}}
\hfill
\adjustbox{valign=t}{\begin{minipage}{0.49\textwidth}
\centering
\includegraphics[width=0.95\textwidth]{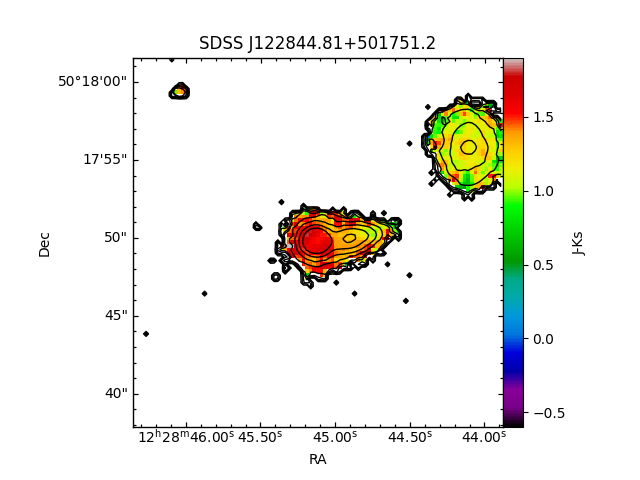}
\caption{\textit{J - $Ks$} colour map of SDSS J122844.81+501751.2.}
\label{fig:j1228color}
\end{minipage}}
\hfill
\end{figure*}

The observed, model, and the residual image of the source can be seen in Fig.~\ref{fig:j1228}. Residuals can be seen on the east side of the galaxy. They could be due to the fact that the S\'{e}rsic 1 component, which mostly models the companion, has quite a large effective radius, in practice also modelling the western part of the NLS1 host. However, since it is not concentric with the NLS1, it also creates a gradient overlaying the NLS1 host, which is probably why we were not able to model the possible disk of the NLS1 host. If the tail is a genuine part of the system, then it is possible that this is an interacting source.


The radial surface brightness profile can be seen in Fig.~\ref{fig:j1228comps}. As can be seen in the plot, the errors remain steady through each curve throughout the entire plot. The observed image and model image curves are very similar. The \textit{J - $Ks$} colour map of the source is shown in Fig.~\ref{fig:j1228color}. The galaxy has a very large red area in the colour map, reaching all the way to the edge. The structure of the galaxy is not distinguishable from the colour map. The nearby source can be seen in the right-hand corner of the colour map. Our source of interest has a colour map that is in-line with what is expected to be seen in a Seyfert 1 galaxy. With the difference in the $J$ and $Ks$ magnitude being above 1.3 for the AGN, this counts as very red with the red possibly signifying dust extinction.

\begin{table*}
\caption[]{Best fit parameters of SDSS J122844.81+501751.2. $\chi^2_{\nu}$ = 1.273 $\substack{+0.03 \\ -0.01}$. }
\centering
\begin{tabular}{l l l l l l l}
\hline\hline
Function       & Mag                             & $r_\text{e}$                          & $n$                              & Axial & PA  & Notes             \\
             &                                   & (kpc)                            &                                  & ratio & (\textdegree) &  \\ \hline
PSF          & 15.55 $\substack{+0.09 \\ -0.09}$ &                                  &                                  &       &               &   \\

S\'{e}rsic 1  & 16.11 $\substack{+0.09 \\ -0.10}$ & 3.91 $\substack{+0.00 \\ -0.16}$ & 1.21 $\substack{+0.00 \\ -0.08}$ & 0.40 $\substack{+0.01 \\ -0.00}$ & 10.71  $\substack{+0.23 \\ -0.00}$ & Tail \\

S\'{e}rsic 2  & 15.58 $\substack{+0.09 \\ -0.11}$ & 1.15 $\substack{+0.03 \\ -0.05}$ & 2.01 $\substack{+0.59 \\ -0.12}$ & 0.72 $\substack{+0.02 \\ -0.00}$ & 80.96  $\substack{+3.24 \\ -0.62}$ & Bulge \\

PSF          & 19.21 $\substack{+0.10 \\ -0.09}$ &                                  &                                  &       &               & Nearby source   \\

S\'{e}rsic 3  & 15.97 $\substack{+0.08 \\ -0.09}$ & 6.28 $\substack{+0.07 \\ -0.05}$ & 0.57 $\substack{+0.02 \\ -0.01}$ & 0.79 $\substack{+0.00 \\ -0.00}$ & -85.09  $\substack{+0.20 \\ -0.10}$ & Nearby source \\

  \hline   
\end{tabular}
\tablefoot{Columns: (1) Function used for modelling, (2) Magnitude, (3) Effective radius, (4) S\'{e}rsic index, (5) Axial ratio, (6) Position angle, (7) Additional notes.}
\label{tab:j1228}
\end{table*}

\subsection{SDSS J123220.11+495721.8}

The detection at 37 GHz suggests the presence of a relativistic jet in this source (see Table~\ref{tab:additional}), however, there are no other radio detections.

We used DAOPHOT to obtain a good PSF model. We obtained the best fit with one PSF and two S\'{e}rsic functions. The best fit values can be found in Table~\ref{tab:j1232}, suggesting disk-like morphology. Based on the best fit parameters of S\'{e}rsic 1, it is possible that S\'{e}rsic 1 is not fully resolved. This is supported by the fact that the axial ratio is almost zero.

\begin{figure*}
\centering
\adjustbox{valign=t}{\begin{minipage}{0.35\textwidth}
\centering
\includegraphics[width=1\textwidth]{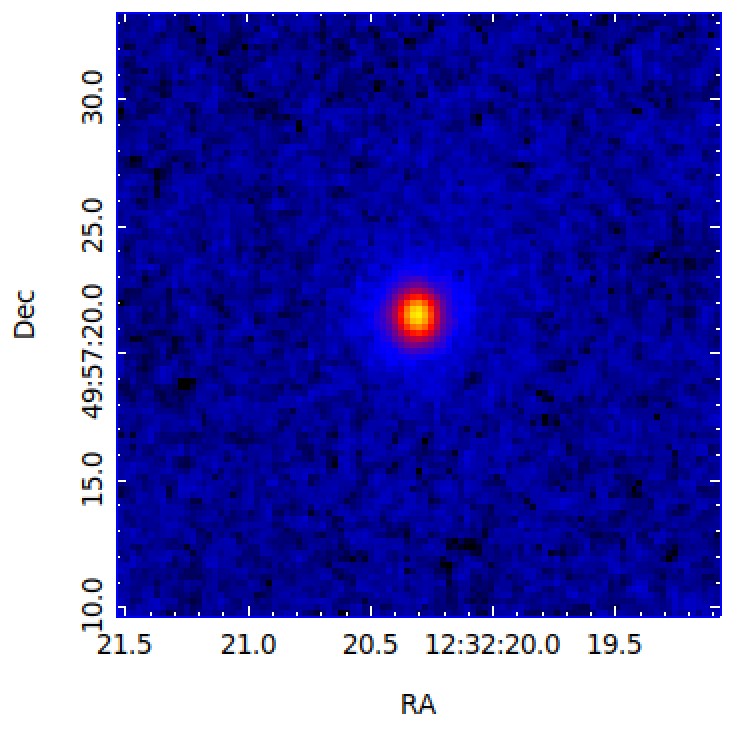}
\end{minipage}}
\adjustbox{valign=t}{\begin{minipage}{0.31\textwidth}
\centering
\includegraphics[width=0.95\textwidth]{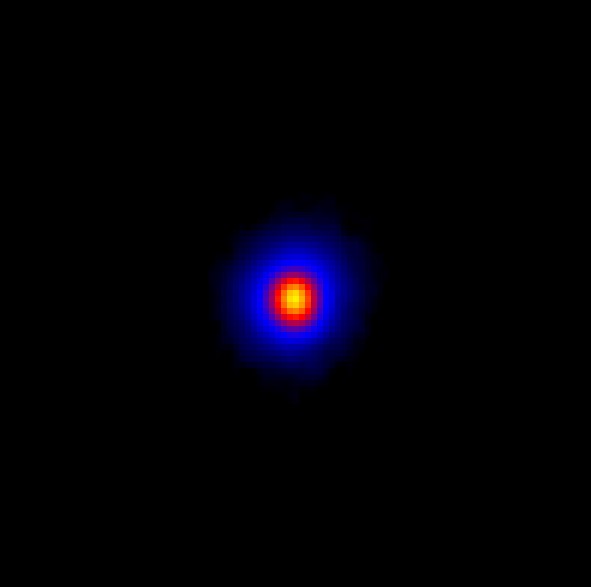}
\end{minipage}}
\adjustbox{valign=t}{\begin{minipage}{0.31\textwidth}
\centering
\includegraphics[width=0.95\textwidth]{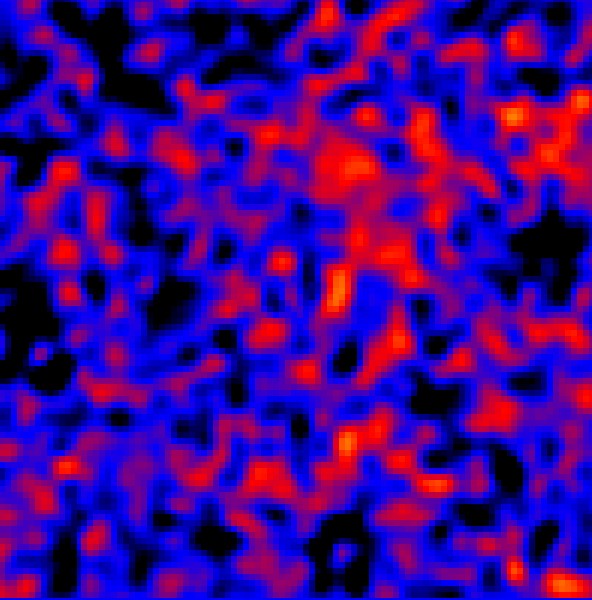}
\end{minipage}}
\hfill
    \caption{\textit{Ks}- band images of SDSS J123220.11+495721.8. The FoV is 23.4\arcsec     $/$ 91.3~kpc in all images. \emph{Left panel:} observed image, \emph{middle panel:} model image, and \emph{right panel:} residual image, smoothed over 3px.}  \label{fig:j1232}

\adjustbox{valign=t}{\begin{minipage}{0.49\textwidth}
\centering
\includegraphics[width=0.95\textwidth]{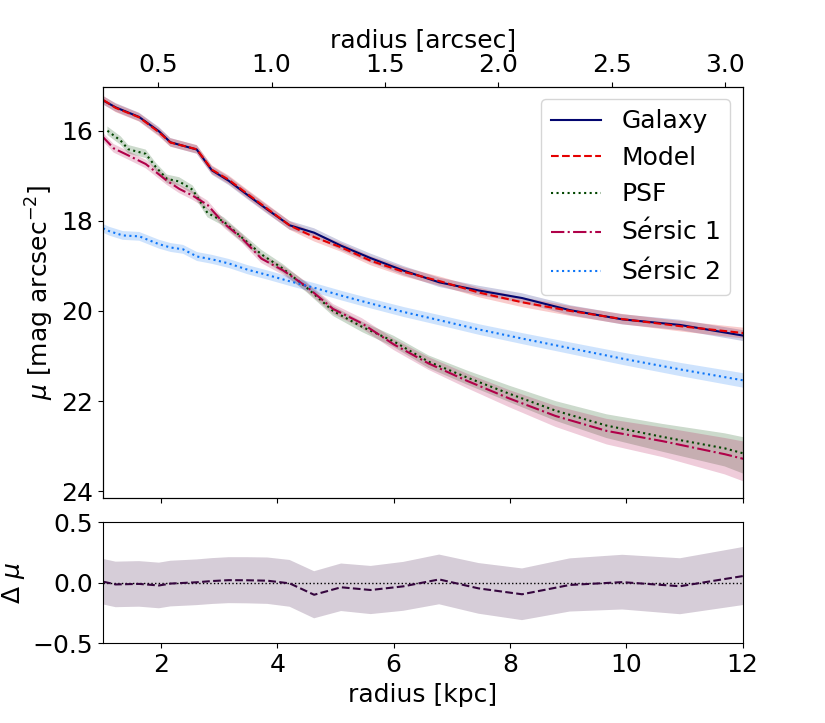}
\caption{Radial surface brightness profile plot of SDSS J123220.11+495721.8. The galaxy component is plotted with a blue line, the model component is plotted with a dashed red line, the PSF component is plotted with a dotted green line, the S\'{e}rsic 1 component is plotted with a dashed pink line, and finally the S\'{e}rsic 2 component is plotted with a dotted light blue line. The shaded area surrounding each profile curve depicts the error linked to each component.}
\label{fig:j1232comps}
\end{minipage}}
\hfill
\adjustbox{valign=t}{\begin{minipage}{0.49\textwidth}
\centering
\includegraphics[width=0.95\textwidth]{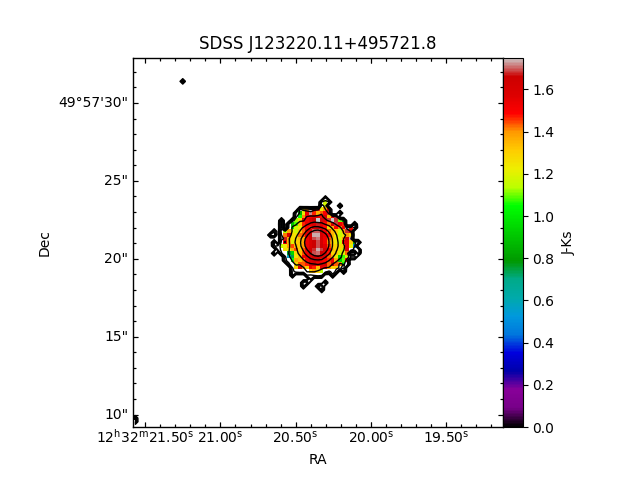}
\caption{\textit{J - $Ks$} colour map of SDSS J123220.11+495721.8.}
\label{fig:j1232color}
\end{minipage}}
\hfill
\end{figure*}

The observed, model, and the residual image of the source can be seen in Fig.~\ref{fig:j1232}. The residual image is quite clean, with only slight residuals remaining.  The reduced $\chi^2_{\nu}$ value is very good. The radial surface brightness profile can be seen in Fig.~\ref{fig:j1232comps}. The model and observed image curves are nearly identical throughout the entire plot, and the errors are very small. The \textit{J - $Ks$} colour map of the source is shown in Fig.~\ref{fig:j1232color}. The map displays a bar-like structure near the central region of the galaxy, however, it is also possible that this structure is simply caused by dust extinction.

\begin{table*}
\caption[]{Best fit parameters of SDSS J123220.11+495721.8. $\chi^2_{\nu}$ = 1.052 $\substack{+0.02 \\ -0.01}$.}
\centering
\begin{tabular}{l l l l l l l}
\hline\hline
Function       & Mag                             & $r_\text{e}$                          & $n$                              & Axial & PA  & Notes             \\
             &                                   & (kpc)                            &                                  & ratio & (\textdegree) &  \\ \hline
PSF          & 16.02 $\substack{+0.09 \\ -0.09}$ &                                  &                                  &       &               &   \\

S\'{e}rsic 1  & 16.10 $\substack{+0.09 \\ -0.09}$ & 0.71 $\substack{+0.00 \\ -0.00}$ & 1.21 $\substack{+0.03 \\ -0.02}$ & 0.10 $\substack{+0.00 \\ -0.00}$ & -34.75  $\substack{+0.01 \\ -0.03}$ &  \\

S\'{e}rsic 2  & 16.29 $\substack{+0.10 \\ -0.09}$ & 5.61 $\substack{+0.11 \\ -0.13}$ & 2.25 $\substack{+0.03 \\ -0.04}$ & 0.85 $\substack{+0.01 \\ -0.01}$ & 59.70  $\substack{+0.21 \\ -0.25}$ &  \\

  \hline   
\end{tabular}
\tablefoot{Columns: (1) Function used for modelling, (2) Magnitude, (3) Effective radius, (4) S\'{e}rsic index, (5) Axial ratio, (6) Position angle, (7) Additional notes.}
\label{tab:j1232}
\end{table*}

\subsection{SDSS J125635.89+500852.4}

This galaxy has been detected at 1.4~GHz in the past, and it has also been detected at 37 GHz  (see Table~\ref{tab:additional}).

We used a DAOPHOT PSF for the fitting. The best fit was achieved with one PSF and one S\'{e}rsic function. The best fit values can be found in Table~\ref{tab:j1256}. The parameters of the S\'{e}rsic function suggest that we are modelling the entire galaxy at once. Because the errors show significant deviation in the S\'{e}rsic index, \textit{n} = 2.46$\substack{+2.31 \\ -0.62}$, we do not find our results conclusive on the morphology of the source. We also fitted a PSF to a nearby source to obtain as clean residuals as possible. This modelled the nearby source completely and is therefore a good indicator that our PSF model is accurate. 

\begin{figure*}
\centering
\adjustbox{valign=t}{\begin{minipage}{0.35\textwidth}
\centering
\includegraphics[width=1\textwidth]{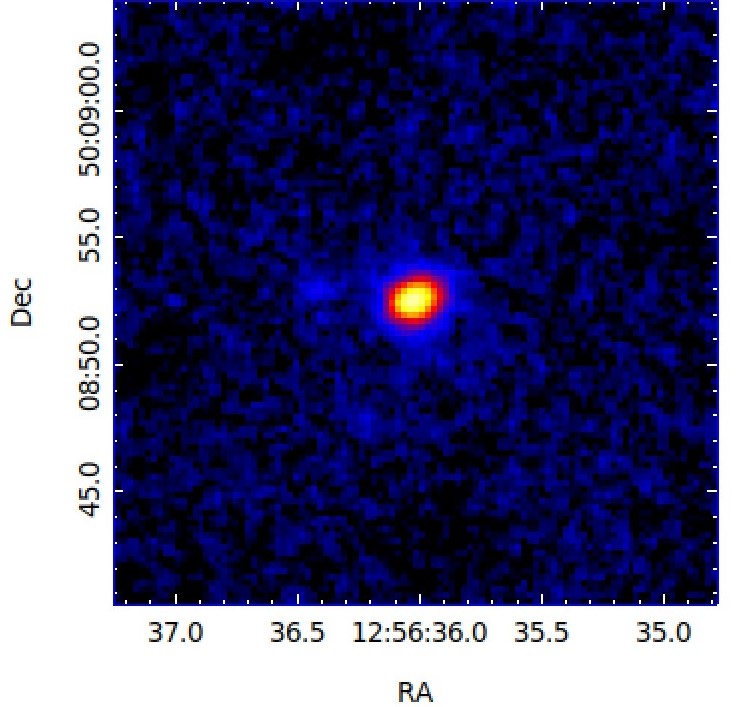}
\end{minipage}}
\adjustbox{valign=t}{\begin{minipage}{0.31\textwidth}
\centering
\includegraphics[width=0.95\textwidth]{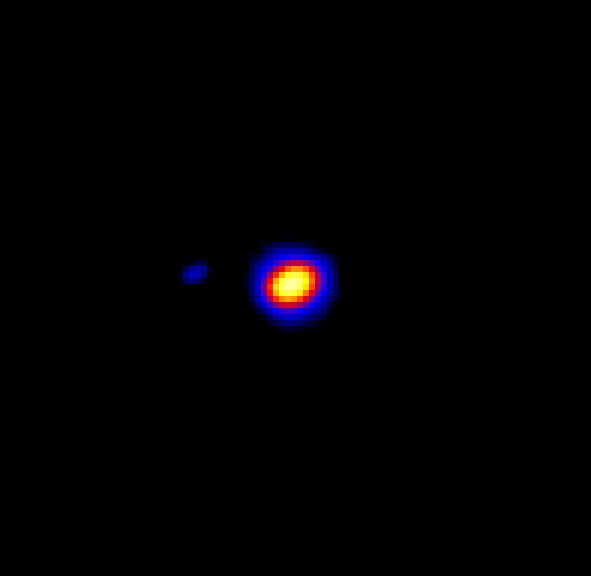}
\end{minipage}}
\adjustbox{valign=t}{\begin{minipage}{0.31\textwidth}
\centering
\includegraphics[width=0.95\textwidth]{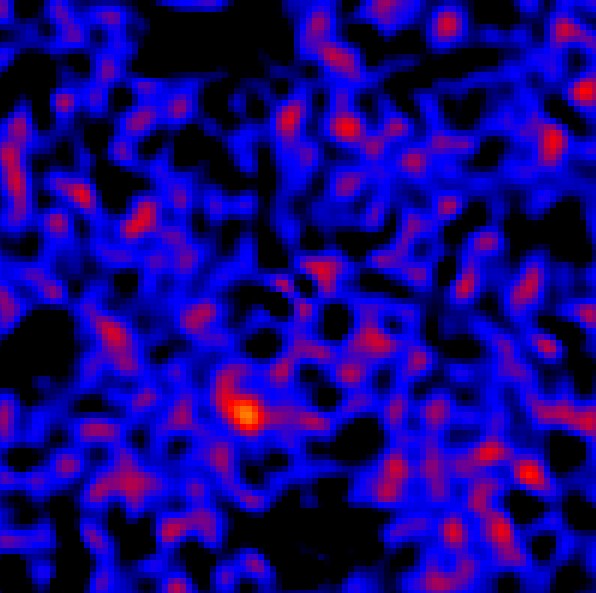}
\end{minipage}}
\hfill
    \caption{\textit{Ks}- band images of SDSS J125635.89+500852.4. The FoV is 23.4\arcsec    $/$ 86.8~kpc in all images. \emph{Left panel:} observed image, \emph{middle panel:} model image, and \emph{right panel:} residual image, smoothed over 3px.}  \label{fig:j1256}
    
\adjustbox{valign=t}{\begin{minipage}{0.49\textwidth}
\centering
\includegraphics[width=0.95\textwidth]{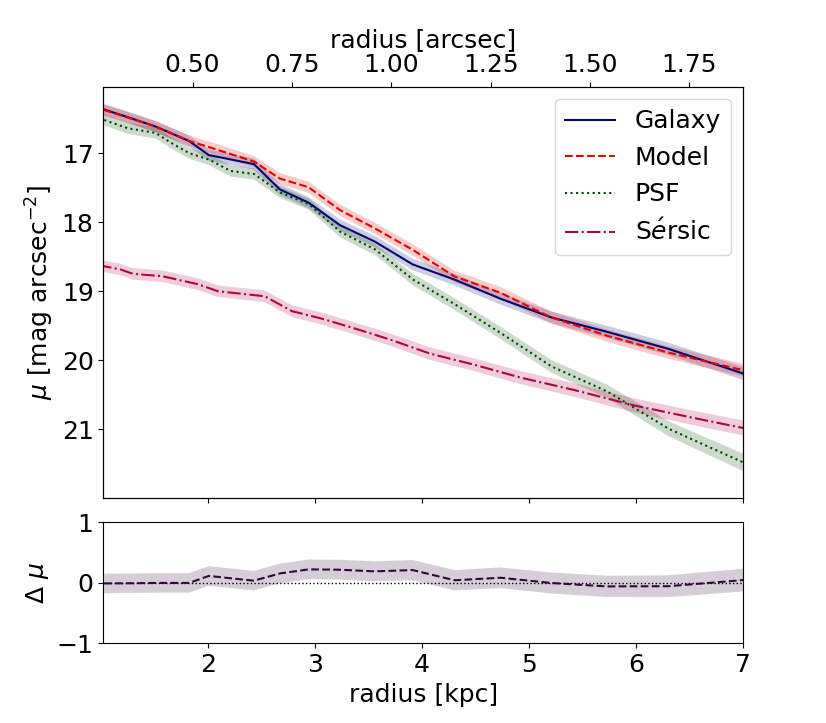}
\caption{Radial surface brightness profile plot of SDSS J125635.89+500852.4.  The galaxy component is plotted with a blue line, the model component is plotted with a dashed red line, the PSF component is plotted with a dotted green line, and the S\'{e}rsic component is plotted with a dashed pink line. The shaded area surrounding each profile curve depicts the error linked to each component.}
\label{fig:j1256comps}
\end{minipage}}
\hfill
\adjustbox{valign=t}{\begin{minipage}{0.49\textwidth}
\centering
\includegraphics[width=0.95\textwidth]{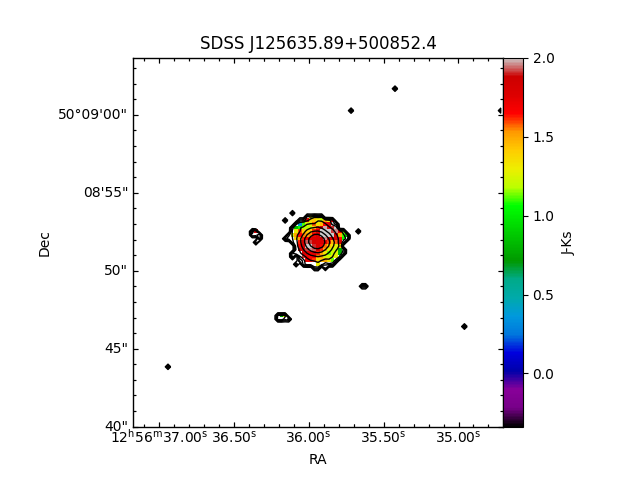}
\caption{\textit{J - $Ks$} colour map of SDSS J125635.89+500852.4.}
\label{fig:j1256color}
\end{minipage}}
\hfill
\end{figure*}

The observed, model, and the residual image of the source can be seen in Fig.~\ref{fig:j1256}. The residual image is clean, with only very small residuals remaining, and the reduced $\chi^2_{\nu}$ value is very good. The best fit parameters suggest that all of the components are fully resolved. The radial surface brightness profile can be seen in Fig.~\ref{fig:j1256comps}. The model and observed image curves are very similar, although not identical. The \textit{J - $Ks$} colour map of the source is shown in Fig.~\ref{fig:j1256color}. With the \textit{J - $Ks$} magnitude being near 2, this is a very red, but small galaxy. The red region near the nucleus suggests there is a bar, this is due to the red region of the source is reaching all the way out to the sides of the galaxy and is not just located at the core of the source. Due to this, it is possible that the host has disk-like morphology. In general the structure is in line with the current expectations of Seyfert 1 galaxies.

\begin{table*}
\caption[]{Best fit parameters of SDSS J125635.89+500852.4. $\chi^2_{\nu}$ = 1.021 $\substack{+0.09 \\ -0.04}$.}
\centering
\begin{tabular}{l l l l l l l}
\hline\hline
Function       & Mag                             & $r_\text{e}$                          & $n$                              & Axial & PA  & Notes             \\
             &                                   & (kpc)                            &                                  & ratio & (\textdegree) &  \\ \hline
PSF 1         & 16.14 $\substack{+0.06 \\ -0.08}$ &                                  &                                  &       &               &   \\

S\'{e}rsic   & 16.97 $\substack{+0.17 \\ -0.42}$ & 4.41 $\substack{+2.48 \\ -0.80}$ & 2.46 $\substack{+2.31 \\ -0.62}$ & 0.73 $\substack{+0.00 \\ -0.03}$ & -42.10  $\substack{+0.00 \\ -0.75}$ &  \\

PSF 2         & 19.72 $\substack{+0.06 \\ -0.07}$ &                                  &                                  &       &               & Nearby source  \\

  \hline   
\end{tabular}
\tablefoot{Columns: (1) Function used for modelling, (2) Magnitude, (3) Effective radius, (4) S\'{e}rsic index, (5) Axial ratio, (6) Position angle, (7) Additional notes.}
\label{tab:j1256}
\end{table*}

\subsection{SDSS J150916.18+613716.7}

This source is jetted with a distinct detection at 37 GHz (see Table~\ref{tab:additional}), even though it has no earlier detections at other radio frequencies.

We used DAOPHOT to model the PSF, and obtained the best fit with one PSF and one S\'{e}rsic function. The best fit values can be found in Table~\ref{tab:j1509}.  Based on the best fit parameters, we believe that we are modelling the entire galaxy. This is supported by the effective radius being in the range of general radii for disk-like galaxies, and by the S\'{e}rsic index around 2 within the errors. 

\begin{figure*}
\centering
\adjustbox{valign=t}{\begin{minipage}{0.35\textwidth}
\centering
\includegraphics[width=1\textwidth]{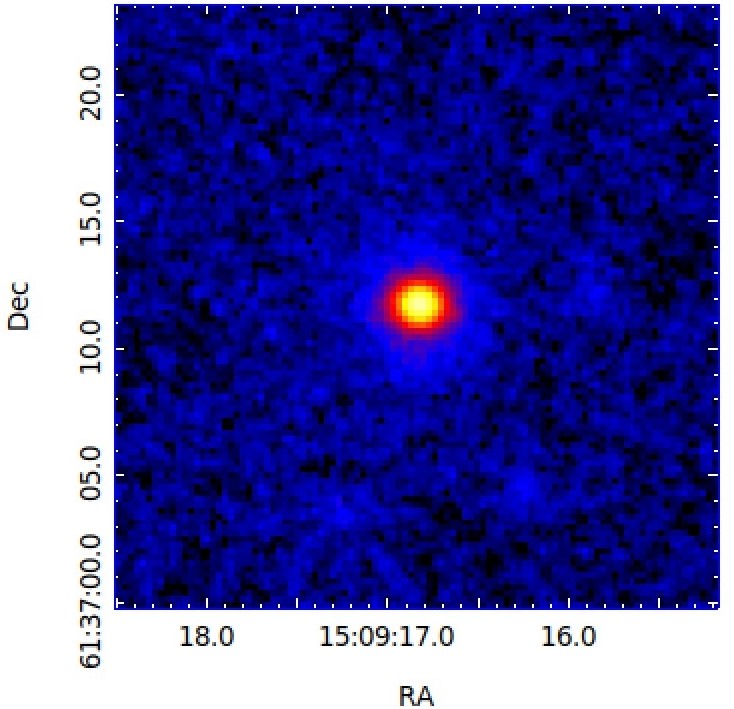}
\end{minipage}}
\adjustbox{valign=t}{\begin{minipage}{0.31\textwidth}
\centering
\includegraphics[width=0.95\textwidth]{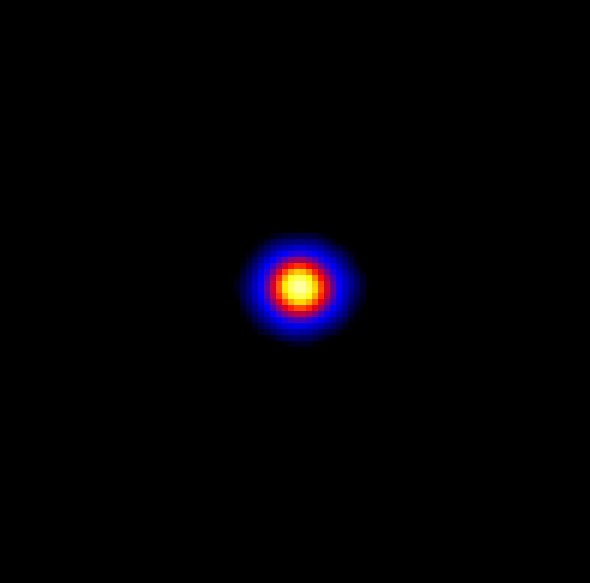}
\end{minipage}}
\adjustbox{valign=t}{\begin{minipage}{0.31\textwidth}
\centering
\includegraphics[width=0.95\textwidth]{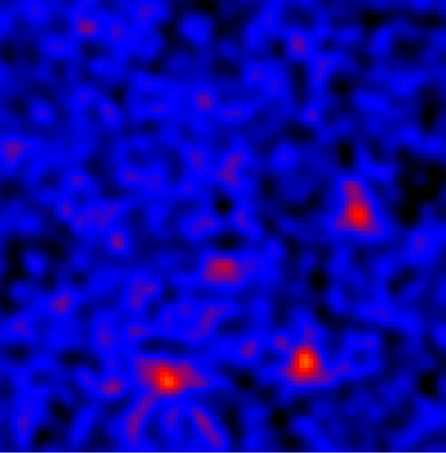}
\end{minipage}}
\hfill
    \caption{\textit{Ks}- band images of SDSS J150916.18+613716.7. The FoV is 23.4\arcsec    $/$ 74.8~kpc in all images. \emph{Left panel:} observed image, \emph{middle panel:} model image, and \emph{right panel:} residual image, smoothed over 3px.}  \label{fig:j1509}

\adjustbox{valign=t}{\begin{minipage}{0.49\textwidth}
\centering
\includegraphics[width=0.95\textwidth]{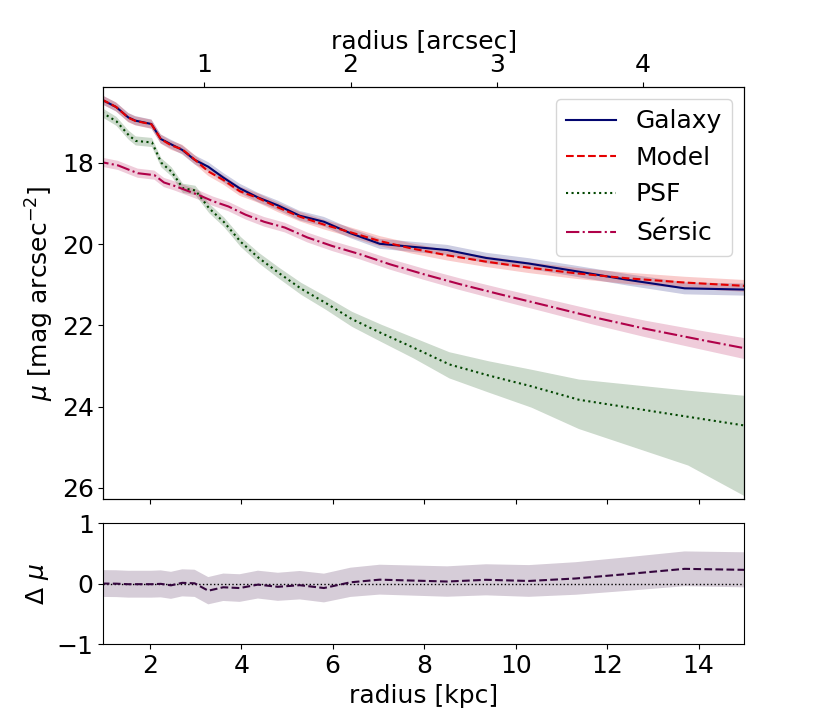}
\caption{Radial surface brightness profile plot of SDSS J150916.18+613716.7.  The galaxy component is plotted with a blue line, the model component is plotted with a dashed red line, the PSF component is plotted with a dotted green line, and the S\'{e}rsic component is plotted with a dashed pink line. The shaded area surrounding each profile curve depicts the error linked to each component.}
\label{fig:j1509comps}
\end{minipage}}
\hfill
\adjustbox{valign=t}{\begin{minipage}{0.49\textwidth}
\centering
\includegraphics[width=0.95\textwidth]{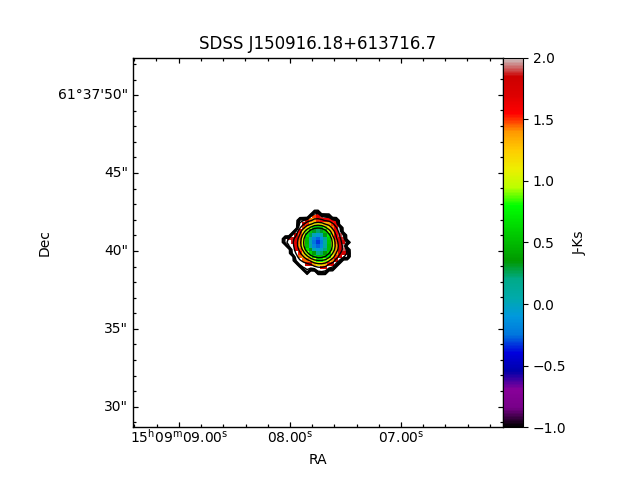}
\caption{ J - $Ks$ colour map of SDSS J150916.18+613716.7.}
\label{fig:j1509color}
\end{minipage}}
\hfill
\end{figure*}

The observed, model, and the residual image of the source can be seen in Fig.~\ref{fig:j1509}. There are virtually no residuals left in the image and the reduced $\chi^2_{\nu}$ is approximately one. These suggest that the fit is good. In the radial surface brightness profile plot, seen in Fig.~\ref{fig:j1509comps}, the model and the observed image curves are approximately identical.  The errors in the other components are very small. The \textit{J - $Ks$} colour map of the source is presented in Fig.~\ref{fig:j1509color}. Unlike the other galaxies in this paper, SDSS J150916.18+613716.7 shows a blue central region with red by the edges of the galaxy. As mentioned earlier, this is probably due to the fact that we were not able to successfully match the PSF sizes of the two images. However, the \textit{J - $Ks$} magnitude is within the accepted range for a Seyfert 1 galaxy. Otherwise, this is a small galaxy and it is not possible to determine the structure based on the colour map.

\begin{table*}
\caption[]{Best fit parameters of SDSS J150916.18+613716.7. $\chi^2_{\nu}$ = 1.052 $\substack{+0.02 \\ -0.01}$.}
\centering
\begin{tabular}{l l l l l l l}
\hline\hline
Function       & Mag                             & $r_\text{e}$                          & $n$                              & Axial & PA  & Notes             \\
             &                                   & (kpc)                            &                                  & ratio & (\textdegree) &  \\ \hline
PSF          & 16.02 $\substack{+0.12 \\ -0.11}$ &                                  &                                  &       &               &   \\

S\'{e}rsic   & 16.10 $\substack{+0.15 \\ -0.14}$ & 4.04 $\substack{+0.43 \\ -0.17}$ & 2.37 $\substack{+0.67 \\ -0.45}$ & 0.81 $\substack{+0.00 \\ -0.00}$ & -2.19  $\substack{+0.15 \\ -0.00}$ &  \\

  \hline   
\end{tabular}
\tablefoot{Columns: (1) Function used for modelling, (2) Magnitude, (3) Effective radius, (4) S\'{e}rsic index, (5) Axial ratio, (6) Position angle, (7) Additional notes.}
\label{tab:j1509}
\end{table*}

\section{Discussion}  
\label{discussion}

We successfully modelled 12 of the 14 sources. The results are summarised in Table~\ref{tab:summary}. We were unable to properly model 6dFGS gJ044739.0-040330 and 6dFGS gJ084510.2-073205, but upon visual inspection it appears clear that 6dFGS gJ084510.2-073205 is a disk-like galaxy. The data for these sources can be found in Appendix~\ref{compromised-fits}. Nine out of the 14 sources appear to have disk-like morphology. Two out of the fourteen sources, SDSS J125635.89+500852.4 and SDSS J102906.68+555625.2, do not show clear disk-like morphology based on the best fit parameters: the S\'{e}rsic indices are close to 2.5 with large errors, pushing the value to be one that can also be found in low-luminosity ellipticals. In addition, it is unclear if FBQS J0744+5149 is disk-like, or possibly a dwarf. All the clear bulge components we were able to model are pseudo-bulges. This is as expected, as disk-like galaxies tend to have pseudo-bulges while elliptical galaxies have classical bulges. The frequency of pseudo-bulges in NLS1 galaxies is also higher than the respective frequency in BLS1 galaxies \citep{2011orban}.


Our results regarding the morphologies of the NLS1 hosts are in line with previous studies, which have concluded that a large majority of NLS1 nuclei are hosted in disk-like or spiral galaxies with pseudo-bulges \citep[e.g.,][]{2003crenshaw,2006deo1,2007ohta1, 2011orban}. What has remained under debate so far is if the host galaxies of jetted NLS1 galaxies are similar to the hosts of their non-jetted counterparts. The number of studied hosts of jetted NLS1 galaxies is finally reaching a number that will allow us to draw significant conclusions, not affected by small number statistics. So far, including this study, in total more than 50 hosts of jetted NLS1 galaxies have been modelled \citep{2008anton, 2014leontavares, 2016kotilainen, 2017olguiglesias, 2018jarvela, 2019berton, 2020olguiglesias, 2021hamilton, 2022vietri}, and only two studies have identified elliptical galaxies \citep{2017dammando, 2018dammando}. There is some debate on the morphology of FBQS J1644+2619 \citep{2017dammando, 2017olguiglesias}, but the fact remains that the host galaxies of jetted NLS1 galaxies seem to preferentially be disk-like.

In the early 2000s, it was speculated that the radio-loudness parameter was directly connected to the black hole mass, and thus that relativistic jets could be exclusively launched by black holes with mass above $10^9$ M$_\odot$ \citep{2000laor1}. This conclusion, however, was based on observations of the Palomar-Green (PG) quasar sample, which included only bright objects with high black hole masses, and was thus biased. It is now clear that relativistic jets are instead a ubiquitous phenomenon that can occur over a wide mass range, and NLS1 galaxies are the most evident proof of this \citep{2011foschini1}. The power of relativistic jets, indeed, scales non-linearly with the black hole mass \citep{2003heinz}, among some other factors. Therefore, low-mass sources have less luminous jets that eluded early surveys. After the discovery of relativistic jets in NLS1 galaxies, some authors tried to reconcile the large mass paradigm by claiming that NLS1 galaxies are not genuinely low-mass sources, but instead that they have rather massive black holes and a flattened BLR observed pole on \citep{2013calderone}. Whereas this could be the case for a few jetted NLS1 galaxies, based on the current evidence it does not seem to be a plausible explanation for all of them. A strong argument against this hypothesis are the host morphologies of jetted NLS1 galaxies, since it is commonly known that disk-like galaxies in general harbour less massive central black holes than elliptical galaxies\citep{2008hu}, wherein the most luminous quasars are found. Among the sources in our sample, the black hole masses vary between 3.3$\times10^{5} M_{\sun}$ and 5.0$\times10^{8} M_{\sun}$, a typical range for NLS1 galaxies, with two sources exceeding $10^{8} M_{\sun}$. The recent studies seem to unquestionably confirm that the jet paradigm is not valid, and that the black hole mass is not a determining factor in launching of the relativistic jets.

In addition to disproving the jet paradigm, the jetted NLS1 galaxies also force us to revise the AGN unification models. Narrow-line Seyfert 1 galaxies were traditionally placed among the radio-quiet, non-jetted population of AGN, but with the number of jetted NLS1 galaxies soon reaching one hundred, they cannot be treated as a curiosity anymore. Since they are not similar to more powerful jetted AGN either, a new unification model for unevolved AGN - with and without jets - was proposed, and it was hypothesised that jetted NLS1 galaxies as high-excitation AGN are similar to the progenitors of FSRQs \citep{2017berton}. However, many questions still remain open, for example, it is still unclear if all NLS1 galaxies should even be treated as one population of sources, and what the future evolution of the subclasses of NLS1 galaxies will be. 

It is known that the environment in which the galaxy resides plays a role in its evolution, and also the level of the nuclear activity. The most powerful and evolved AGN, such as blazars, usually reside in the densest environments, such as superclusters, whereas less evolved AGN, including Seyfert galaxies, favour the less dense parts, residing in voids and intermediate-density regions \citep[e.g.,][]{2011lietzen1}. It has also been speculated that within the NLS1 population the jetted NLS1 galaxies prefer more dense large-scale environments than their non-jetted counterparts \citep{2017jarvela}, highlighting the possibility that mergers and interaction, which are more likely to take place in a denser environment, might be important for the launching of the jets. We have the large-scale environment data for seven of our sources, obtained from \citep{2017jarvela} (see Table~\ref{tab:summary}). Of these seven sources, two sources reside in a void, four sources in a supercluster, and one source in an intermediate-density region. All of the sources that we have the large-scale environment data for most likely harbour a powerful relativistic jet. The data is not conclusive, as the sample sizes are still too small, but it seems like whereas the jetted NLS1 galaxies might slightly favour denser environments, the large-scale environment is not a determining factor in the triggering of the jets.

More important for the evolution of an AGN than the large-scale environment is its nearby, group- and galaxy-scale, environment, as it can efficiently reshape the galaxy and its dynamics. Whereas the role of interaction and mergers in the grand AGN scheme is still unclear, it was earlier speculated that there could be a connection between mergers/interaction and jetted NLS1 galaxies \citep[e.g.,][]{2018jarvela, 2020olguiglesias, 2021berton}. However, only one of our jetted sources exhibit possible interaction. The lack of interacting sources might be partly due to the limitations of resolution in our study, and especially minor mergers, also able to trigger nuclear activity, could easily go unnoticed. A noteworthy example of this is the case of IRAS 17020+4544, a jetted NLS1 which does not show any evidence of merging in optical/NIR imaging, but revealed signs of a small companion galaxy only with observations of its molecular gas \citep{2021salome, 2022jarvela}. Furthermore, \citet{2020olguiglesias} find signs of minor mergers in a larger fraction of their $\gamma$-ray-emitting NLS1 galaxies (75\%) than non-$\gamma$ radio-loud NLS1 galaxies (53\%). They also find that the bulges of $\gamma$-NLS1 galaxies show red NIR colours ($J - K$ $\sim$ 2), clearly distinct from their hosts, whereas the bulge colours of the non-$\gamma$ sources are comparable to their hosts. They attribute the red nuclei to a dust-embedded AGN, or, alternatively, the bulge exhibiting enhanced star formation. A similar trend is seen in our jetted NLS1 galaxies as all but one of them show a clearly redder nucleus than the rest of the galaxy. If the hypothesis of \citet{2020olguiglesias} is correct, this might be a sign of past minor mergers in these sources. However, in our case, also the nuclei of the likely non-jetted NLS1 galaxies are redder than their host galaxies. The role of mergers in the evolution of jetted NLS1 galaxies remains unclear, but based on the results so far, it looks like at least major mergers are not required for triggering relativistic jets in NLS1 galaxies. In the future, higher-resolution data of larger samples are required to investigate the impact of mergers in the development of the NLS1 population.

\begin{table*}
\caption[]{Summary of the results.}
\centering
\begin{tabular}{l l l l l l}
\hline\hline
Source              & Morphology & Components     & Jetted & Large-scale  & Notes   \\
                    &            &                &        & environment  &              \\ \hline
6dFGS gJ042021.7-053054  &   Disk-like            & Disk    & Yes &   & \\ 
6dFGS gJ044739.0-040330  &   Unclear              &                   &     &   & \\
FBQS J0744+5149          &   Disk-like/Dwarf      &                   &     &   & \\
SDSS J080535.16+302201.7 &   Disk-like            & Bulge or disk     & Yes &   & \\
6dFGS gJ084510.2-073205  &   Disk-like            &                   &     &   & Visual inspection \\
SDSS J084516.17+421129.9 &   Disk-like            &                   & Yes &   & \\
SDSS J090113.23+465734.7 &   Disk-like            &                   & Yes & Supercluster  & \\
SDSS J093712.32+500852.1 &   Disk-like            & Bulge and bar             & Yes &   & \\
SDSS J102906.68+555625.2 &   Unclear/Elliptical   &                   & Yes & Supercluster  & \\
SDSS J103123.73+423439.2 &   Disk-like            &                   & Yes &  Void & \\
SDSS J122844.81+501751.2 &   Disk-like            & Bulge (and tail)  & Yes & Supercluster  & \\
SDSS J123220.11+495721.8 &   Disk-like            &                   & Yes & Supercluster  & \\
SDSS J125635.89+500852.4 &   Unclear/Elliptical   &                   & Yes & Intermediate  & \\
SDSS J150916.18+613716.7 &   Disk-like            &                   & Yes & Void  & \\\hline
\end{tabular}
\tablefoot{Columns: (1) Source name, (2) Probable morphology of the host galaxy, (3) Components of the model, (4) Presence of a relativistic jet based on the Metsähovi 37 GHz detections, (5) Large-scale environment parameter, (6) Additional notes.}
\label{tab:summary}
\end{table*}

\section{Conclusions}
\label{conclusions}

We observed 14 NLS1 galaxies in $J-$ and $Ks$-bands with NOT with the aim to determine their host galaxy morphologies, and investigate if they exhibit any signs of interaction. In addition, we obtained $J - Ks$ colour maps for 13 sources observed in both bands to study mainly the properties of the nucleus, and any additional hints regarding the composition of the host galaxies. Our main conclusions are: 


\begin{enumerate}
    \item The large majority of jetted NLS1 galaxies -- in this study, and in general -- are proven to be hosted in disk-like galaxies.
    \item  Our results confirm that large elliptical galaxies are not needed for launching and maintaining relativistic jets. As proven by this and previous studies, the jet paradigm is undoubtedly inaccurate and needs to be reassessed based on recent evidence. 
    \item Although previously speculated otherwise, major mergers might not be the driving factor in the launching of powerful relativistic jets -- at least in NLS1 galaxies. 
    \item The cores of the jetted NLS1 galaxies are, as expected, red, possibly due to dust extinction, suggesting that these sources might have recently undergone minor mergers.
\end{enumerate}


With this and previous studies, the sample size of jetted NLS1 galaxies with known host morphologies is finally reaching a number suitable for statistically significant and reliable results. However, higher-resolution observations, for example, using the Very Large Telescope or the James Webb Space Telescope, are still needed to properly model all the components of the hosts, and to study the sources at higher redshift. In addition, observations at different wavelengths, such as sub-mm observations with the Atacama Large Millimeter Array, might help in assessing the frequency and role of interaction and mergers in jetted NLS1 galaxies. With NLS1 galaxies being in an early evolutionary stage, they can be a crucial factor in understanding the evolution of jetted AGN and an ideal laboratory for studying AGN feedback. Furthermore, if jetted NLS1 galaxies are similar to the progenitors of FSRQs, further studies of NLS1 galaxies can help estimate the impact of kinematically young jetted AGN on the galaxy evolution in the early Universe.



\begin{acknowledgements}
IRAF is distributed by the National Optical Astronomy Observatories, which are operated by the Association of Universities for Research in Astronomy, Inc., under cooperative agreement with the National Science Foundation. Our observations were made with the Nordic Optical Telescope, operated by the Nordic Optical Telescope Scientific Association at the Observatorio del Roque de los Muchachos, La Palma, Spain, of the Instituto de Astrofisica de Canarias. The data presented here were obtained in part with ALFOSC, which is provided by the Instituto de Astrofisica de Andalucia (IAA) under a joint agreement with the University of Copenhagen and NOT. This research has made use of GALFIT. This publication made use of the SIMBAD database, operated at CDS, Strasbourg, France. This research has made use of the NASA/IPAC Extragalactic Database (NED), which is operated by the Jet Propulsion Laboratory, California Institute of Technology, under contract with the National Aeronautics and Space Administration. This research made use of Astropy,\footnote{http://www.astropy.org} a community-developed core Python package for Astronomy \citep{astropy:2013, astropy:2018}. This publication makes use of data obtained at Metsähovi Radio Observatory, operated by Aalto University, Finland. I.B. would like to thank both the Finnish Cultural Foundation and the Vilho, Yrjö and Kalle Väisälä Foundation of the Finnish Academy of Science and Letters for their support. E.J. is a current European Space Agency Science Research Fellow. M.B. is an ESO fellow. E.C. acknowledges support from ANID BASAL projects ACE210002 and FB210003.

\end{acknowledgements}

\bibliographystyle{aa}
\bibliography{articles.bib}

\begin{thebibliography}{86}
\expandafter\ifx\csname natexlab\endcsname\relax\def\natexlab#1{#1}\fi

\bibitem[{{Abdo} {et~al.}(2009{\natexlab{a}}){Abdo}, {Ackermann}, {Ajello},
  {Axelsson}, {Baldini}, {Ballet}, {Barbiellini}, {Bastieri}, {Battelino},
  {Baughman}, {Bechtol}, {Bellazzini}, {Bloom}, {Bonamente}, {Borgland},
  {Bregeon}, {Brez}, {Brigida}, {Bruel}, {Caliandro}, {Cameron}, {Caraveo},
  {Casandjian}, {Cavazzuti}, {Cecchi}, {Chekhtman}, {Cheung}, {Chiang},
  {Ciprini}, {Claus}, {Cohen-Tanugi}, {Collmar}, {Conrad}, {Costamante},
  {Dermer}, {de Angelis}, {de Palma}, {Digel}, {Silva}, {Drell}, {Dubois},
  {Dumora}, {Farnier}, {Favuzzi}, {Focke}, {Foschini}, {Frailis}, {Fuhrmann},
  {Fukazawa}, {Funk}, {Fusco}, {Gargano}, {Gehrels}, {Germani}, {Giebels},
  {Giglietto}, {Giordano}, {Giroletti}, {Glanzman}, {Grenier}, {Grondin},
  {Grove}, {Guillemot}, {Guiriec}, {Hanabata}, {Harding}, {Hartman},
  {Hayashida}, {Hays}, {Hughes}, {J{\'o}hannesson}, {Johnson}, {Johnson},
  {Johnson}, {Kamae}, {Katagiri}, {Kataoka}, {Kerr}, {Kn{\"o}dlseder}, {Kuehn},
  {Kuss}, {Lande}, {Latronico}, {Lemoine-Goumard}, {Longo}, {Loparco}, {Lott},
  {Lovellette}, {Lubrano}, {Madejski}, {Makeev}, {Max-Moerbeck}, {Mazziotta},
  {McConville}, {McEnery}, {Meurer}, {Michelson}, {Mitthumsiri}, {Mizuno},
  {Monte}, {Monzani}, {Morselli}, {Moskalenko}, {Murgia}, {Nolan}, {Norris},
  {Nuss}, {Ohsugi}, {Omodei}, {Orlando}, {Ormes}, {Paneque}, {Panetta},
  {Parent}, {Pavlidou}, {Pearson}, {Pepe}, {Pesce-Rollins}, {Piron}, {Porter},
  {Rain{\`o}}, {Rando}, {Razzano}, {Readhead}, {Reimer}, {Reimer}, {Reposeur},
  {Richards}, {Ritz}, {Rodriguez}, {Romani}, {Ryde}, {Sadrozinski}, {Sambruna},
  {Sanchez}, {Sander}, {Parkinson}, {Scargle}, {Schalk}, {Sgr{\`o}}, {Smith},
  {Spandre}, {Spinelli}, {Starck}, {Stevenson}, {Strickman}, {Suson},
  {Tagliaferri}, {Takahashi}, {Tanaka}, {Thayer}, {Thompson}, {Tibaldo},
  {Tibolla}, {Torres}, {Tosti}, {Tramacere}, {Uchiyama}, {Usher}, {Vilchez},
  {Vitale}, {Waite}, {Winer}, {Wood}, {Ylinen}, {Zensus}, {Ziegler}, {Fermi/LAT
  Collaboration}, {Ghisellini}, {Maraschi}, {Tavecchio}, \&
  {Angelakis}}]{2009abdo2}
{Abdo}, A.~A., {Ackermann}, M., {Ajello}, M., {et~al.} 2009{\natexlab{a}},
  \apj, 699, 976

\bibitem[{{Abdo} {et~al.}(2009{\natexlab{b}}){Abdo}, {Ackermann}, {Ajello},
  {Axelsson}, {Baldini}, {Ballet}, {Barbiellini}, {Bastieri}, {Baughman},
  {Bechtol}, \& et~al.}]{2009abdo1}
{Abdo}, A.~A., {Ackermann}, M., {Ajello}, M., {et~al.} 2009{\natexlab{b}},
  \apj, 707, 727

\bibitem[{{Ant{\'o}n} {et~al.}(2008){Ant{\'o}n}, {Browne}, \&
  {March{\~a}}}]{2008anton}
{Ant{\'o}n}, S., {Browne}, I.~W.~A., \& {March{\~a}}, M.~J. 2008, \aap, 490,
  583

\bibitem[{{Astropy Collaboration} {et~al.}(2018){Astropy Collaboration},
  {Price-Whelan}, {Sip{\H{o}}cz}, {G{\"u}nther}, {Lim}, {Crawford}, {Conseil},
  {Shupe}, {Craig}, {Dencheva}, {Ginsburg}, {VanderPlas}, {Bradley},
  {P{\'e}rez-Su{\'a}rez}, {de Val-Borro}, {Aldcroft}, {Cruz}, {Robitaille},
  {Tollerud}, {Ardelean}, {Babej}, {Bach}, {Bachetti}, {Bakanov}, {Bamford},
  {Barentsen}, {Barmby}, {Baumbach}, {Berry}, {Biscani}, {Boquien}, {Bostroem},
  {Bouma}, {Brammer}, {Bray}, {Breytenbach}, {Buddelmeijer}, {Burke},
  {Calderone}, {Cano Rodr{\'\i}guez}, {Cara}, {Cardoso}, {Cheedella}, {Copin},
  {Corrales}, {Crichton}, {D'Avella}, {Deil}, {Depagne}, {Dietrich}, {Donath},
  {Droettboom}, {Earl}, {Erben}, {Fabbro}, {Ferreira}, {Finethy}, {Fox},
  {Garrison}, {Gibbons}, {Goldstein}, {Gommers}, {Greco}, {Greenfield},
  {Groener}, {Grollier}, {Hagen}, {Hirst}, {Homeier}, {Horton}, {Hosseinzadeh},
  {Hu}, {Hunkeler}, {Ivezi{\'c}}, {Jain}, {Jenness}, {Kanarek}, {Kendrew},
  {Kern}, {Kerzendorf}, {Khvalko}, {King}, {Kirkby}, {Kulkarni}, {Kumar},
  {Lee}, {Lenz}, {Littlefair}, {Ma}, {Macleod}, {Mastropietro}, {McCully},
  {Montagnac}, {Morris}, {Mueller}, {Mumford}, {Muna}, {Murphy}, {Nelson},
  {Nguyen}, {Ninan}, {N{\"o}the}, {Ogaz}, {Oh}, {Parejko}, {Parley}, {Pascual},
  {Patil}, {Patil}, {Plunkett}, {Prochaska}, {Rastogi}, {Reddy Janga},
  {Sabater}, {Sakurikar}, {Seifert}, {Sherbert}, {Sherwood-Taylor}, {Shih},
  {Sick}, {Silbiger}, {Singanamalla}, {Singer}, {Sladen}, {Sooley},
  {Sornarajah}, {Streicher}, {Teuben}, {Thomas}, {Tremblay}, {Turner},
  {Terr{\'o}n}, {van Kerkwijk}, {de la Vega}, {Watkins}, {Weaver}, {Whitmore},
  {Woillez}, {Zabalza}, \& {Astropy Contributors}}]{astropy:2018}
{Astropy Collaboration}, {Price-Whelan}, A.~M., {Sip{\H{o}}cz}, B.~M., {et~al.}
  2018, \aj, 156, 123

\bibitem[{{Astropy Collaboration} {et~al.}(2013){Astropy Collaboration},
  {Robitaille}, {Tollerud}, {Greenfield}, {Droettboom}, {Bray}, {Aldcroft},
  {Davis}, {Ginsburg}, {Price-Whelan}, {Kerzendorf}, {Conley}, {Crighton},
  {Barbary}, {Muna}, {Ferguson}, {Grollier}, {Parikh}, {Nair}, {Unther},
  {Deil}, {Woillez}, {Conseil}, {Kramer}, {Turner}, {Singer}, {Fox}, {Weaver},
  {Zabalza}, {Edwards}, {Azalee Bostroem}, {Burke}, {Casey}, {Crawford},
  {Dencheva}, {Ely}, {Jenness}, {Labrie}, {Lim}, {Pierfederici}, {Pontzen},
  {Ptak}, {Refsdal}, {Servillat}, \& {Streicher}}]{astropy:2013}
{Astropy Collaboration}, {Robitaille}, T.~P., {Tollerud}, E.~J., {et~al.} 2013,
  \aap, 558, A33

\bibitem[{{Barth} {et~al.}(2008){Barth}, {Bentz}, {Greene}, \&
  {Ho}}]{2008barth1}
{Barth}, A.~J., {Bentz}, M.~C., {Greene}, J.~E., \& {Ho}, L.~C. 2008, \apjl,
  683, L119

\bibitem[{{Berton} {et~al.}(2019){Berton}, {Congiu}, {Ciroi}, {Komossa},
  {Frezzato}, {Di Mille}, {Ant{\'o}n}, {Antonucci}, {Caccianiga}, {Coppi},
  {J{\"a}rvel{\"a}}, {Kotilainen}, {L{\"a}hteenm{\"a}ki}, {Mathur}, {Chen},
  {Cracco}, {La Mura}, \& {Rafanelli}}]{2019berton}
{Berton}, M., {Congiu}, E., {Ciroi}, S., {et~al.} 2019, \aj, 157, 48

\bibitem[{{Berton} {et~al.}(2018){Berton}, {Congiu}, {J{\"a}rvel{\"a}},
  {Antonucci}, {Kharb}, {Lister}, {Tarchi}, {Caccianiga}, {Chen}, {Foschini},
  {L{\"a}hteenm{\"a}ki}, {Richards}, {Ciroi}, {Cracco}, {Frezzato}, {La Mura},
  \& {Rafanelli}}]{2018berton}
{Berton}, M., {Congiu}, E., {J{\"a}rvel{\"a}}, E., {et~al.} 2018, \aap, 614,
  A87

\bibitem[{{Berton} {et~al.}(2017){Berton}, {Foschini}, {Caccianiga}, {Ciroi},
  {Congiu}, {Cracco}, {Frezzato}, {La Mura}, \& {Rafanelli}}]{2017berton}
{Berton}, M., {Foschini}, L., {Caccianiga}, A., {et~al.} 2017, Frontiers in
  Astronomy and Space Sciences, 4, 8

\bibitem[{{Berton} \& {J{\"a}rvel{\"a}}(2021)}]{2021berton2}
{Berton}, M. \& {J{\"a}rvel{\"a}}, E. 2021, Universe, 7, 188

\bibitem[{{Berton} {et~al.}(2020){Berton}, {J{\"a}rvel{\"a}}, {Crepaldi},
  {L{\"a}hteenm{\"a}ki}, {Tornikoski}, {Congiu}, {Kharb}, {Terreran}, \&
  {Vietri}}]{2020berton}
{Berton}, M., {J{\"a}rvel{\"a}}, E., {Crepaldi}, L., {et~al.} 2020, \aap, 636,
  A64

\bibitem[{{Berton} {et~al.}(2021){Berton}, {Peluso}, {Marziani}, {Komossa},
  {Foschini}, {Ciroi}, {Chen}, {Congiu}, {Gallo}, {Bj{\"o}rklund}, {Crepaldi},
  {Di Mille}, {J{\"a}rvel{\"a}}, {Kotilainen}, {Kreikenbohm}, {Morrell},
  {Romano}, {Sani}, {Terreran}, {Tornikoski}, {Vercellone}, \&
  {Vietri}}]{2021berton}
{Berton}, M., {Peluso}, G., {Marziani}, P., {et~al.} 2021, \aap, 654, A125

\bibitem[{{Boroson}(2005)}]{2005boroson}
{Boroson}, T. 2005, \aj, 130, 381

\bibitem[{{Boroson} \& {Green}(1992)}]{1992boroson}
{Boroson}, T.~A. \& {Green}, R.~F. 1992, \apj, 80, 109

\bibitem[{{Caccianiga} {et~al.}(2015){Caccianiga}, {Ant{\'o}n}, {Ballo},
  {Foschini}, {Maccacaro}, {Della Ceca}, {Severgnini}, {March{\~a}}, {Mateos},
  \& {Sani}}]{2015caccianiga}
{Caccianiga}, A., {Ant{\'o}n}, S., {Ballo}, L., {et~al.} 2015, \mnras, 451,
  1795

\bibitem[{{Calderone} {et~al.}(2013){Calderone}, {Ghisellini}, {Colpi}, \&
  {Dotti}}]{2013calderone}
{Calderone}, G., {Ghisellini}, G., {Colpi}, M., \& {Dotti}, M. 2013, \mnras,
  431, 210

\bibitem[{{Carniani} {et~al.}(2016){Carniani}, {Marconi}, {Maiolino},
  {Balmaverde}, {Brusa}, {Cano-D{\'\i}az}, {Cicone}, {Comastri}, {Cresci},
  {Fiore}, {Feruglio}, {La Franca}, {Mainieri}, {Mannucci}, {Nagao}, {Netzer},
  {Piconcelli}, {Risaliti}, {Schneider}, \& {Shemmer}}]{2016carniani}
{Carniani}, S., {Marconi}, A., {Maiolino}, R., {et~al.} 2016, \aap, 591, A28

\bibitem[{{Chen} {et~al.}(2018){Chen}, {Berton}, {La Mura}, {Congiu},
  {Foschini}, H., {Ciroi}, {Rafanelli}, \& {Bastieri}}]{2018chen}
{Chen}, S., {Berton}, M., {La Mura}, G., {et~al.} 2018, \aap, 615, A167

\bibitem[{{Cisternas} {et~al.}(2011){Cisternas}, {Jahnke}, {Inskip},
  {Kartaltepe}, {Koekemoer}, {Lisker}, {Robaina}, {Scodeggio}, {Sheth},
  {Trump}, {Andrae}, {Miyaji}, {Lusso}, {Brusa}, {Capak}, {Cappelluti},
  {Civano}, {Ilbert}, {Impey}, {Leauthaud}, {Lilly}, {Salvato}, {Scoville}, \&
  {Taniguchi}}]{2011cisternas1}
{Cisternas}, M., {Jahnke}, K., {Inskip}, K.~J., {et~al.} 2011, \apj, 726, 57

\bibitem[{{Corbin}(2000)}]{2000corbin1}
{Corbin}, M.~R. 2000, \apjl, 536, L73

\bibitem[{{Cracco} {et~al.}(2016){Cracco}, {Ciroi}, {Berton}, {Di Mille},
  {Foschini}, {La Mura}, \& {Rafanelli}}]{2016cracco1}
{Cracco}, V., {Ciroi}, S., {Berton}, M., {et~al.} 2016, \mnras, 462, 1256

\bibitem[{{Crenshaw} {et~al.}(2003){Crenshaw}, {Kraemer}, \&
  {Gabel}}]{2003crenshaw}
{Crenshaw}, D.~M., {Kraemer}, S.~B., \& {Gabel}, J.~R. 2003, \aj, 126, 1690

\bibitem[{{D'Ammando} {et~al.}(2018){D'Ammando}, {Acosta-Pulido}, {Capetti},
  {Baldi}, {Orienti}, {Raiteri}, \& {Ramos Almeida}}]{2018dammando}
{D'Ammando}, F., {Acosta-Pulido}, J.~A., {Capetti}, A., {et~al.} 2018, \mnras,
  478, L66

\bibitem[{{D'Ammando} {et~al.}(2017){D'Ammando}, {Acosta-Pulido}, {Capetti},
  {Raiteri}, {Baldi}, {Orienti}, \& {Ramos Almeida}}]{2017dammando}
{D'Ammando}, F., {Acosta-Pulido}, J.~A., {Capetti}, A., {et~al.} 2017, \mnras,
  469, L11

\bibitem[{{Decarli} {et~al.}(2008){Decarli}, {Dotti}, {Fontana}, \&
  {Haardt}}]{2008decarli}
{Decarli}, R., {Dotti}, M., {Fontana}, M., \& {Haardt}, F. 2008, \mnras, 386,
  L15

\bibitem[{{Deo} {et~al.}(2006){Deo}, {Crenshaw}, \& {Kraemer}}]{2006deo1}
{Deo}, R.~P., {Crenshaw}, D.~M., \& {Kraemer}, S.~B. 2006, \aj, 132, 321

\bibitem[{{Du} {et~al.}(2018){Du}, {Brotherton}, {Wang}, {Huang}, {Hu},
  {Kasper}, {Chick}, {Nguyen}, {Maithil}, {Hand}, {Li}, {Ho}, {Bai}, {Bian},
  {Wang}, \& {MAHA Collaboration}}]{2018du}
{Du}, P., {Brotherton}, M.~S., {Wang}, K., {et~al.} 2018, \apj, 869, 142

\bibitem[{{Du} \& {Wang}(2019)}]{2019du1}
{Du}, P. \& {Wang}, J.-M. 2019, \apj, 886, 42

\bibitem[{{Elmegreen} \& {Elmegreen}(2014)}]{2014elmegreen}
{Elmegreen}, D.~M. \& {Elmegreen}, B.~G. 2014, \apj, 781, 11

\bibitem[{{Fabian}(2012)}]{2012fabian1}
{Fabian}, A.~C. 2012, \araa, 50, 455

\bibitem[{{Fisher} \& {Drory}(2008)}]{2008fisher}
{Fisher}, D.~B. \& {Drory}, N. 2008, \aj, 136, 773

\bibitem[{{Foschini}(2011)}]{2011foschini1}
{Foschini}, L. 2011, in Narrow-Line Seyfert 1 Galaxies and their Place in the
  Universe

\bibitem[{{Foschini} {et~al.}(2015){Foschini}, {Berton}, {Caccianiga}, {Ciroi},
  {Cracco}, {Peterson}, {Angelakis}, {Braito}, {Fuhrmann}, {Gallo}, {Grupe},
  {J{\"a}rvel{\"a}}, {Kaufmann}, {Komossa}, {Kovalev}, {L{\"a}hteenm{\"a}ki},
  {Lisakov}, {Lister}, {Mathur}, {Richards}, {Romano}, {Sievers},
  {Tagliaferri}, {Tammi}, {Tibolla}, {Tornikoski}, {Vercellone}, {La Mura},
  {Maraschi}, \& {Rafanelli}}]{2015foschini1}
{Foschini}, L., {Berton}, M., {Caccianiga}, A., {et~al.} 2015, \aap, 575, A13

\bibitem[{{Foschini} {et~al.}(2017){Foschini}, {Berton}, {Caccianiga}, {Ciroi},
  {Cracco}, {Peterson}, {Angelakis}, {Braito}, {Fuhrmann}, {Gallo}, {Grupe},
  {J{\"a}rvel{\"a}}, {Kaufmann}, {Komossa}, {Kovalev}, {L{\"a}hteenm{\"a}ki},
  {Lisakov}, {Lister}, {Mathur}, {Richards}, {Romano}, {Sievers},
  {Tagliaferri}, {Tammi}, {Tibolla}, {Tornikoski}, {Vercellone}, {La Mura},
  {Maraschi}, \& {Rafanelli}}]{2017foschini}
{Foschini}, L., {Berton}, M., {Caccianiga}, A., {et~al.} 2017, \aap, 603, C1

\bibitem[{{Foschini} {et~al.}(2010){Foschini}, {Fermi/Lat Collaboration},
  {Ghisellini}, {Maraschi}, {Tavecchio}, \& {Angelakis}}]{2010foschini}
{Foschini}, L., {Fermi/Lat Collaboration}, {Ghisellini}, G., {et~al.} 2010, in
  Accretion and Ejection in AGN: A Global View

\bibitem[{{Ganci} {et~al.}(2019){Ganci}, {Marziani}, {D'Onofrio}, {del Olmo},
  {Bon}, {Bon}, \& {Negrete}}]{2019ganci}
{Ganci}, V., {Marziani}, P., {D'Onofrio}, M., {et~al.} 2019, \aap, 630, A110

\bibitem[{{Goodrich}(1989)}]{1989goodrich1}
{Goodrich}, R.~W. 1989, \apj, 342, 224

\bibitem[{{Graham} \& {Driver}(2005)}]{2005graham1}
{Graham}, A.~W. \& {Driver}, S.~P. 2005, \pasa, 22, 118

\bibitem[{{Hamilton} {et~al.}(2021){Hamilton}, {Berton}, {Ant{\'o}n}, {Busoni},
  {Caccianiga}, {Ciroi}, {G{\"a}ssler}, {Georgiev}, {J{\"a}rvel{\"a}},
  {Komossa}, {Mathur}, \& {Rabien}}]{2021hamilton}
{Hamilton}, T.~S., {Berton}, M., {Ant{\'o}n}, S., {et~al.} 2021, \mnras, 504,
  5188

\bibitem[{{Heinz} \& {Sunyaev}(2003)}]{2003heinz}
{Heinz}, S. \& {Sunyaev}, R.~A. 2003, \mnras, 343, L59

\bibitem[{{Hu}(2008)}]{2008hu}
{Hu}, J. 2008, \mnras, 386, 2242

\bibitem[{{Jarrett}(2000)}]{2000jarrett}
{Jarrett}, T.~H. 2000, \pasp, 112, 1008

\bibitem[{{J{\"a}rvel{\"a}} {et~al.}(2022){J{\"a}rvel{\"a}}, {Dahale},
  {Crepaldi}, {Berton}, {Congiu}, \& {Antonucci}}]{2022jarvela}
{J{\"a}rvel{\"a}}, E., {Dahale}, R., {Crepaldi}, L., {et~al.} 2022, \aap, 658,
  A12

\bibitem[{{J{\"a}rvel{\"a}} {et~al.}(2018){J{\"a}rvel{\"a}},
  {L{\"a}hteenm{\"a}ki}, \& {Berton}}]{2018jarvela}
{J{\"a}rvel{\"a}}, E., {L{\"a}hteenm{\"a}ki}, A., \& {Berton}, M. 2018, \aap,
  619, A69

\bibitem[{{J{\"a}rvel{\"a}} {et~al.}(2017){J{\"a}rvel{\"a}},
  {L{\"a}hteenm{\"a}ki}, {Lietzen}, {Poudel}, {Hein{\"a}m{\"a}ki}, \&
  {Einasto}}]{2017jarvela}
{J{\"a}rvel{\"a}}, E., {L{\"a}hteenm{\"a}ki}, A., {Lietzen}, H., {et~al.} 2017,
  \aap, 606, A9

\bibitem[{{J{\"a}rvel{\"a}} {et~al.}(2015){J{\"a}rvel{\"a}},
  {L{\"a}hteenm{\"a}ki}, \& {Léon-Tavares}}]{2015jarvela}
{J{\"a}rvel{\"a}}, E., {L{\"a}hteenm{\"a}ki}, A., \& {Léon-Tavares}, J. 2015,
  \aap, 573, A76

\bibitem[{Järvelä(2018)}]{2018jarvelaphd}
Järvelä, E. 2018, Doctoral thesis, School of Electrical Engineering

\bibitem[{{Kellermann} {et~al.}(1989){Kellermann}, {Sramek}, {Schmidt},
  {Shaffer}, \& {Green}}]{1989kellermann1}
{Kellermann}, K.~I., {Sramek}, R., {Schmidt}, M., {Shaffer}, D.~B., \& {Green},
  R. 1989, \aj, 98, 1195

\bibitem[{{Komossa} {et~al.}(2006){Komossa}, {Voges}, {Xu}, {Mathur}, {Adorf},
  {Lemson}, {Duschl}, \& {Grupe}}]{2006komossa1}
{Komossa}, S., {Voges}, W., {Xu}, D., {et~al.} 2006, \aj, 132, 531

\bibitem[{{Kotilainen} {et~al.}(2016){Kotilainen}, {Le{\'o}n-Tavares},
  {Olgu{\'\i}n-Iglesias}, {Baes}, {An{\'o}rve}, {Chavushyan}, \&
  {Carrasco}}]{2016kotilainen}
{Kotilainen}, J.~K., {Le{\'o}n-Tavares}, J., {Olgu{\'\i}n-Iglesias}, A.,
  {et~al.} 2016, \apj, 832, 157

\bibitem[{{Krongold} {et~al.}(2001){Krongold}, {Dultzin-Hacyan}, \&
  {Marziani}}]{2001krongold}
{Krongold}, Y., {Dultzin-Hacyan}, D., \& {Marziani}, P. 2001, \aj, 121, 702

\bibitem[{{L{\"a}hteenm{\"a}ki} {et~al.}(2017){L{\"a}hteenm{\"a}ki},
  {J{\"a}rvel{\"a}}, {Hovatta}, {Tornikoski}, {Harrison}, {L{\'o}pez- Caniego},
  {Max-Moerbeck}, {Mingaliev}, {Pearson}, {Ramakrishnan}, {Readhead}, {Reeves},
  {Richards}, {Sotnikova}, \& {Tammi}}]{2017lahteenmaki1}
{L{\"a}hteenm{\"a}ki}, A., {J{\"a}rvel{\"a}}, E., {Hovatta}, T., {et~al.} 2017,
  \aap, 603, A100

\bibitem[{{L{\"a}hteenm{\"a}ki} {et~al.}(2018){L{\"a}hteenm{\"a}ki},
  {J{\"a}rvel{\"a}}, {Ramakrishnan}, {Tornikoski}, {Tammi}, {Vera}, \&
  {Chamani}}]{2018lahteenmaki}
{L{\"a}hteenm{\"a}ki}, A., {J{\"a}rvel{\"a}}, E., {Ramakrishnan}, V., {et~al.}
  2018, \aap, 614, L1

\bibitem[{{Laor}(2000)}]{2000laor1}
{Laor}, A. 2000, \apjl, 543, L111

\bibitem[{{Le{\'o}n Tavares} {et~al.}(2014){Le{\'o}n Tavares}, {Kotilainen},
  {Chavushyan}, {A{\~n}orve}, {Puerari}, {Cruz-Gonz{\'a}lez},
  {Pati{\~n}o-Alvarez}, {Ant{\'o}n}, {Carrami{\~n}ana}, {Carrasco}, {Guichard},
  {Karhunen}, {Olgu{\'\i}n-Iglesias}, {Sanghvi}, \& {Valdes}}]{2014leontavares}
{Le{\'o}n Tavares}, J., {Kotilainen}, J., {Chavushyan}, V., {et~al.} 2014,
  \apj, 795, 58

\bibitem[{{Lietzen} {et~al.}(2011){Lietzen}, {Hein{\"a}m{\"a}ki}, {Nurmi},
  {Liivam{\"a}gi}, {Saar}, {Tago}, {Takalo}, \& {Einasto}}]{2011lietzen1}
{Lietzen}, H., {Hein{\"a}m{\"a}ki}, P., {Nurmi}, P., {et~al.} 2011, \aap, 535,
  A21

\bibitem[{{MacLeod} {et~al.}(2012){MacLeod}, {Ivezi{\'c}}, {Sesar}, {de Vries},
  {Kochanek}, {Kelly}, {Becker}, {Lupton}, {Hall}, {Richards}, {Anderson}, \&
  {Schneider}}]{2012macleod}
{MacLeod}, C.~L., {Ivezi{\'c}}, {\v{Z}}., {Sesar}, B., {et~al.} 2012, \apj,
  753, 106

\bibitem[{{Marziani} {et~al.}(2018){Marziani}, {Dultzin}, {Sulentic}, {Del
  Olmo}, {Negrete}, {Mart{\'\i}nez-Aldama}, {D'Onofrio}, {Bon}, {Bon}, \&
  {Stirpe}}]{2018marziani}
{Marziani}, P., {Dultzin}, D., {Sulentic}, J.~W., {et~al.} 2018, Frontiers in
  Astronomy and Space Sciences, 5, 6

\bibitem[{{Mathur}(2000)}]{2000mathur}
{Mathur}, S. 2000, \mnras, 314, L17

\bibitem[{{Mathur} {et~al.}(2012){Mathur}, {Fields}, {Peterson}, \&
  {Grupe}}]{2012mathur}
{Mathur}, S., {Fields}, D., {Peterson}, B.~M., \& {Grupe}, D. 2012, \apj, 754,
  146

\bibitem[{{Ohta} {et~al.}(2007){Ohta}, {Aoki}, {Kawaguchi}, \&
  {Kiuchi}}]{2007ohta1}
{Ohta}, K., {Aoki}, K., {Kawaguchi}, T., \& {Kiuchi}, G. 2007, \apjs, 169, 1

\bibitem[{{Olgu{\'\i}n-Iglesias} {et~al.}(2020){Olgu{\'\i}n-Iglesias},
  {Kotilainen}, \& {Chavushyan}}]{2020olguiglesias}
{Olgu{\'\i}n-Iglesias}, A., {Kotilainen}, J., \& {Chavushyan}, V. 2020, \mnras,
  492, 1450

\bibitem[{{Olgu{\'\i}n-Iglesias} {et~al.}(2017){Olgu{\'\i}n-Iglesias},
  {Kotilainen}, {Le{\'o}n Tavares}, {Chavushyan}, \&
  {A{\~n}orve}}]{2017olguiglesias}
{Olgu{\'\i}n-Iglesias}, A., {Kotilainen}, J.~K., {Le{\'o}n Tavares}, J.,
  {Chavushyan}, V., \& {A{\~n}orve}, C. 2017, \mnras, 467, 3712

\bibitem[{{Orban de Xivry} {et~al.}(2011){Orban de Xivry}, {Davies},
  {Schartmann}, {Komossa}, {Marconi}, {Hicks}, {Engel}, \&
  {Tacconi}}]{2011orban}
{Orban de Xivry}, G., {Davies}, R., {Schartmann}, M., {et~al.} 2011, \mnras,
  417, 2721

\bibitem[{{Osterbrock} \& {Pogge}(1985)}]{1985osterbrock1}
{Osterbrock}, D.~E. \& {Pogge}, R.~W. 1985, \apj, 297, 166

\bibitem[{{Paliya} {et~al.}(2018){Paliya}, {Ajello}, {Rakshit}, {Mandal},
  {Baughmann}, {Stalin}, {Kaur}, \& {Hartmann}}]{2018paliya}
{Paliya}, V.~S., {Ajello}, M., {Rakshit}, S., {et~al.} 2018, \apjl, 853, L2

\bibitem[{{Paliya} {et~al.}(2019){Paliya}, {Parker}, {Jiang}, {Fabian},
  {Brenneman}, {Ajello}, \& {Hartmann}}]{2019paliya}
{Paliya}, V.~S., {Parker}, M.~L., {Jiang}, J., {et~al.} 2019, \apj, 872, 169

\bibitem[{{Peng} {et~al.}(2010){Peng}, {Ho}, {Impey}, \& {Rix}}]{2010peng1}
{Peng}, C.~Y., {Ho}, L.~C., {Impey}, C.~D., \& {Rix}, H.-W. 2010, \aj, 139,
  2097

\bibitem[{{Peterson}(2011)}]{2011peterson}
{Peterson}, B.~M. 2011, ArXiv:1109.4181 [\eprint[arXiv]{1109.4181}]

\bibitem[{{Povi{\'c}} {et~al.}(2012){Povi{\'c}}, {S{\'a}nchez-Portal},
  {P{\'e}rez Garc{\'{\i}}a}, {Bongiovanni}, {Cepa}, {Huertas-Company},
  {Lara-L{\'o}pez}, {Fern{\'a}ndez Lorenzo}, {Ederoclite}, {Alfaro},
  {Casta{\~n}eda}, {Gallego}, {Gonz{\'a}lez-Serrano}, \&
  {Gonz{\'a}lez}}]{2012povic1}
{Povi{\'c}}, M., {S{\'a}nchez-Portal}, M., {P{\'e}rez Garc{\'{\i}}a}, A.~M.,
  {et~al.} 2012, \aap, 541, A118

\bibitem[{{Romano} {et~al.}(2018){Romano}, {Vercellone}, {Foschini},
  {Tavecchio}, {Landoni}, \& {Kn{\"o}dlseder}}]{2018romano}
{Romano}, P., {Vercellone}, S., {Foschini}, L., {et~al.} 2018, \mnras, 481,
  5046

\bibitem[{{Salo} {et~al.}(2015){Salo}, {Laurikainen}, {Laine}, {Comer{\'o}n},
  {Gadotti}, {Buta}, {Sheth}, {Zaritsky}, {Ho}, {Knapen}, {Athanassoula},
  {Bosma}, {Laine}, {Cisternas}, {Kim}, {Mu{\~n}oz-Mateos}, {Regan}, {Hinz},
  {Gil de Paz}, {Menendez-Delmestre}, {Mizusawa}, {Erroz-Ferrer}, {Meidt}, \&
  {Querejeta}}]{2015salo}
{Salo}, H., {Laurikainen}, E., {Laine}, J., {et~al.} 2015, \apjs, 219, 4

\bibitem[{{Salom{\'e}} {et~al.}(2021){Salom{\'e}}, {Longinotti}, {Krongold},
  {Feruglio}, {Chavushyan}, {Vega}, {Garc{\'\i}a-Burillo}, {Fuente},
  {Olgu{\'\i}n-Iglesias}, {Pati{\~n}o-{\'A}lvarez}, {Puerari}, \&
  {Robleto-Or{\'u}s}}]{2021salome}
{Salom{\'e}}, Q., {Longinotti}, A.~L., {Krongold}, Y., {et~al.} 2021, \mnras,
  501, 219

\bibitem[{{Sani} {et~al.}(2010){Sani}, {Lutz}, {Risaliti}, {Netzer}, {Gallo},
  {Trakhtenbrot}, {Sturm}, \& {Boller}}]{2010sani1}
{Sani}, E., {Lutz}, D., {Risaliti}, G., {et~al.} 2010, \mnras, 403, 1246

\bibitem[{{Sbarrato} {et~al.}(2021){Sbarrato}, {Ghisellini}, {Giovannini}, \&
  {Giroletti}}]{2021sbarrato}
{Sbarrato}, T., {Ghisellini}, G., {Giovannini}, G., \& {Giroletti}, M. 2021,
  \aap, 655, A95

\bibitem[{{Spergel} {et~al.}(2007){Spergel}, {Bean}, {Dor{\'e}}, {Nolta},
  {Bennett}, {Dunkley}, {Hinshaw}, {Jarosik}, {Komatsu}, {Page}, {Peiris},
  {Verde}, {Halpern}, {Hill}, {Kogut}, {Limon}, {Meyer}, {Odegard}, {Tucker},
  {Weiland}, {Wollack}, \& {Wright}}]{2007spergel1}
{Spergel}, D.~N., {Bean}, R., {Dor{\'e}}, O., {et~al.} 2007, \apjs, 170, 377

\bibitem[{{Stetson}(1987)}]{1987stetson}
{Stetson}, P.~B. 1987, \pasp, 99, 191

\bibitem[{{Tortosa} {et~al.}(2022){Tortosa}, {Ricci}, {Tombesi}, {Ho}, {Du},
  {Inayoshi}, {Wang}, {Shangguan}, \& {Li}}]{2022tortosa}
{Tortosa}, A., {Ricci}, C., {Tombesi}, F., {et~al.} 2022, \mnras, 509, 3599

\bibitem[{{Urrutia} {et~al.}(2008){Urrutia}, {Lacy}, \&
  {Becker}}]{2008urrutia1}
{Urrutia}, T., {Lacy}, M., \& {Becker}, R.~H. 2008, \apj, 674, 80

\bibitem[{{van de Ven} \& {Fathi}(2010)}]{2010vandeven1}
{van de Ven}, G. \& {Fathi}, K. 2010, \apj, 723, 767

\bibitem[{{Vietri} {et~al.}(2022){Vietri}, {J{\"a}rvel{\"a}}, {Berton},
  {Ciroi}, {Congiu}, {Chen}, \& {Di Mille}}]{2022vietri}
{Vietri}, A., {J{\"a}rvel{\"a}}, E., {Berton}, M., {et~al.} 2022, \aap, 662,
  A20

\bibitem[{{Wang} {et~al.}(2016){Wang}, {Du}, {Hu}, {Bai}, {Wang}, {Yi}, {Wang},
  {Zhang}, {Xin}, {Lun}, {Chang}, \& {Fan}}]{2016wang}
{Wang}, F., {Du}, P., {Hu}, C., {et~al.} 2016, \apj, 824, 149

\bibitem[{{Winkel} {et~al.}(2022){Winkel}, {Husemann}, {Davis},
  {Smirnova-Pinchukova}, {Bennert}, {Combes}, {Gaspari}, {Jahnke}, {Neumann},
  {O'Dea}, {P{\'e}rez-Torres}, {Singha}, {Tremblay}, \& {Rix}}]{2022winkel}
{Winkel}, N., {Husemann}, B., {Davis}, T.~A., {et~al.} 2022, \aap, 663, A104

\bibitem[{{Yuan} {et~al.}(2008){Yuan}, {Zhou}, {Komossa}, {Dong}, {Wang}, {Lu},
  \& {Bai}}]{2008yuan}
{Yuan}, W., {Zhou}, H.~Y., {Komossa}, S., {et~al.} 2008, \apj, 685, 801

\bibitem[{{Zamanov} {et~al.}(2002){Zamanov}, {Marziani}, W., {Calvani},
  {Dultzin-Hacyan}, \& {Bachev}}]{2002zamanov}
{Zamanov}, R., {Marziani}, P., W., S.~J., {et~al.} 2002, \apj, 576, 9

\bibitem[{{Zhou} {et~al.}(2006){Zhou}, {Wang}, {Yuan}, {Lu}, {Dong}, {Wang}, \&
  {Lu}}]{2006zhou}
{Zhou}, H., {Wang}, T., {Yuan}, W., {et~al.} 2006, \apj, 166, 128

\end{thebibliography}

\begin{appendix}

\section{Compromised fits}
\label{compromised-fits}
We were unable to properly model two of our sources: 6dFGS gJ044739.0-040330 and 6dFGS gJ084510.2-073205. Both of these sources suffered from at least one of the following: high redshift or high complexity. We plotted a radial surface brightness plot for those sources we were able to obtain a physically meaningful plot of. An intensity image is given of the possible components. 

\subsection{6dFGS gJ044739.0-040330}

This is one of the three sources that have not been detected at 37~GHz or any other radio frequency. For the PSF we chose a star directly from the image. We tried to model the source with one PSF and two S\'{e}rsic functions. The best fit values can be found in Table~\ref{tab:j0447}. We also modelled a nearby source inside the fitting region with a PSF and a S\'{e}rsic component. 

Our source and the nearby source are so close to each other that it appeared this could be a merger. To study the possible companion galaxy, we applied for additional observing time with Andalucia Faint Object Spectrograph and Camera (ALFOSC) at NOT to obtain its spectrum and redshift (proposal ID 63-409, PI I. Björklund). It turned out that the companion is actually a star in the Galaxy and not a physical companion of 6dFGS gJ044739.0-040330.

\begin{figure}
\centering
\includegraphics[width=0.5\textwidth]{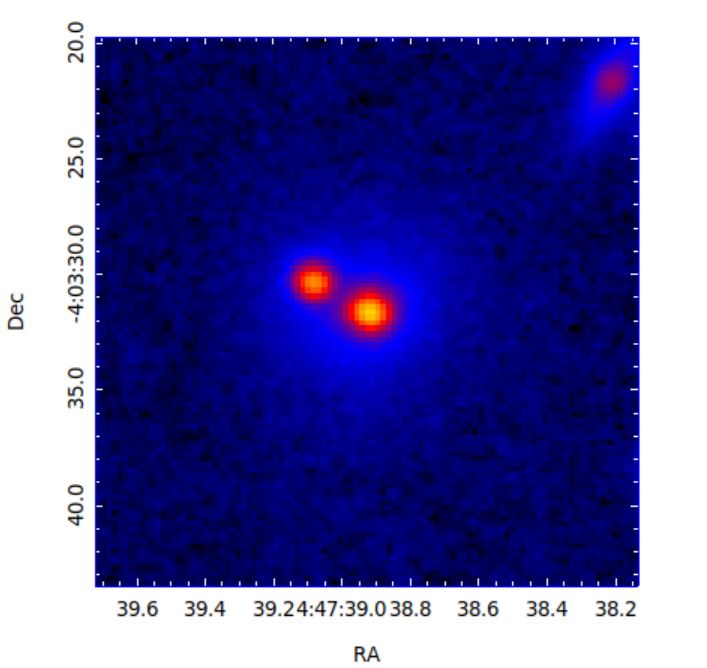}
\caption{\textit{Ks}- band images of SDSS 6dFGS gJ044739.0-040330. Observed image of the source. The FoV is 23.4\arcsec $/$ 34.5~kpc in the image.}
\label{fig:j0447comps}
\end{figure}

\begin{figure}
\centering
\includegraphics[width=0.5\textwidth]{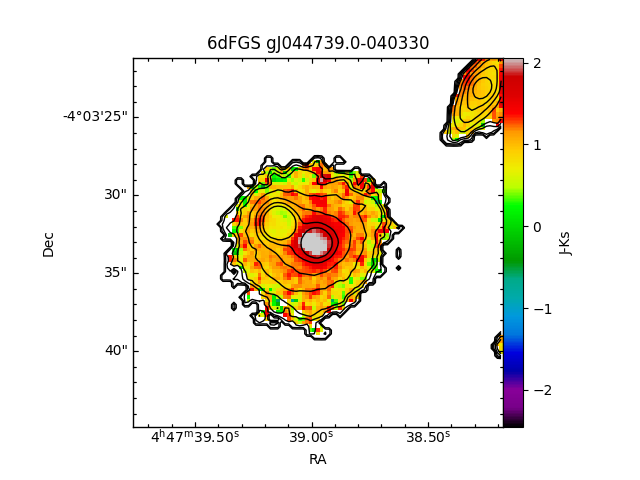}
\caption{\textit{J - $Ks$} colour map of 6dFGS gJ044739.0-040330.}
\label{fig:j0447color}
\end{figure}

Because our source is too close to the star, we were unable to obtain a proper fit, and therefore only present the observed image, seen in Fig.~\ref{fig:j0447comps}. The \textit{J - $Ks$} colour map of the source is shown in Fig.~\ref{fig:j0447color}. The central region of 6dFGS gJ044739.0-040330 has \textit{J - $Ks$} magnitude of approximately 2, meaning that it is very red, suggesting dust extinction. Rest of the galaxy has \textit{J - $Ks$} magnitude close to 1. There is no evident structure visible in the map. The companion source is, on the other hand, very yellow (\textit{J - $Ks$}$\sim$0.5), unlike the AGN, supporting the conclusion that it is a star and not part of the galaxy.

\begin{table}[H]
\caption[]{Best fit parameters of 6dFGS gJ044739.0-040330. $\chi^2_{\nu}$ = 1.816.}
\centering
\begin{tabular}{l l l l l l l}
\hline\hline
Function       & Mag                             & $r_\text{e}$                          & $n$                              & Axial & PA  & Notes             \\
             &                                   & (kpc)                            &                                  & ratio & (\textdegree) &  \\ \hline
PSF 1          & 13.93  &                                  &                                  &       &               &   \\

S\'{e}rsic 1  & 14.12 & 4.84  & 0.52  & 0.82  & 33.50   \\

S\'{e}rsic 2  & 13.70 & 0.60  & 0.01  & 0.45  & 5.49    \\

S\'{e}rsic 3  & 15.43 & 15.65 & 0.24 & 0.27  & -85.08 & Nearby source \\

PSF  2        & 13.84  &                                  &                                  &       &               & Nearby source  \\

  \hline   
\end{tabular}
\tablefoot{Columns: (1) Function used for modelling, (2) Magnitude, (3) Effective radius, (4) S\'{e}rsic index, (5) Axial ratio, (6) Position angle, (7) Additional notes.}
\label{tab:j0447}
\end{table}

\subsection{6dFGS gJ084510.2-073205}

\begin{figure}
\centering
\adjustbox{valign=t}{\begin{minipage}{0.35\textwidth}
\centering
\includegraphics[width=1\textwidth]{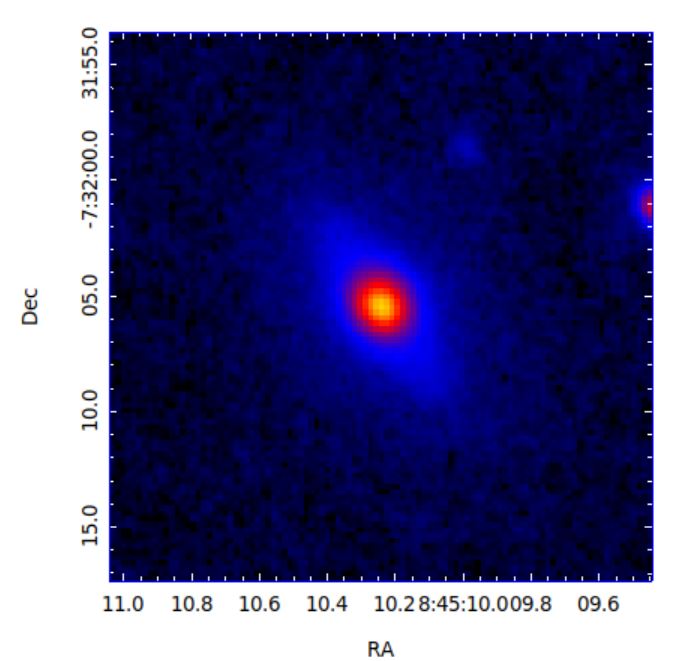}
\end{minipage}}
\adjustbox{valign=t}{\begin{minipage}{0.31\textwidth}
\centering
\includegraphics[width=0.95\textwidth]{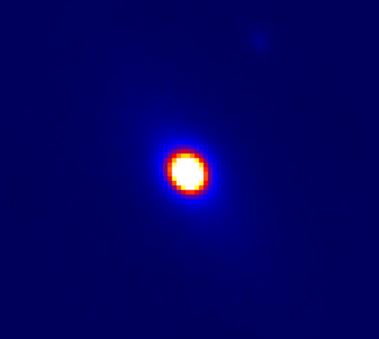}
\end{minipage}}
\adjustbox{valign=t}{\begin{minipage}{0.31\textwidth}
\centering
\includegraphics[width=0.95\textwidth]{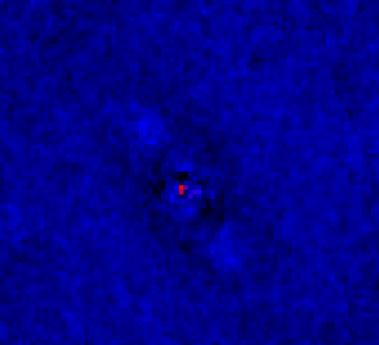}
\end{minipage}}
\hfill
    \caption{\textit{Ks}- band images of 6dFGS gJ084510.2-073205. The FoV is 23.4\arcsec     $/$ 42.8~kpc in all images. \emph{Top panel:} observed image, \emph{middle panel:} model image, and \emph{bottom panel:} residual image, smoothed over 3px.}  \label{fig:j0845}
\end{figure}

\begin{table}[H]
\caption[]{Best fit parameters of 6dFGS gJ084510.2-073205. $\chi^2_{\nu}$ = 0.808.}
\centering
\begin{tabular}{l l l l l l l}
\hline\hline
Function       & Mag                             & $r_\text{e}$                          & $n$                              & Axial & PA  & Notes             \\
             &                                   & (kpc)                            &                                  & ratio & (\textdegree) &  \\ \hline
PSF           & 14.50  &                                  &                                  &       &               &   \\

S\'{e}rsic 1  & 14.59 & 5.14  & 0.96 & 0.38  & -56.03  \\

S\'{e}rsic 2  & 14.98 & 26.51 & 0.04  & 0.25  & -88.71   \\

  \hline   
\end{tabular}
\tablefoot{Columns: (1) Function used for modelling, (2) Magnitude, (3) Effective radius, (4) S\'{e}rsic index, (5) Axial ratio, (6) Position angle, (7) Additional notes.}
\label{tab:j0845}
\end{table}

This is another of the three sources not detected at 37~GHz, or, in fact, any other radio frequency either. A visual inspection of this source suggests that the source has at least a disk and a bulge component. The PSF model for this source was created from a star directly from the image. The best fit we were able to obtain had one PSF component and two S\'{e}rsic components. The values are listed in Table~\ref{tab:j0845}. The obtained parameters, however, were not physically meaningful. This was unexpected, as the redshift of this source is only $\sim$0.10 and the seeing was tolerable, even though not particularly good. 

\begin{figure}
\centering
\includegraphics[width=0.5\textwidth]{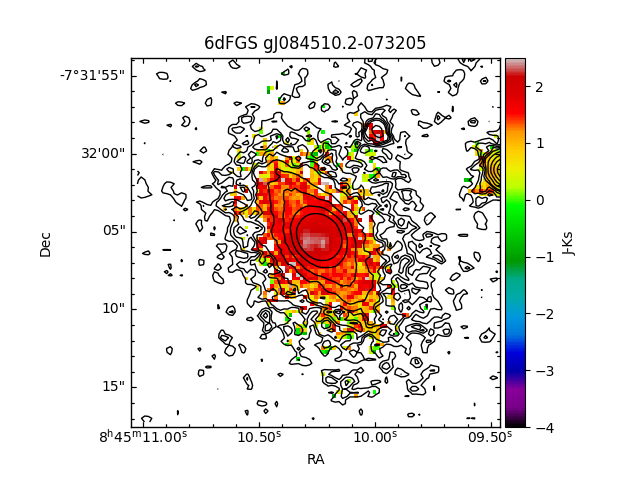}
\caption{\textit{J - $Ks$} colour map of 6dFGS gJ084510.2-073205. }
\label{fig:j0845color}
\end{figure}

\begin{figure}
\centering
\includegraphics[width=0.5\textwidth]{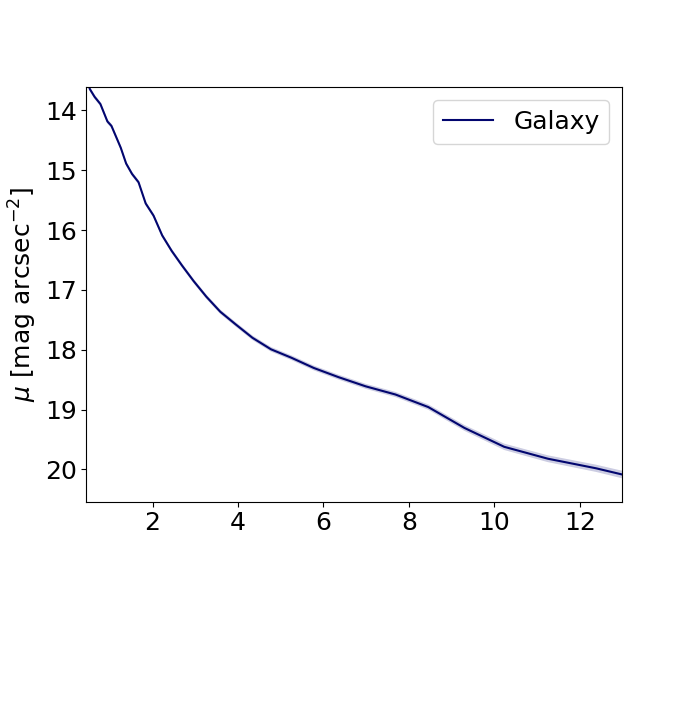}
\caption{Radial surface brightness profile plot of 6dFGS gJ084510.2-073205. The galaxy component is modelled with a blue line. The shaded area surrounding the profile curve depicts the error linked to the component.}
\label{fig:j0845comps}
\end{figure}

The observed, model, and the residual image of the source can be seen in Fig.~\ref{fig:j0845}. There are a lot of residuals left and the model image is very simplified compared to the observed image, meaning that we were unable to model the source adequately. The residuals, especially in the nucleus, and the fit parameters hint at a possible issue with the PSF. The radial surface brightness profile plot of the galaxy is shown in Fig.~\ref{fig:j0845comps} and the \textit{J - $Ks$} colour map in Fig.~\ref{fig:j0845color}. This galaxy is extremely red (\textit{J - $Ks$} magnitude $\sim$2), with very little of anything else visible in the map. Although this galaxy is large, the poor data quality makes it difficult to distinguish any clear structures in the colour map. 

\end{appendix}

\end{document}